\documentclass[twocolumn,twocolappendix]{aastex631}

\usepackage{csvsimple}
\usepackage{amsmath}
\usepackage{booktabs}
\usepackage[flushleft]{threeparttable}
\usepackage{hyperref}
\usepackage{xcolor}
\usepackage{comment}
\usepackage{enumitem}
\setcitestyle{authoryear,open={(},close={)}}

\usepackage{tikz} 

\def\HI{\rm H\,\textsc{i}}
\def\HIspace{\rm H\,\textsc{i} }

\def\Haspace{H$\alpha$ }
\def\kms{$\rm km~s^{-1}$}

\graphicspath{{./}{figures/}}

\begin{document}

\title{Baryonic Masses and Properties of Gaseous Satellite Galaxies}

\correspondingauthor{Jingyao Zhu}
\email{jingyao.zhu@columbia.edu}

\author[0000-0002-9001-6713]{Jingyao Zhu}
\affiliation{Department
of Astronomy, Columbia University, New York, NY 10027, USA}

\author[0000-0002-8320-2198]{Yasmeen Asali}
\affiliation{Department of Astronomy, Yale University, New Haven, CT 06520, USA}

\author[0000-0002-1129-1873]{Mary E. Putman}
\affiliation{Department
of Astronomy, Columbia University, New York, NY 10027, USA}

\author[0000-0002-5300-2486]{Tobias Westmeier}
\affiliation{ICRAR, The University of Western Australia, 35~Stirling Highway, Crawley WA 6009, Australia}

\author[0000-0001-8957-4518]{W.J.G. de Blok}
\affiliation{Netherlands Institute for Radio Astronomy (ASTRON), Oude Hoogeveensedijk 4, 7991 PD Dwingeloo, the Netherlands}
\affiliation{Dept.\ of Astronomy, Univ.\ of Cape Town, Private Bag X3, Rondebosch 7701, South Africa}
\affiliation{Kapteyn Astronomical Institute, University of Groningen, PO Box 800, 9700 AV Groningen, The Netherlands}

\author[0000-0002-7625-562X]{Barbara Catinella}
\affiliation{ICRAR, The University of Western Australia, 35~Stirling Highway, Crawley WA 6009, Australia}

\author[0000-0003-3523-7633]{Nathan Deg}
\affiliation{Department of Physics, Engineering Physics and Astronomy, Queen's University, Kingston, ON, K7l 3N6, Canada}

\author[0000-0002-0196-5248]{Bi-Qing For}
\affiliation{ICRAR, The University of Western Australia, 35~Stirling Highway, Crawley WA 6009, Australia}

\author[0000-0002-7573-555X]{Dane Kleiner}
\affiliation{Netherlands Institute for Radio Astronomy (ASTRON), Oude Hoogeveensedijk 4, 7991 PD Dwingeloo, the Netherlands}

\author[0000-0003-4844-8659]{Karen Lee-Waddell}
\affiliation{Australian SKA Regional Centre (AusSRC), The University of Western Australia, 35~Stirling Highway, Crawley WA 6009, Australia}
\affiliation{CSIRO Space and Astronomy, PO Box 1130, Bentley WA 6102, Australia}
\affiliation{ICRAR, Curtin University, Bentley, WA 6102, Australia}

\author[0000-0002-9930-1844]{Filippo M. Maccagni}
\affiliation{Osservatorio Astronomico di Cagliari, via della Scienza 5, 09047 Selargius (CA), Italy}
\affiliation{Wits Centre for Astrophysics, School of Physics, University of the Witwatersrand, 1 Jan Smuts Avenue, 2000, Johannesburg, South Africa}

\author[0000-0001-7996-7860]{D.J. Pisano}
\affiliation{University of Cape Town}

\author[0009-0002-5445-7770]{Austin X. Shen}
\affiliation{CSIRO Space and Astronomy, PO Box 1130, Bentley WA 6102, Australia}

\author[0000-0002-0956-7949]{Kristine Spekkens}
\affiliation{Department of Physics, Engineering Physics and Astronomy, Queen's University, Kingston, ON, K7l 3N6, Canada}

\author[0000-0002-8057-0294]{Lister Staveley-Smith}
\affiliation{ICRAR, The University of Western Australia, 35~Stirling Highway, Crawley WA 6009, Australia}

\begin{abstract}
We present a sample of 127 gas-bearing dwarf galaxies around 56 late-type host galaxies within 30 Mpc using 21-cm \HIspace data from the WALLABY, MHONGOOSE, and ALFALFA surveys. We characterize the environment of each dwarf galaxy based on its host galaxy halo and derive optical properties using the DESI Legacy Surveys for 110. The gaseous satellites span $\log (M_{\HI}/M_{\odot}) = 5.8-9.7$ and $\log (M_{\star}/M_{\odot}) = 5.6-10.0$, with a median velocity line-width of $W_{50}=40$ \kms, comparable to the Local Group gaseous dwarf galaxies. We assess the \HIspace mass sensitivity of the data by injecting model dwarf galaxies and find $M_{\HI,lim} \approx 10^{6.7} M_{\odot}$ for WALLABY and $M_{\HI,lim} \approx 10^{5.4} M_{\odot}$ for MHONGOOSE at 10 Mpc. Our sample has lower average atomic gas fractions ($M_{\HI}/M_{\star}$) than previous gaseous dwarf samples in field and satellite environments, although this offset partly reflects differences in \HIspace sensitivity. The abundance of gaseous satellites per host is low and increases with host mass: $0-2$ for dwarf centrals and $0-5$ for Milky Way-mass spiral hosts. These numbers are consistent with the Milky Way, M31, and star-forming satellite abundances from recent deep optical surveys. The inferred quenched fractions and gas-depleted satellites indicate that environmental quenching is effective in Milky Way-mass hosts, likely driven by gas stripping processes.

\end{abstract}

\section{Introduction}\label{sec:intro}

The lowest-mass ``dwarf" ($M_{\star} \leq 10^{9}~M_{\odot}$) galaxies are the most abundant type of galaxies in the Universe \citep{binggeli_abundance_1990} and the building blocks for larger systems 
\citep{blumenthal_formation_1984,white_galaxy_1991,navarro_assembly_1995}. Dwarf galaxies in low-density environments often contain more gas -- the fuel for star formation -- than stars \citep{schombert_gas_2001,geha_baryon_2006,scholte_atomic_2024}; and collectively, the baryonic matter makes up only a small percentage of the total mass compared to the dark matter \citep{garrison-kimmel_organized_2017,read_stellar_2017,behroozi_universemachine_2019,munshi_quantifying_2021,lelli_gas_2022}. This suggests that dwarf galaxies are inefficient in forming stars, which emphasizes the need to include baryonic processes, such as stellar feedback and photoionization, in dwarf galaxy evolution models \citep{somerville_physical_2015,wright_baryon_2024}. Dwarf galaxies are particularly sensitive to baryonic processes because of their low masses and shallow gravitational potentials (see reviews by \citealt{bullock_small-scale_2017,sales_baryonic_2022}).

External processes, such as galaxy environment, also play a crucial role in shaping a dwarf galaxy's baryonic content. This is clearly demonstrated by dwarf galaxies in our own Local Group (LG; \citealt{mcconnachie_observed_2012}). Over 90\% of the satellite dwarf galaxies of the Milky Way (MW) or M31 are devoid of gas \citep{grcevich_h_2009,spekkens_dearth_2014,putman_gas_2021} and quenched of star formation \citep{wetzel_rapid_2015}. Within the MW/M31 dark matter halos, only $2-4$ (usually the most massive) dwarf satellites still contain gas, while dozens more are gas-poor \citep{putman_gas_2021}. This is in direct contrast with the Local Group ``field" dwarf galaxies unassociated with either host \citep{mcconnachie_observed_2012} and the isolated dwarf galaxies outside the Local Group \citep{geha_stellar_2012,karachentsev_morphological_2018}, which tend to be gas-rich and star-forming. The MW/M31 environment removes gas in the dwarf satellites and ultimately quenches their star formation.

Several physical mechanisms can lead to gas stripping and the eventual quenching of satellite galaxies (see review by \citealt{cortese_dawes_2021}), and their relative strengths have been examined in cosmological hydrodynamical simulations of MW-like systems (e.g., \citealt{fillingham_taking_2015,simpson_quenching_2018,akins_quenching_2021,simons_figuring_2020,samuel_extinguishing_2022,engler_satellites_2023,christensen_environment_2024,rodriguez-cardoso_agora_2025}). Across all simulations, the leading mechanism for dwarf satellite gas loss is ram pressure stripping (RPS; \citealt{gunn_infall_1972}), the direct removal of satellite gas by the host gaseous halo, quenching the dwarf satellites within $1-2$ Gyr timescales. 
Gas loss via RPS is often accompanied by tidal forces from the host galaxy, which, under close and/or repeated encounters, can remove gas, stars, and dark matter outside the tidal radii of the satellites \citep{mateo_velocity_2008,penarrubia_tidal_2008,riley_auriga_2025} and disturb the internal structure of the satellites \citep{mayer_tidal_2001,mayer_simultaneous_2006,serra_meerkat_2023}. Independent of environment, cosmic reionization \citep{gnedin_cosmological_2000,somerville_can_2002} can directly quench the smallest ``ultra-faint" ($M_{\star} \leq 10^{5}~M_{\odot}$) dwarf galaxies \citep{brown_quenching_2014,weisz_star_2014-1,bullock_small-scale_2017}.

Theoretical models are often tailored to MW-like systems because of the strong observational constraints available. This motivates the question of how representative the MW satellite population is, which is investigated by a large and growing body of literature on satellites around external galaxies. Such observations include targeted deep searches of individual host galaxies (e.g., \citealt{spencer_survey_2014,smercina_lonely_2018,smercina_saga_2020,muller_dwarf_2019,muller_too-many-dwarf-galaxy-satellites_2024,bennet_m101_2019,garling_search_2021,mutlu-pakdil_hubble_2022,mutlu-pakdil_faint_2024,davis_lbt_2024}), as well as large samples that enable statistical comparisons across host galaxies, e.g., the Satellites Around Galactic Analogs (SAGA) survey \citep{mao_saga_2024,geha_saga_2024,wang_saga_2024} and the Exploration of Local VolumE Satellites (ELVES) survey \citep{carlsten_exploration_2022}. These studies have found that the Local Group and its satellite properties are typical (i.e., within $1\sigma$ host-to-host scatter) among external analogs \citep{carlsten_exploration_2022,mao_saga_2024}, although some tension remains in the satellite quenched fraction and its dependence on satellite stellar mass \citep{karunakaran_quenched_2023,geha_saga_2024}. Host-to-host scatter in satellite populations is non-negligible, which can originate from the mass \citep{carlsten_exploration_2022}, merger history \citep{smercina_relating_2022}, environment \citep{mutlu-pakdil_faint_2024}, and morphological type of the host galaxy \citep{trentham_dwarf_2009,crnojevic_faint_2019}. On a distinctively smaller mass scale, new surveys have also started to characterize satellites of Magellanic-mass dwarf hosts \citep{hunter_identifying_2025,li_elves-dwarf_2025}, which constrains the dependence of satellite quenching on host masses across a broader range \citep{jahn_effects_2022,nadler_symphony_2023}.

It remains unclear how satellite baryons, particularly the gaseous ones, are shaped by environment beyond the Local Group. Gas is directly subject to RPS; it is an ideal tracer for galaxy environment \citep{haynes_influence_1984} that is unconstrained in optical satellite studies. Previous work has found a low number of gas-rich dwarf satellites per spiral host \citep{karunakaran_h_2022,zhu_census_2023} or Local Group analog \citep{pisano_h_2011}, indicating an effective satellite stripping picture similar to that in the MW/M31 \citep{putman_gas_2021}. However, the data were insufficient to simultaneously constrain the baryonic (\HIspace and stellar) masses and properties of these satellites. In this work, we present a new sample of gaseous satellite galaxies around late-type hosts using data from two recent \HIspace surveys, WALLABY (the Widefield ASKAP L-band Legacy All-sky Blind surveY; \citealt{koribalski_wallaby_2020,westmeier_wallaby_2022,murugeshan_wallaby_2024}) and MHONGOOSE (MeerKAT HI Observations of Nearby Galactic Objects - Observing Southern Emitters; \citealt{de_blok_mhongoose_2024}). We derive the baryonic properties of the dwarf galaxies, assess the impact of environment on dwarf galaxy baryons, and compare the satellite population with those of the MW, M31, and external analogs.

The paper is organized as follows. Section \ref{sec:data} describes the \HIspace surveys used in this work. Section \ref{sec:methods} presents the host selection, \HIspace source finding, sensitivity assessment, stellar mass calibration, and satellite classification methodology. Section \ref{sec:results} presents the results, covering the satellite \HIspace properties (\S \ref{subsec:sat_hi_properties}), gas-and-stellar mass relations (\S \ref{subsec:sat_gas_star_properties}), gas depletion as a function of host mass (\S \ref{subsec:sat_deplete_host_mass}), and the abundance of gaseous satellites per host (\S \ref{sec:host_sat_abundance}). Section \ref{subsec:discussion_literature_compare} compares our \HIspace sample with deep optical studies, and Section \ref{sec:gas_loss_discuss} compares with environmental gas stripping theory. Finally, Section \ref{sec:summary} summarizes our findings. We adopt $H_{0}=70~ \rm km/s/Mpc$ throughout.

\section{Data}\label{sec:data}

This paper uses two recent 21-cm \HIspace surveys, WALLABY and MHONGOOSE, as the data sources for nearby \HI-bearing dwarf satellite galaxies. WALLABY \citep{koribalski_wallaby_2020} is an ASKAP wide-field survey. So far, Phase 1 \citep{westmeier_wallaby_2022} and Phase 2 \citep{murugeshan_wallaby_2024} of the WALLABY pilot surveys have been completed, each covering $\sim$180 deg$^{2}$ regions of the sky (in a few $\sim 30-120$ deg$^{2}$ footprints), and collectively detected over 2000 \HIspace sources up to $z \approx 0.09$. The data we use have a spatial resolution of 30\arcsec~and a channel spacing of 3.91 \kms. The typical rms noise per channel for the WALLABY pilot surveys is $\sigma \approx 1.75 \rm ~mJy$ per beam \citep{westmeier_wallaby_2022,murugeshan_wallaby_2024}. The ongoing full survey observations will operate at a similar sensitivity to Pilot Phase 2, but will eventually cover 14000 deg$^{2}$ of the southern sky. In this work, we use data from the two publicly available pilot surveys. We describe our selection method for nearby spiral host galaxies in Section \ref{subsec:host_sample} below.

MHONGOOSE is a deep MeerKAT survey of thirty star-forming galaxies in the southern sky within distances of 30 Mpc \citep{de_blok_mhongoose_2024}. Each target galaxy has a uniform angular coverage of 1.5$\deg$ in diameter. With the primary scientific goal of studying low column density accretion, the target galaxies in MHONGOOSE were selected to have a range of masses and not to reside within the densest environments. This selection makes the MHONGOOSE targets ideal host galaxies for satellite studies; their medium- to low-density environments reduce complications from galaxy clusters or massive groups, 
while the deep sensitivity allows the detection of low-mass gaseous satellites. In this work, we utilize the full-depth MHONGOOSE data cubes with a 55-hour integration time on each host galaxy field, which contain new low-mass companion galaxies relative to the 5.5-hour single-track data listed in Table 7 of \cite{de_blok_mhongoose_2024}. We describe our \HIspace source finding method to uncover these low-mass sources in Section \ref{subsec:method_source_finding}. MHONGOOSE data cubes are available\footnote{\url{https://mhongoose.astron.nl}} in a range of spatial resolutions; the cubes we use in this work (\texttt{r15\_t00}\footnote{At this resolution, the beam size is most comparable to WALLABY, and the high column sensitivity is ideal for the detection of faint dwarf galaxies; see Table 4 of \cite{de_blok_mhongoose_2024}.}) have a spatial resolution of 30\arcsec~and a channel spacing of 1.38 \kms. The typical rms noise per channel at this resolution is $\sigma \approx 0.154 \rm ~mJy$ per beam \citep{de_blok_mhongoose_2024}.

Because of the scope and strategies of the individual surveys, WALLABY and MHONGOOSE differ by approximately an order of magnitude in \HIspace sensitivity (see, e.g., Fig. 2 in \citealt{de_blok_mhongoose_2024}). This leads to varying abilities in recovering the lowest-mass dwarf galaxy population around each host galaxy. In Section \ref{subsec:method_sensitivity} below, we describe our method to systematically assess the dwarf satellite mass completeness for all host galaxies in our sample across the different surveys.

When presenting results, we often include the sample from our previous work of gaseous dwarf satellites from the ALFALFA survey (\citealt{zhu_census_2023}; 46 \HI-bearing dwarf galaxies around 15 spiral hosts). The ALFALFA survey is a wide-field, single-dish survey of $\sim$31500 extragalactic \HIspace sources, mapping $\sim$7000 deg$^{2}$ of the northern sky accessible to the Arecibo telescope \citep{giovanelli_arecibo_2005-1,haynes_arecibo_2011,haynes_arecibo_2018}. The average rms noise per channel from the final survey catalog is $\sigma \approx 2.41 \rm ~mJy$ per beam \citep{haynes_arecibo_2018}. The ALFALFA survey has a spatial resolution of 4\arcmin\ and a spectral resolution of 10 \kms\ after Hanning smoothing, both coarser than those of WALLABY and MHONGOOSE. We also extensively compare with the \HI-bearing dwarf galaxies within 2 Mpc in the Local Group \citep{putman_gas_2021}, and with gaseous dwarf galaxy samples in various environments in the literature (Section \ref{subsec:sat_gas_star_properties}).

\section{Methodology} \label{sec:methods}

\subsection{The Host Galaxy Sample}\label{subsec:host_sample}

\begin{figure}[!htb]
    \centering
    \includegraphics[width=1.0\linewidth]{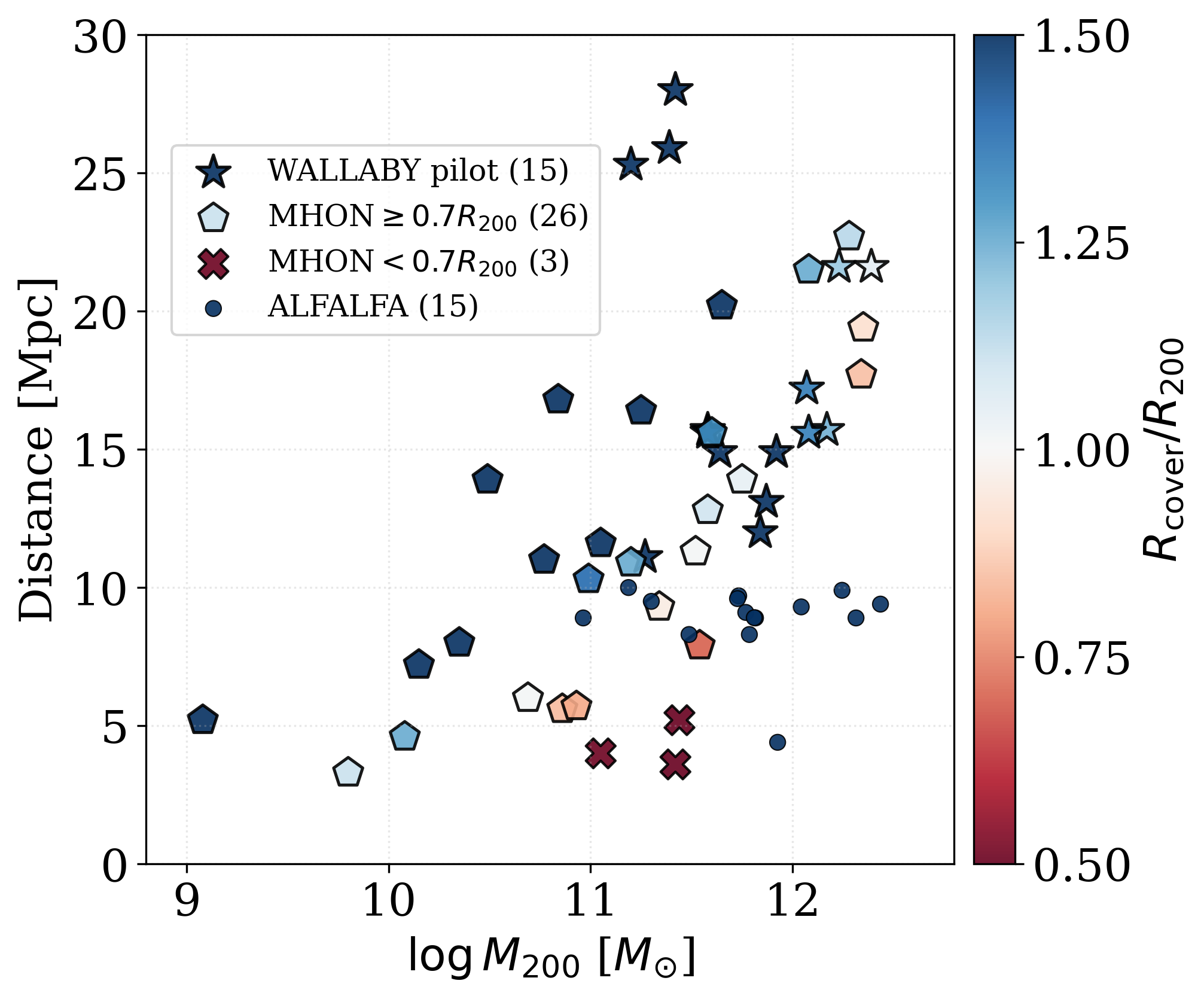}
    \caption{Summary of the nearby \HI-bearing host galaxies in this study. Data points show the distances versus halo masses of the host galaxies (Table \ref{table:hosts}), and in parentheses the host count from the survey: WALLABY pilot (stars; \citealt{westmeier_wallaby_2022,murugeshan_wallaby_2024}), MHONGOOSE (pentagons; \citealt{de_blok_mhongoose_2024}), and ALFALFA from \cite{zhu_census_2023} (small circles; \HIspace data from \citealt{haynes_arecibo_2018}). Colors show the fractional halo coverage ($R_{\rm cover}/R_{200}$; see Section \ref{subsec:host_sample}). We excluded three MHONGOOSE hosts with the lowest projected halo coverages ($R_{\rm cover}/R_{200}<0.7$; see the red crosses), which are intermediate-mass spirals with no gaseous satellites in their covered inner halos (NGC 625, NGC 7793, and NGC 5068).}
    \label{fig:host_m200_halo_rcover}
\end{figure}

This section describes our host galaxy sample, first on how we selected the host galaxies from the WALLABY and MHONGOOSE surveys (Section \ref{sec:data}), and then on the derivation of their physical properties. The host sample contains 56 \HI-bearing galaxies within 30 Mpc: 41 from WALLABY and MHONGOOSE, and 15 from ALFALFA in our previous work \citep{zhu_census_2023}. The host galaxies span several orders of magnitude in mass, from dwarf centrals to M31-mass spiral galaxies, as summarized in Figure \ref{fig:host_m200_halo_rcover} and Table \ref{table:hosts}.

From the WALLABY pilot survey catalogs \citep{westmeier_wallaby_2022,murugeshan_wallaby_2024}, we select nearby spiral host galaxies for which WALLABY is sensitive enough to detect their satellite populations. We apply the following cuts: (i) $M_{\HI} \geq 10^{8} M_{\odot}$, a host gas mass threshold, and $W_{50} \geq 120$ \kms\ ($W_{50}$: the width of the \HIspace line at $50\%$ peak flux density), a simple dynamical mass threshold; combined, these aim to exclude gas-rich dwarf centrals and retain only spiral hosts, and (ii) a velocity-flow distance of $D_{\rm flow} = V_{\rm los}/H_{0} \leq 30$ Mpc. The distance cut provides a reasonable sample size by covering a large enough volume, while maintaining an \HIspace sensitivity down to dwarf galaxy masses. After the initial selection, we manually inspect each galaxy and exclude those whose halo region, approximated by a radius of 300 kpc around the galaxy\footnote{The 300 kpc radius is an upper limit and always greater than $R_{200}$ of the WALLABY hosts; see Table \ref{table:hosts}.}, is not fully covered by the WALLABY footprints. This step excludes about 20 galaxies close to the field edge with partial halo coverages that are subject to significantly higher noise as a result (see noise variation maps in Appendix A of \citealt{westmeier_wallaby_2022}). We also excluded one galaxy in an extremely close binary system, PGC 46579 (equivalently MCG -03-34-054), which is $\sim$1.1 arcmin away from its companion, MCG -03-34-053. This results in 15 host galaxies from WALLABY, as summarized in the first section of Table \ref{table:hosts}. Some host galaxies are in Local Group-like pairs, while others contain massive satellites analogous to the MW-LMC system.

The MHONGOOSE survey maps the \HIspace emission in fields of 1.5 $\deg$ in diameter around 30 target galaxies (see Section \ref{sec:data} and \citealt{de_blok_mhongoose_2024}). We select host galaxies out of these 30. First, we note that the target galaxies span a large range in mass, from low-mass dwarfs ($\log M_{200}/M_{\odot} < 10.5$) to MW-mass spirals ($\log M_{200}/M_{\odot} \approx 12$). As we later demonstrate in Section \ref{subsec:method_sensitivity}, the \HIspace sensitivity around the MHONGOOSE dwarf target galaxies is deep enough for the detection of the lowest-mass gaseous dwarf galaxies ($\log M_{\HI,\rm lim}/M_{\odot} \approx 5.4$). We thus include the dwarf targets in our host sample, as long as they are central galaxies in medium- and low-density environments, to study the gaseous satellites of low-mass centrals. One dwarf target, UGCA 015 (equivalently J0049-20), is itself a satellite of a nearby spiral galaxy, NGC 247 \citep{de_blok_mhongoose_2024}, and is excluded from our sample.

Under the uniform 1.5 $\deg$ field coverage, the projected halo areas of the MHONGOOSE target galaxies are not always fully covered. We further excluded three galaxies with low halo coverage ($R_{\rm cover}/R_{200} < 0.7$; see Figure \ref{fig:host_m200_halo_rcover}) from our analyses. In some cases, the MHONGOOSE cube captures another galaxy of comparable or higher (e.g., NGC 4808 in the field of UGCA 307) stellar mass relative to the target galaxy. These represent pair-like environments, where the non-target galaxy typically has low halo coverage; we exclude all such non-target galaxies in our host sample. In total, 26 MHONGOOSE target galaxies are included as hosts, as listed in the second part of Table \ref{table:hosts}. Most (20 out of 26) have full projected virial radius coverage ($R_{\rm cover}/R_{200} \geq 1$); we indicate hosts with partial halo coverage in the results (Section \ref{sec:results}).

With this \HI-selected host galaxy sample, we cross-match with optical surveys to obtain their stellar masses ($M_{*}$) and distances ($D$; typically redshift independent). Our WALLABY hosts are well-studied nearby galaxies, all of which are covered in the $z=0$ Multiwavelength Galaxy Synthesis project (z0MGS; \citealt{leroy_z_2019}). We thus obtain their distances\footnote{Distances from z0MGS are compiled from the Extragalactic Distance Database (EDD; \citealt{tully_extragalactic_2009}, incorporating high-precision TRGB distances where available), CosmicFlows \citep{tully_cosmicflows-3_2016}, and the Lyon Extragalactic Database (LEDA; \citealt{paturel_hyperleda_2003}).} and stellar masses from \cite{leroy_z_2019}, unless targeted high-precision distance measurements exist in the literature (NGC 4666 from \citealt{shappee_young_2016}). For the MHONGOOSE hosts, we follow the published distances and stellar masses in \cite{de_blok_mhongoose_2024} (see their Table 1), where distances are also from z0MGS \citep{leroy_z_2019} wherever available, and the stellar masses are derived from WISE.

Satellite galaxies reside in the dark matter halos of the host galaxies. We derive the host halo masses using the median stellar mass-halo mass (SMHM) relation for star-forming central galaxies at $z \approx 0$ from UniverseMachine \citep{behroozi_universemachine_2019}. Given the literature stellar masses of the host galaxies (Table \ref{table:hosts}), we solve for their halo masses $M_{\rm halo,peak}$ and assume $M_{\rm halo,peak} \approx M_{200}$ ($M_{\rm halo,peak}$ follows the virial mass definition of \cite{bryan_statistical_1998}, which is approximately consistent with $M_{200}$ at $z\approx 0$; see \citealt{white_mass_2001}). We note that this relation is extrapolated for the lowest-mass dwarf centrals in our sample, where the SMHM relation is more uncertain. Given $M_{200}$, we calculate the virial radius ($R_{200}$) based on the standard definition, $M_{200} = \frac{4 \pi}{3} \cdot 200 \rho_{\rm crit} R_{200}^{3}$. Finally, we define the host scaling velocity as the point-mass escape velocity evaluated at $r=R_{200}$, $V_{\rm esc}(R_{200}) = \sqrt{\frac{2 G M_{200}}{R_{200}}}$.

\begin{deluxetable*}{lccccccccccccc}\label{table:hosts}
\tablecaption{Properties of the \HI-containing host galaxies.} 
\tablewidth{0pt}
\tablehead{\colhead{Name} & \colhead{PGC} & \colhead{RA} & \colhead{Dec} & \colhead{$V_{\rm los}$} & \colhead{$D$} & \colhead{Inc.} & \colhead{$W_{50}$} & \colhead{$\log M_{\HI}$} & \colhead{$\log M_{*}$} & \colhead{$\log M_{200}$} & \colhead{$R_{200}$} & \colhead{$\frac{R_{\rm cover}}{R_{200}}$} & \colhead{$M_{\HI, lim}$} \\
\colhead{} & \colhead{} & \colhead{(deg)}  & \colhead{(deg)} & \colhead{(km/s)}  & \colhead{(Mpc)} & \colhead{($\deg$)} & \colhead{(km/s)} & \colhead{($M_{\odot}$)}  & \colhead{($M_{\odot}$)}  & \colhead{($M_{\odot}$)} & \colhead{(kpc)} & \colhead{} & \colhead{($M_{\odot}$)}}
\colnumbers
\startdata
NGC3137 & 29530 & 152.266 & -29.044 & 1110 & 14.9 & 73 & 250 & 9.87 & 9.75 & 11.64 & 160 & 1.87 & 7.14$^{a}$ \\
NGC3175 & 29892 & 153.674 & -28.873 & 1106 & 14.9 & 82 & 300 & 8.84 & 10.21 & 11.92 & 198 & 1.51 & 7.14 \\
NGC4592 & 42336 & 189.822 & -0.531 & 1070 & 11.1 & 79 & 198 & 9.71 & 9.05 & 11.27 & 121 & 2.48 & 6.89 \\
UGC07841 & 42542 & 190.298 & 1.41 & 1691 & 25.9 & 53 & 147 & 9.04 & 9.27 & 11.39 & 132 & 2.27 & 7.57 \\
NGC4632 & 42689 & 190.636 & -0.083 & 1715 & 15.7 & 73 & 217 & 9.44 & 9.64 & 11.58 & 153 & 1.96 & 7.14 \\
NGC4629 & 42692 & 190.635 & -1.353 & 1107 & 28 & 41 & 132 & 9.7 & 9.33 & 11.42 & 135 & 2.22 & 7.58 \\
NGC4666 & 42975 & 191.285 & -0.463 & 1530 & 15.7 & 80 & 379 & 9.71 & 10.53 & 12.17 & 241 & 1.24 & 7.11 \\
NGC6221 & 59175 & 253.202 & -59.219 & 1472 & 15.6 & 54 & 289 & 9.57 & 10.42 & 12.08 & 224 & 1.34 & 7.07 \\
ESO138-G010 & 59373 & 254.755 & -60.212 & 1141 & 13.1 & 43 & 209 & 9.78 & 10.13 & 11.87 & 190 & 1.58 & 6.96 \\
NGC4984 & 45585 & 197.241 & -15.517 & 1250 & 12 & 28 & 267 & 7.94 & 10.09 & 11.84 & 187 & 1.61 & 6.92 \\
NGC5247 & 48171 & 204.512 & -17.882 & 1352 & 21.6 & 34 & 133 & 9.77 & 10.59 & 12.23 & 253 & 1.19 & 7.36 \\
NGC5170 & 47394 & 202.449 & -17.963 & 1515 & 21.6 & 89 & 507 & 10 & 10.72 & 12.39 & 285 & 1.05 & 7.34 \\
ESO576-G037 & 46550 & 200.121 & -19.842 & 1711 & 25.3 & 77 & 142 & 8.38 & 8.9 & 11.2 & 114 & 2.63 & 7.53 \\
NGC5054 & 46247 & 199.245 & -16.633 & 1737 & 17.2 & 56 & 312 & 9.17 & 10.41 & 12.07 & 222 & 1.35 & 7.26 \\
NGC4808 & 44086 & 193.952 & 4.302 & 757 & 15.6 & 72 & 256 & 9.64 & 9.63 & 11.58 & 153 & 1.97 & 7.14 \\
\hline
ESO349-G031 & 621 & 2.056 & -34.578 & 220 & 3.3 & 35 & 26 & 7.11 & 6.13 & 9.8 & 39 & 1.11 & $<$5.4 \\
KK98-195 & 166163 & 200.284 & -31.529 & 571 & 5.2 & 63 & 26 & 7.6 & 4.71 & 9.08 & 22 & 3.03 & $<$5.4 \\
ESO300-G016 & 11842 & 47.544 & -40.003 & 708 & 8 & 36 & 27 & 7.8 & 7.22 & 10.35 & 59 & 1.77 & $<$5.4 \\
ESO473-G024 & 1920 & 7.844 & -22.766 & 538 & 7.2 & 57 & 34 & 7.95 & 6.84 & 10.15 & 51 & 1.84 & $<$5.4 \\
ESO444-G084 & 48111 & 204.333 & -28.045 & 587 & 4.6 & 40 & 53 & 8.02 & 6.69 & 10.08 & 48 & 1.25 & $<$5.4 \\
NGC1311 & 12460 & 50.029 & -52.186 & 570 & 5.6 & 74 & 84 & 8.05 & 8.23 & 10.86 & 88 & 0.84 & $<$5.4 \\
NGC1705 & 16282 & 73.556 & -53.361 & 632 & 5.7 & 42 & 108 & 8.07 & 8.37 & 10.93 & 93 & 0.81 & $<$5.4 \\
NGC1592 & 15292 & 67.417 & -27.409 & 943 & 10.3 & 60 & 54 & 8.1 & 8.49 & 10.99 & 97 & 1.39 & 5.44 \\
KKS2000-23 & 3097702 & 166.55 & -14.407 & 1036 & 13.9 & 70 & 78 & 8.74 & 7.51 & 10.49 & 66 & 2.74 & 5.71 \\
UGCA307 & 43851 & 193.489 & -12.106 & 822 & 11 & 60 & 67 & 8.88 & 8.05 & 10.77 & 82 & 1.76 & 5.5 \\
ESO300-G014 & 11812 & 47.408 & -41.03 & 954 & 10.9 & 59 & 129 & 8.89 & 8.9 & 11.2 & 114 & 1.25 & 5.49 \\
IC1954 & 13090 & 52.881 & -51.905 & 1057 & 12.8 & 62 & 216 & 8.9 & 9.63 & 11.58 & 153 & 1.1 & 5.68 \\
ESO302-G014 & 13985 & 57.92 & -38.452 & 870 & 16.8 & 28 & 67 & 8.9 & 8.19 & 10.84 & 86 & 2.55 & 5.88 \\
IC4951 & 64181 & 302.382 & -61.85 & 811 & 11.6 & 80 & 118 & 8.93 & 8.6 & 11.05 & 101 & 1.5 & 5.53 \\
UGCA320 & 45084 & 195.82 & -17.423 & 740 & 6 & 83 & 108 & 8.97 & 7.91 & 10.69 & 77 & 1.01 & $<$5.4 \\
NGC2101 & 17793 & 86.601 & -52.089 & 1187 & 16.4 & 47 & 92 & 9.22 & 9 & 11.25 & 119 & 1.81 & 5.9 \\
NGC4781 & 43902 & 193.613 & -10.508 & 1256 & 11.3 & 65 & 230 & 9.22 & 9.52 & 11.52 & 146 & 1.01 & 5.54 \\
ESO362-G011 & 17027 & 79.162 & -37.103 & 1337 & 15.6 & 81 & 272 & 9.57 & 9.68 & 11.6 & 156 & 1.31 & 5.8 \\
NGC1744 & 16517 & 74.991 & -26.022 & 740 & 9.3 & 57 & 192 & 9.54 & 9.18 & 11.34 & 127 & 0.96 & $<$5.4 \\
NGC3511 & 33385 & 165.849 & -23.087 & 1099 & 13.9 & 70 & 288$^{b}$ & 9.54 & 9.94 & 11.75 & 174 & 1.04 & 5.7 \\
NGC7424 & 70096 & 344.327 & -41.071 & 937 & 7.9 & 32 & 153 & 9.6 & 9.56 & 11.54 & 148 & 0.7 & $<$5.4 \\
UGCA250 & 37271 & 178.35 & -28.553 & 1696 & 20.2 & 82 & 269 & 9.84 & 9.76 & 11.65 & 161 & 1.64 & 5.97 \\
NGC1371 & 13255 & 53.756 & -24.933 & 1456 & 22.7 & 46 & 383 & 9.97 & 10.63 & 12.28 & 261 & 1.14 & 6.12 \\
NGC0289 & 3089 & 13.177 & -31.206 & 1620 & 21.5 & 45 & 267 & 10.35 & 10.43 & 12.08 & 225 & 1.25 & 6.08 \\
NGC1566 & 14897 & 65.002 & -54.938 & 1496 & 17.7 & 37 & 198 & 10.08 & 10.68 & 12.34 & 274 & 0.85 & 5.92 \\
NGC1672 & 15941 & 71.427 & -59.247 & 1327 & 19.4 & 34 & 255 & 10.29 & 10.69 & 12.35 & 276 & 0.92 & 5.95 \\
\enddata
\tablecomments{Host galaxies from the WALLABY and MHONGOOSE surveys, separated by a line. Columns: (1) Common name. (2) Principal Galaxies Catalogue (PGC) identification. (3--4) Right ascension and declination coordinates in $\deg$ (J2000). (5) Heliocentric (line-of-sight) velocity. (6) Distance from \cite{leroy_z_2019} unless specified in Section \ref{subsec:host_sample}. (7) Indicative inclination (NED and SIMBAD). (8) \HIspace linewidth. (9) \HIspace mass evaluated at $D$ (Column 6). (10) Stellar mass from \cite{leroy_z_2019} (WALLABY galaxies) or WISE photometry (MHONGOOSE galaxies; \citealt{de_blok_mhongoose_2024}). (11) Dark matter halo mass derived from the SMHM relation \citep{behroozi_universemachine_2019}. (12) Virial radius (halo size). (13) Fraction of host halo covered ($R_{\rm cover}/R_{200}$) for satellite search. Three galaxies in MHONGOOSE with a low coverage are excluded (Figure \ref{fig:host_m200_halo_rcover}). $R_{\rm cover}=300$ kpc for the WALLABY galaxies, and 0.75 $\deg$ in angular size for the MHONGOOSE galaxies. (14) \HIspace mass sensitivity to low-mass dwarf galaxies of $W_{50}=25$ km/s from our injection tests (Appendix \ref{subsubsec:injection_in_sensitivity_test}).\\
$^{a}$ Pair with NGC 3175. $^{b}$ Interacting pair with NGC 3513, connected by an \HIspace bridge \citep{de_blok_mhongoose_2024}.}
\end{deluxetable*}

Table \ref{table:hosts} summarizes the basic properties of the 41 host galaxies (15 from WALLABY, 26 from MHONGOOSE). The distances versus halo masses of these galaxies are additionally displayed in Figure \ref{fig:host_m200_halo_rcover}, color-coding the fractional projected radius searched in this satellite study. WALLABY is a wide-field survey, and we created cutouts of $R_{\rm cover} =300$ kpc centered at the host galaxy locations for our satellite search (see Section \ref{subsec:method_source_finding}). Properties of the 15 host galaxies from the ALFALFA survey are described in our previous work (\citealt{zhu_census_2023}; $R_{\rm cover}/R_{200} \approx 2$). Here, we recalculated their halo masses (and related halo quantities) using \cite{behroozi_universemachine_2019} for consistency.

\subsection{\HIspace source finding}\label{subsec:method_source_finding}

We conducted untargeted source finding on the host galaxy fields (Section \ref{subsec:host_sample}) to search for \HI-bearing satellite candidates. For WALLABY, we created cutout data cubes covering a 300 kpc radius at the distance of the host galaxy and $\pm 300$ km/s around the host's systemic velocity. For MHONGOOSE, we used the full-sensitivity cubes at \texttt{r15\_t00} resolution that cover 0.75 deg in radius and $\pm 500$ km/s around the host's systemic velocity (see Section \ref{sec:data}), but restricted the satellite search to $\pm 300$ km/s (see \S \ref{subsec:method_sat_environment}). We iteratively ran version 2 of the \HIspace Source Finding Application (SoFiA-2; \citealt{serra_sofia_2015,westmeier_sofia_2021}) on the host galaxy cubes, starting from the parameter sets used by the WALLABY and MHONGOOSE surveys (see \citealt{westmeier_wallaby_2022,murugeshan_wallaby_2024,de_blok_mhongoose_2024}). We then adjusted several key parameters to optimize the detection for the lowest-mass \HIspace sources\footnote{This optimization primarily targets the wide-field WALLABY data, as source finding for lowest-mass galaxies has been extensively tested in MHONGOOSE \citep{de_blok_mhongoose_2024}.}, yielding 7 new detections not reported in the WALLABY pilot fields and 16 new detections from the full-sensitivity MHONGOOSE data. These new detections are small, gas-bearing dwarf galaxies that are spatially unresolved (comparable to the telescope beam size). The modified parameters we used and their effects are summarized below.

\begin{verbatim}
SoFiA Parameter               Value
_____________________________________________
scfind.kernelsXY         =  0, 3, 5, 10
scfind.kernelsZ          =  0, 3, 7, 15
scfind.threshold         =  3.8
linker.radiusXY          =  2
linker.radiusZ           =  1
linker.minSizeXY         =  5
linker.minSizeZ          =  3 (WALLABY cubes)
reliability.threshold    =  0.5 (test 0.3, 0.7)
reliability.autoKernel   =  true
reliability.minSNR       =  3.5
\end{verbatim}

These parameters belong to three modules in SoFiA-2, the smooth-and-clip finder (\texttt{scfind}), \texttt{linker}, and \texttt{reliability} modules. The units for the kernels are pixels for XY and velocity channel number for Z. Below, we explain the effects of the modified parameters in comparison with those used in the WALLABY pilot surveys (Table 2 in \citealt{westmeier_wallaby_2022}, Table 5 in \citealt{murugeshan_wallaby_2024}).

\begin{itemize}
    \item \texttt{scfind}: Decreasing the smallest spatial kernel size (the first number after 0, where 0 stands for no smoothing) helps detect marginally resolved dwarf galaxies. Here, we adopt the smallest kernel size of 3 pixels, which is smaller than the beam FWHM of $\sim$5 pixels for WALLABY and MHONGOOSE.
    Lowering the SNR threshold (\texttt{scfind.threshold}), on the other hand, only tends to introduce false detections, so we use the WALLABY standard value of 3.8.
    
    \item \texttt{linker}: We adopted small values for the linker radii in both spatial (XY) and velocity (Z) dimensions to minimize the confusion of close-by sources. This has little effect on dwarf galaxy detection but helps prevent source merging in the injection tests in Section \ref{subsec:method_sensitivity}.
    We set \texttt{linker.minSizeZ} to 3 for WALLABY cubes to filter out objects with $W_{50} < 3 v_{\rm chan} \approx 12$ km/s. This limit is reduced from typical thresholds of $W_{50} \lesssim 20$ km/s to include dwarf galaxies with low velocity widths.
    
    \item \texttt{reliability}: We enable \texttt{auto.Kernel} to optimize for the reliability kernel size, which can vary from field to field. The reliability module uses statistical estimations to identify candidates for true detections, rather than generating a deterministic list \citep{serra_using_2012}. We iterated through a range of \texttt{reliability.threshold} values (0.3, 0.5, 0.7) for each host galaxy field and found that empirically, the threshold value of 0.5 is optimal in detecting faint dwarfs. Further lowering the \texttt{reliability.threshold} or \texttt{reliability.minSNR} value rarely results in the detection of new, real sources.
    
\end{itemize}

Optimizing for completeness comes at the cost of increased occurrence of false detections captured by the source finder. We examine all \HIspace sources generated from the SoFiA-2 runs via two metrics: (i) good quality of the \HIspace data, including a high signal-to-noise ratio (SNR) and a plausible spectrum and morphology for a low-mass galaxy, (ii) the existence of an optical counterpart. An \HIspace source is required to satisfy both metrics to be considered a dwarf galaxy in our sample (Table \ref{table:sats}). For the optical counterparts, we visually cross-matched with the DESI Legacy Imaging Surveys DR10 \citep{dey_2019AJ....157..168D}, and the Digitized Sky Survey II\footnote{\url{https://archive.stsci.edu/cgi-bin/dss_form}} (DSS2), where DESI imaging is unavailable. Optical maps from Legacy Surveys DR10 are later used to calibrate the stellar masses of the dwarf galaxies (Section \ref{subsec:method_stellar_mass}). This step excluded genuine \HIspace sources with no optical counterpart, which are likely gas clouds from interactions, and sources with relatively low \HIspace SNR, which are likely false detections.

\subsection{Sensitivity and Completeness Assessment}\label{subsec:method_sensitivity}

Rigorous assessment of \HIspace sensitivity is required to distinguish sensitivity-limited non-detections from true gas loss. Our host galaxy sample (\S \ref{subsec:host_sample}) collates sources from different surveys and at a range of distances. For wide-field surveys like the WALLABY pilot surveys \citep{westmeier_wallaby_2022,murugeshan_wallaby_2024} and ALFALFA \citep{haynes_arecibo_2011,haynes_arecibo_2018}, field-to-field variations also contribute to inhomogeneities in the local noise levels. As a result, the \HIspace sensitivity and completeness of the detected gaseous dwarf population can vary substantially within our host sample.

\begin{figure}[!htb]
    \centering
    \includegraphics[width=1.0\linewidth]{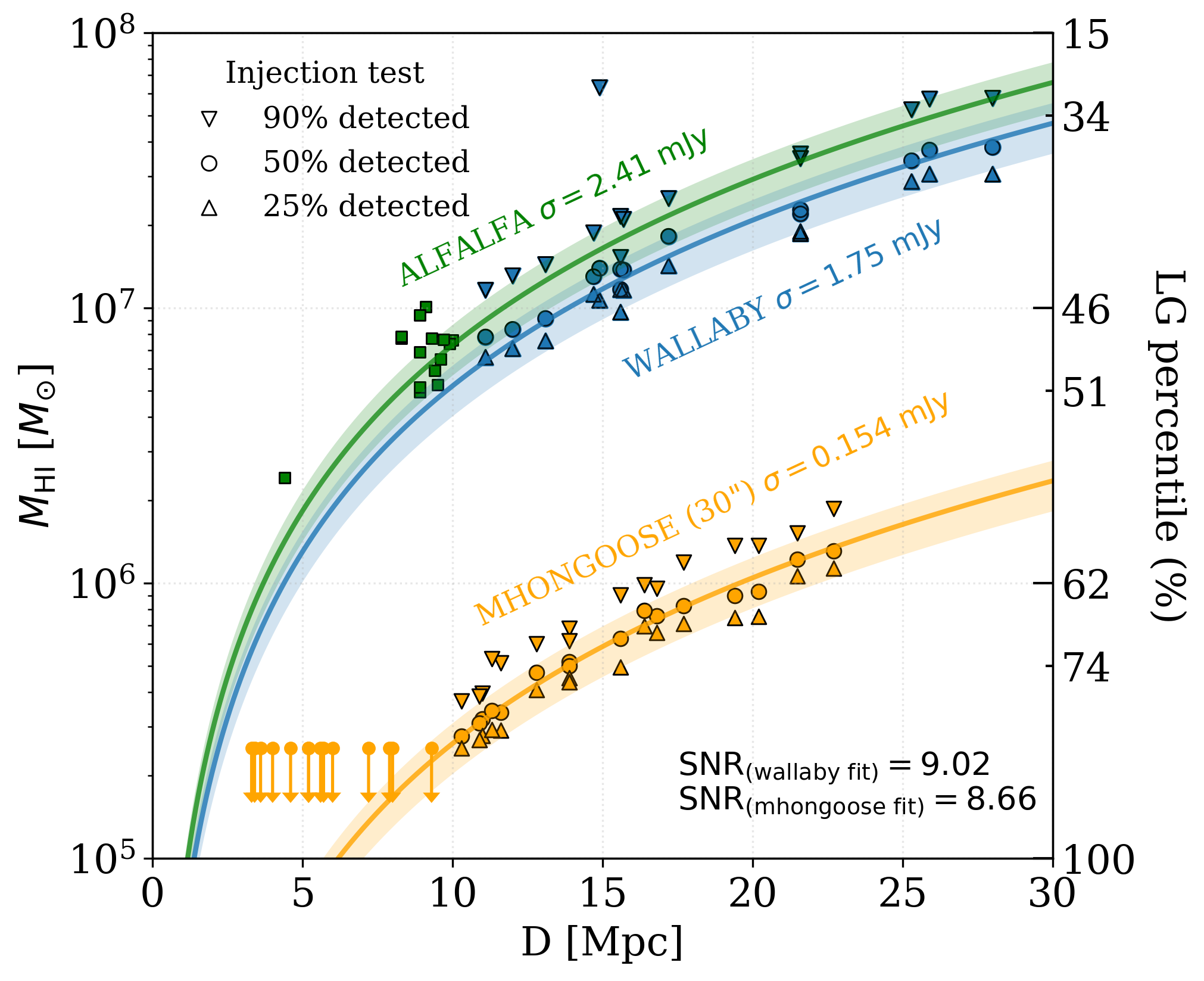}
    \caption{The \HIspace sensitivity limits across surveys. Scatter points show the 25\%, 50\%, and 90\% detectability thresholds (upward triangles, circles, downward triangles) from dwarf injection tests for WALLABY and MHONGOOSE, and from empirical estimates of \cite{haynes_arecibo_2011} for ALFALFA (squares). Downward arrows mark MHONGOOSE hosts with sensitivities deeper than the lowest mass injection model (Appendix \ref{subsubsec:injection_in_sensitivity_test}). Lines and shaded regions show the sensitivity in Equation \ref{eqn:mhi_sensitivity_compre} at the best-fit SNRs for $W_{50}=25$ and $15-35$ km/s, respectively. The right-hand y-axis gives the corresponding completeness relative to Local Group (LG) \HI-bearing dwarfs \citep{putman_gas_2021}.}
    \label{fig:dwarf_mhi_sensitivity_compre}
\end{figure}

We have developed a flexible injection pipeline, detailed in Appendix \ref{appendix:inject}, to assess the \HIspace mass sensitivity ($M_{\HI,lim}$) of a given data cube. Using the injection pipeline, we added the \HIspace fluxes of modeled low-mass dwarf galaxies into the WALLABY and MHONGOOSE host galaxy data cubes (\S \ref{subsec:method_source_finding}) and derived \HIspace mass thresholds at 25\%, 50\%, 90\% recovery rates (Appendix \ref{subsubsec:injection_in_sensitivity_test}). We adopt the 50\% detectability thresholds as the sensitivity limits for each host galaxy ($M_{\HI, \rm lim}$ in Table \ref{table:hosts}; circles in Figure \ref{fig:dwarf_mhi_sensitivity_compre}). The average sensitivities of the WALLABY (wide-field; in blue) and MHONGOOSE (deep targeted; in orange) host galaxies are separated by $\sim$1.6 dex: the 15 WALLABY hosts have a median sensitivity of $\log M_{\HI,lim}/M_{\odot} \approx 7.1$, and the 26 MHONGOOSE hosts have a median of $\log M_{\HI,lim}/M_{\odot} \approx 5.5$.

Sensitivity can be converted to completeness relative to a reference population. Here, we choose the Local Group (LG) gaseous dwarf population from \cite{putman_gas_2021} as the reference population, as its proximity enables exceptional data depth. There are 26 LG dwarf galaxies within 2 Mpc with \HIspace detections, with a median \HIspace mass of $\log M_{\HI}/M_{\odot} \approx 6.9$. We use the cumulative $M_{\HI}$ distribution of these 26 to convert from an \HIspace mass limit to the percentile of LG gaseous dwarf galaxies above that mass, as shown on the right-hand y-axis of Figure \ref{fig:dwarf_mhi_sensitivity_compre}. Our WALLABY host galaxies' median sensitivity ($\log M_{\HI,lim}/M_{\odot} \approx 7.1$) would detect $\sim 38\%$ of the LG gaseous dwarfs in \cite{putman_gas_2021}, while the MHONGOOSE hosts' median sensitivity ($\log M_{\HI,lim}/M_{\odot} \approx 5.5$) would detect $\sim$83\%. These numbers are dictated by the distances to our hosts, and future WALLABY galaxies can be selected at smaller distances to reach deeper effective sensitivities.

We further characterize the survey depths by fitting the empirical sensitivity limits from the injection tests to analytical scaling relations. Given that the minimum detectable \HIspace mass ($M_{\HI,lim}$) scales with distance ($D_{\rm Mpc}$) and the integrated flux limit ($S_{V,\rm lim}$),
\begin{equation}\label{eqn:mhi_sensitivity_vflux}
    M_{\HI,lim}(M_{\odot}) = 2.356 \times 10^{5} \cdot D_{\rm Mpc}^{2} \cdot \big(\frac{S_{V,\rm lim}}{\rm Jy \cdot km~s^{-1}} \big) 
\end{equation}
we characterize an integrated signal-to-noise ratio (SNR) to represent the survey sensitivity in units of the typical noise levels ($\sigma$ in Jy/beam/channel; see Section \ref{sec:data}) of the surveys. The SNR is given by the summation of fluxes divided by the rms noise in each channel $i$ (assuming the source is spatially unresolved, which is mostly true for the low-mass dwarf galaxies),
\begin{equation}\label{eqn:mhi_sensitivity_snr}
    \begin{aligned}
    \rm SNR & = \frac{\sum_{i=1}^{N_{\rm chan}} {S_{i}}}{\sqrt{\sum_{i=1}^{N_{\rm chan}} \sigma_{i}^{2}}} \approx  \frac{(S_{V,\rm lim} /V_{\rm res})}{\sigma \sqrt{N_{\rm chan}}}
    \approx \frac{S_{V, \rm lim}}{\sigma \sqrt{2 W_{50} V_{\rm res}}} 
    \end{aligned}
\end{equation}
where $V_{\rm res}$ is the velocity resolution (in \kms), $N_{\rm chan}$ is the number of velocity channels, and $W_{50}$ (in \kms) is the velocity FWHM of the source. Here we have assumed (i) the noise per channel $\sigma_{i}$ averages to $\sigma$ such that $\sum \sigma_{i}^2 \approx N_{\rm chan} \sigma^{2}$, and (ii) $N_{\rm chan} \approx 2W_{50}/V_{\rm res}$, a reasonable approximation for dwarf galaxies harboring Gaussian-like emission profiles. 

Combining equations \ref{eqn:mhi_sensitivity_vflux} and \ref{eqn:mhi_sensitivity_snr}, we obtain the final form of a survey's \HIspace sensitivity,
\begin{equation}\label{eqn:mhi_sensitivity_compre}
    M_{\HI,lim}(M_{\odot}) = 2.356 \times 10^{5} \cdot D_{\rm Mpc}^{2} \cdot \sqrt{2 W_{50} V_{\rm res}} \cdot \rm SNR \cdot \sigma
\end{equation}
We fit for the effective SNR of each survey (Equation \ref{eqn:mhi_sensitivity_compre}) taking the following inputs: $M_{\HI,lim}$ and $D_{\rm Mpc}$ from the host injection tests at the $50\%$ detectability threshold (circles in Figure \ref{fig:dwarf_mhi_sensitivity_compre}), $W_{50,\rm inject}=25$ km/s from our model assumption (\S \ref{subsubsec:model_in_sensitivity_test}), and $V_{\rm res}$ and $\sigma$ from each survey (see Section \ref{sec:data}). We obtained effective SNRs of 9.02 for WALLABY and 8.66 for MHONGOOSE via least-squares fitting; note that despite similar effective SNRs, the MHONGOOSE noise level is over 10 times lower. In Figure \ref{fig:dwarf_mhi_sensitivity_compre}, solid lines and shaded regions show Equation \ref{eqn:mhi_sensitivity_compre} at the best-fit SNR values for WALLABY and MHONGOOSE, adopting velocity widths $W_{50}=25$ km/s and $W_{50}=15-35$ km/s, respectively, based on typical values of LG gaseous dwarfs \citep{putman_gas_2021}. For ALFALFA, we adopt the empirical 50\% (median) sensitivity thresholds from \cite{haynes_arecibo_2011}, and separately estimate the median sensitivity of our 15 ALFALFA hosts (green squares; \citealt{zhu_census_2023}), whose deviation from the survey median line reflects local noise variations around $\sigma \approx 2.41$ mJy.

To conclude, the injection tests provide host-specific \HIspace mass sensitivities ($M_{\HI,lim}$; Table \ref{table:hosts}), which we use to evaluate completeness in the satellite abundance analysis (Section \ref{sec:host_sat_abundance}). At 10 Mpc, the WALLABY pilot surveys (with $\log M_{\HI,lim}/M_{\odot} \approx 6.7$) are only $0.15$ dex more sensitive than ALFALFA, and the MHONGOOSE survey (with $\log M_{\HI,lim}/M_{\odot} \approx 5.4$) is more than one dex deeper than both wide-field surveys (Figure \ref{fig:dwarf_mhi_sensitivity_compre}).

\subsection{Optical photometry and stellar mass calibration}\label{subsec:method_stellar_mass}


This section describes the stellar mass calibration for the dwarf galaxies. For candidate \HIspace detections with visible optical counterparts (\S \ref{subsec:method_source_finding}), we measure optical photometry in the $g$- and $r$-bands from Legacy DR10 \citep{dey_2019AJ....157..168D, lang_2016ascl.soft04008L}. We calculate dereddened optical magnitudes using the total Sérsic flux in each band and the \citet{schlegel98} dust reddening map with updated Galactic Extinction coefficients for the DECam filters from the Legacy documentation following \citet{Schlafly2011}. Sérsic fluxes are measured using an updated version of the Bayesian Sérsic profile fitting tool \texttt{pysersic} \citep{pasha_2023JOSS....8.5703P, asali_2025arXiv250925335A}. We use the \texttt{pysersic} autoprior function to generate a prior on the total flux and $R_\mathrm{eff}$ for each source, and we set uniform priors between $0<e<0.9$ for ellipticity, $0<\phi<2\pi$ for position angle, and $0.3<n<8.0$ for Sérsic index for all sources. 

While most galaxy brightness profiles can be reasonably approximated as Sérsic profiles, we explicitly check how our measured photometry is affected by this choice of model by comparing to aperture photometry. Aperture and Sérsic-derived fluxes can differ because the Sérsic profile integrates light to large radii and can capture extended low–surface-brightness emission whereas our aperture, defined at 3$R_{\mathrm{eff}}$ from our best fit Sérsic model, can capture bright star-forming clumps or substructure not well described by the smooth Sérsic fit. When we translate these fluxes into stellar mass estimates, we find typical residuals between the two methods of within 0.3 dex in stellar mass. Stellar masses are calculated using the optical color calibration from \citet{de_los_reyes_stellar_2025}, assuming a Chabrier IMF. We ensure all literature relations compared against in this work use a Chabrier IMF, since the offset due to different IMF assumptions is expected to dominate the systematic uncertainty between mass calibrations \citep{bell_2001ApJ...550..212B, bell_2003ApJS..149..289B}. Beyond IMF differences, stellar mass calibrations can still vary at the $\sim$0.1–0.2 dex level depending on methodology (e.g., SED fitting vs. color calibrations; \citealt{pacifici_2023ApJ...944..141P, de_los_reyes_stellar_2025}), but these differences are minor and do not affect our main results.

We manually inspected all cases where the Sérsic- and aperture-derived stellar masses differed by more than 0.3 dex. In four such cases, we report the optical magnitudes and stellar masses from aperture photometry, as these galaxies exhibit strong non-Sérsic features or significant residuals relative to the best-fit profile. These include bright systems with bars, prominent bulges, or intense star-forming clumps.

\subsection{Characterizing satellite environment}\label{subsec:method_sat_environment}

We characterized the environment of gaseous dwarf galaxies identified in Section \ref{subsec:method_source_finding}, considering only systems within $\pm 300$ km/s of their host galaxies' systemic velocity. Galaxies with larger line-of-sight velocity differences ($|\Delta V_{\rm los}| > 300$ km/s) are extremely unlikely to be bound satellites of MW-mass or lower hosts \citep{mcconnachie_observed_2012,mao_saga_2024} and are excluded from the environment analysis.

Satellites are defined as galaxies gravitationally bound to the host galaxy, requiring the satellite-host relative velocity ($\Delta \vec{V}$) at the satellite location ($r$) to be lower than the local escape velocity, $|\Delta \vec{V}(r)| < V_{\rm esc}(r)$. Given the host halo mass $M_{200}$, the escape velocity can be approximated under a point-mass assumption as $V_{\rm esc}(r) = \sqrt{2GM_{200}/r}$. In our sample, host masses vary considerably, so we scale the gravitationally bound condition by host halo properties,

\begin{equation}\label{eqn:grav_bound_theory}
    \frac{|\Delta \vec{V}(r)|}{V_{\rm esc}(R_{200})} < \frac{V_{\rm esc}(r)}{V_{\rm esc}(R_{200})} = \left(\frac{r}{R_{200}}\right)^{-1/2}
\end{equation}

In Equation \ref{eqn:grav_bound_theory}, the halo mass term on the right-hand side cancels out, leaving only a distance term scaled by the host's virial radius ($r/R_{200}$). In observations, however, only projections of the three-dimensional (3D) separations are available: the on-sky distance between a dwarf galaxy and its host ($d_{\rm proj}$) is a 2D projection of the spatial separation $r$, and the line-of-sight velocity difference ($\Delta V_{\rm los}$) is a 1D projection of the velocity difference $|\Delta \vec{V}|$. We can express Equation \ref{eqn:grav_bound_theory} in the observable plane $(d_{\rm proj}, \Delta V_{\rm los})$ as

\begin{equation}\label{eqn:grav_bound_observable}
    \frac{|\Delta V_{\rm los}|}{V_{\rm esc}(R_{200})} < \frac{1}{R_{V} \sqrt{R_{d}}} \left(\frac{d_{\rm proj}}{R_{200}}\right)^{-1/2}
\end{equation}

where $R_{d} = r/ d_{\rm proj} \geq 1$ and $R_{V} = |\Delta \vec{V}|/ \Delta V_{\rm los} \geq 1$. We adopt the lower limits for the geometric factors, $R_{d} = R_{V} = 1$, and classify galaxies that meet this condition (Equation \ref{eqn:grav_bound_observable}) as satellite candidates; the remainder are considered unbound. This inclusive definition of satellites prioritizes sample completeness at the cost of including potential interlopers (unassociated galaxies that appear close in projection; see \S \ref{sec:gas_loss_discuss}). The resulting classifications and projected phase-space distributions are presented below.

\begin{figure*}[!htb]
    \centering
    \includegraphics[width=0.9\linewidth]{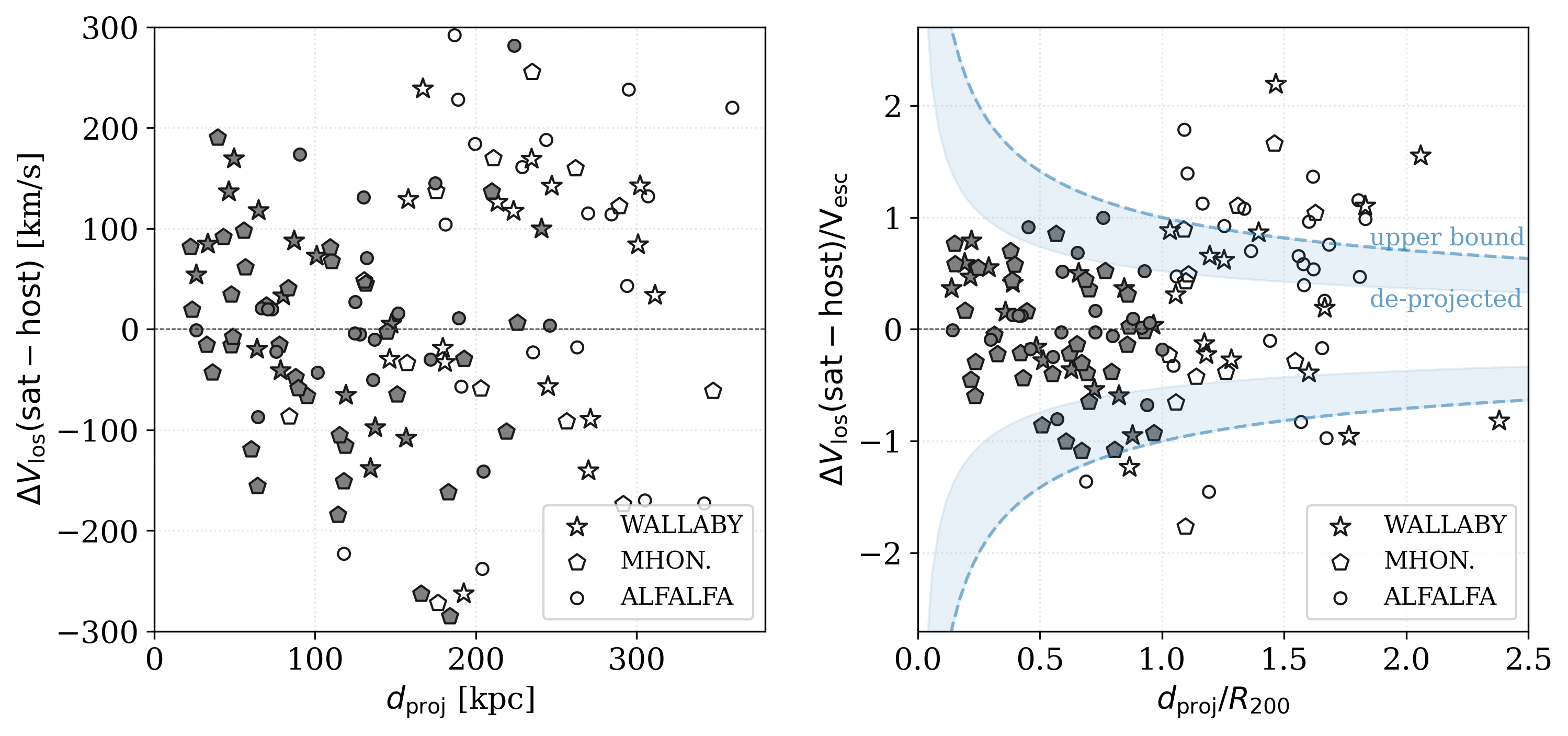}
    \caption{The spatial and kinematic separations between the gaseous dwarf galaxies and their hosts. Left: projected distance ($d_{\rm proj}$) and line-of-sight velocity separation ($\Delta V_{\rm los}$) are in physical units (Table \ref{table:sats}); different symbols distinguish the \HIspace data sources (Section \ref{sec:data}). Right: the same quantities scaled by the host $R_{200}$ and $V_{\rm esc}$ (Section \ref{subsec:host_sample}). Blue curves show the point-mass escape velocity, with dashed lines denoting our satellite-unbound classification (Equation \ref{eqn:grav_bound_observable}; taking $R_{V}, R_{d}=1$) and shaded regions showing the expected range under projection effects. Filled symbols in both panels highlight the ``close" satellite candidates within host projected $R_{200}$.}
    \label{fig:sat_kinematic_distribution}
\end{figure*}

Figure \ref{fig:sat_kinematic_distribution} shows the separations across the observable plane, $d_{\rm proj}-\Delta V_{\rm los}$, where the left-hand panel is in physical units (Table \ref{table:sats}), and the right-hand panel is scaled by host halo properties ($V_{\rm esc}$ and $R_{200}$; see Section \ref{subsec:host_sample}). Our satellite-versus-unbound classification (Equation \ref{eqn:grav_bound_observable}) is denoted by the escape velocity contours in the right-hand panel (dashed blue lines). Galaxies within the escape velocity contours, i.e., with $|\Delta V_{\rm los}| \leq V_{\rm esc}(d_{\rm proj})$, are classified as satellite candidates ($N_{\rm sat,tot}=102$). Otherwise, they are unbound to the hosts ($N_{\rm unbound,tot}=25$). The fraction of unbound sources is low because we only searched for galaxies in the vicinity of the hosts ($\sim$300 kpc in radius and $\pm$300 km/s in velocity; \S \ref{subsec:method_source_finding}). The unbound galaxies are not isolated; their environment lies between that of satellites and isolated ``field" galaxies.

Among the satellite candidates, the average $R_{d}$ and $R_{V}$ for true satellites under randomized viewing angles can be estimated assuming spherical symmetry: $<R_{d}> = \sqrt{3/2}$ \citep{zhu_census_2023} and $<R_{V}> = \sqrt{3}$ \citep{mcconnachie_observed_2012}, respectively. The shaded regions in the right-hand panel illustrate how the escape velocity curves change from our classification (upper bound dashed lines, $R_{V}=R_{d}=1$) down to the average de-projection assuming spherical symmetry (lower bound; $R_{d} = \sqrt{3/2}$, $R_{V}=\sqrt{3}$). Galaxies within the shaded region are more likely to be interlopers than those with lower velocity separations.

Filled symbols in Figure \ref{fig:sat_kinematic_distribution} annotate 73 ``close" satellite candidates that reside within the host projected virial radius ($d_{\rm proj}/R_{200} \leq 1$). The close satellite candidates occupy a projected phase space that is (i) least contaminated by interlopers (\citealt{oman_satellite_2016,rhee_phase-space_2017}; see \S \ref{sec:gas_loss_discuss}), and (ii) almost uniformly covered across our sample, as 50/56 of the hosts have full halo coverage and the remaining six also have a significant coverage ($R_{\rm cover}/R_{200} \geq 0.7$ and full velocity range coverage; Figure \ref{fig:host_m200_halo_rcover}). We will focus on this population when later presenting, e.g., the abundance of satellites per host galaxy (\S \ref{sec:host_sat_abundance}). The close satellite population is denoted with $s_{\rm flag}=1$ in the dwarf galaxy table (Table \ref{table:sats}), the satellite candidates outside of $R_{200}$ as $s_{\rm flag}=2$, and the unbound population as $s_{\rm flag}=3$.

\section{Results}\label{sec:results}

We have built a sample of 127 gaseous dwarf galaxies located near 56 nearby host galaxies -- 41 from WALLABY and MHONGOOSE (as described in \S \ref{subsec:host_sample}) and 15 from ALFALFA \citep{zhu_census_2023}. Among these, 102 are classified as satellite candidates (including possible interlopers) and 25 as gravitationally unbound (\S \ref{subsec:method_sat_environment}). We emphasize that the following results characterize \HI-bearing satellites and therefore do not include gas-poor/quenched satellites around the same hosts. The properties of these dwarfs are listed in Table \ref{table:sats} in Appendix \ref{appendix:dwarfs}. The following sections present results on the dwarf galaxy \HIspace masses and kinematics (\S \ref{subsec:sat_hi_properties}), stellar masses and atomic gas fractions (\S \ref{subsec:sat_gas_star_properties}), satellite depletion versus host mass (\S \ref{subsec:sat_deplete_host_mass}), and the abundance of gaseous satellites per host galaxy (\S \ref{sec:host_sat_abundance}).

The dwarf galaxy masses we present in Table \ref{table:sats} ($M_{\HI}$ and $M_{\star}$) are scaled to their host galaxy distances ($D$ in Table \ref{table:hosts}). Accurate distances are crucial for reliable mass estimations, yet they are costly and often unavailable for individual dwarf galaxies. Velocity-flow distances ($D_{\rm flow}=V_{\odot}/H_{0}$) can carry large uncertainties in the local volume due to peculiar motions. Models that account for large-scale velocity flows \citep{tully_cosmicflows-2_2013,kourkchi_cosmicflows-4_2020} also rely heavily on assumptions of group association. We therefore adopt host distances as a reasonable approximation for true satellites, which typically reside within $\sim 2 R_{200}$ of their hosts (implying sub-Mpc uncertainties). The unbound dwarf galaxies, however, may reside in the host's foreground or background, leading to larger uncertainties in their distance and mass estimations.

\subsection{\HIspace masses and kinematics}\label{subsec:sat_hi_properties}

In this section, we present the \HIspace masses ($M_{\HI}$) and velocity line-widths ($W_{50}$) of our gaseous dwarf sample (Table \ref{table:sats}). In dwarf galaxies, the \HIspace line-widths serve as a proxy for the combined rotational and turbulent motions of neutral gas (see review by \citealt{lelli_gas_2022}), which trace the total masses in galaxies enclosed within \HIspace extent. The line-widths are not corrected for inclination, as such corrections require assumptions about the intrinsic shape of dwarf galaxies \citep{kado-fong_tracing_2020,carlsten_structures_2021}, which remain highly uncertain for low-mass irregular galaxies. The apparent axis ratios of the stellar counterparts are listed in Table \ref{table:sats}. More accurate kinematic modeling (e.g., \citealt{oh_high-resolution_2015,lelli_sparc_2016,deg_wallaby_2022}) can be performed for a spatially resolved subsample and will be explored in future work, since many of our detections are unresolved (smaller than three beams). 

\begin{figure*}[!htb]
    \centering
    \includegraphics[width=1.0\linewidth]{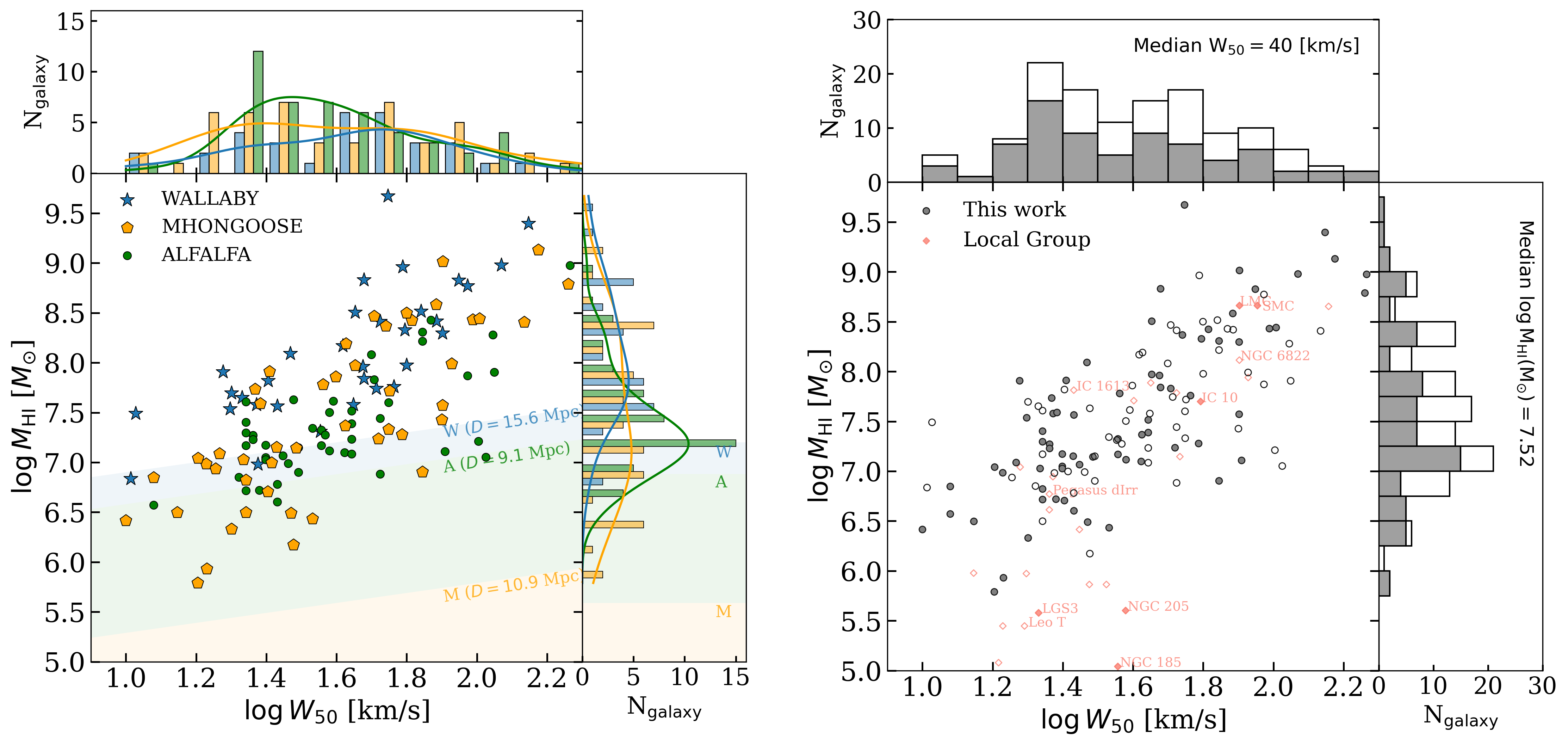}
    \caption{The \HIspace mass ($M_{\HI}$) versus velocity line-width ($W_{50}$) distribution for the gaseous dwarf sample (Table \ref{table:sats}). Scatter points in the central panels represent individual \HI-emitting dwarf galaxies; histograms in the corner panels show the marginal distribution. Left: symbols distinguish the \HIspace surveys, and shadings in the central and right-hand ($M_{\HI}$ marginal) panels show the corresponding \HIspace mass sensitivities. Right: the full sample (circles; combining three surveys) compared with the Local Group gaseous dwarf population in the central panel (light red diamond symbols; \citealt{putman_gas_2021}). Close satellites within the host galaxies' projected $R_{200}$ are shown with filled symbols as in Figure \ref{fig:sat_kinematic_distribution}, or shaded histograms in the marginal panels.}
    \label{fig:sat_HI_mass_width}   
\end{figure*}

Figure \ref{fig:sat_HI_mass_width} summarizes the $M_{\HI} - W_{50}$ distribution of our sample, showing a comparison among the three data sources (left panel), and the dependence on galaxy environment (right panel). We first focus on the left panel. Overall, the gas mass increases with the velocity line width. Across the three surveys, the distribution of dwarf galaxy $M_{\HI}$ closely follows the sensitivity of each dataset, indicated by the colored shaded regions in the joint-distribution panel. The upper bounds of these shaded regions are calculated from Equation \ref{eqn:mhi_sensitivity_compre} for WALLABY (in blue) and MHONGOOSE (in orange), and from the empirical limits of \cite{haynes_arecibo_2011} for ALFALFA (in green), see Section \ref{subsec:method_sensitivity} for details. Although the WALLABY survey is intrinsically deeper than ALFALFA (Figure \ref{fig:dwarf_mhi_sensitivity_compre}), its pilot fields include relatively few nearby host galaxies, making the \textit{effective} $M_{\HI,lim}$ of the WALLABY host galaxies higher than ALFALFA. Most dwarf galaxies in our sample lie above the average sensitivity limits of their respective surveys.

The marginal distribution of $M_{\HI}$ and $W_{50}$ is shown in the corner panels. For the $M_{\HI}$ marginal, the colored shaded regions again indicate the survey sensitivity limits (as in the joint-distribution panel), here averaged at the median velocity width of the full sample ($W_{50}=40$ km/s). Each survey's subsample is heavily truncated below its corresponding $M_{\HI}$ limit. The median$\pm34\%$ range is $\log (M_{\HI}/M_{\odot}) = 7.56_{-0.52}^{+0.81}$ for WALLABY and ALFALFA, and $\log (M_{\HI}/M_{\odot}) = 7.35_{-0.85}^{+1.08}$ for MHONGOOSE. The velocity width marginal distribution is relatively flat ($W_{50} = 40_{-18}^{+40} ~\rm km~s^{-1}$). The velocity resolutions of WALLABY ($V_{\rm res} \approx 4$ km/s) and MHONGOOSE ($V_{\rm res} \approx 1.4$ km/s) are sufficient to spectrally resolve the lowest-mass gas-rich galaxies (e.g., Leo T has $W_{50} \approx 20$ km/s; \citealt{adams_deep_2018}), while ALFALFA ($V_{\rm res}$ smoothed to 10 km/s) is approaching the resolution limit, and therefore the ALFALFA subsample is truncated at low $W_{50}$ values (green histograms). However, we note that the detections with the lowest $W_{50}$ values are faint, marginally resolved sources at low SNR, for which $W_{50}$ is likely underestimated (e.g., \citealt{westmeier_sofia_2021}).

In the right panel of Figure \ref{fig:sat_HI_mass_width}, we present the distribution of our full gaseous dwarf sample, combining the subsamples from different surveys. Here, we distinguish the environment of these galaxies: close satellite candidates within the host's projected $R_{200}$ are shown as filled symbols and histograms, while those outside of $R_{200}$ -- including the gravitationally unbound nearby galaxies -- are shown as open symbols and unfilled histograms. These two populations, separated by the environment, are nearly indistinguishable across the $M_{\HI}-W_{50}$ plane. The mean and standard deviation are $W_{50} = 50\pm41$ km/s, $\log M_{\HI}/M_{\odot}=7.57\pm0.83$ for the close satellites, and $W_{50} = 50 \pm 36$ km/s, $\log M_{\HI}/M_{\odot}=7.59\pm0.75$ for the full sample. The median values of the full samples are additionally annotated in the marginal panels.

We also show the Local Group gaseous dwarf population (diamond symbols; 26 within 2 Mpc from \citealt{putman_gas_2021}) for comparison in the right panel of Figure \ref{fig:sat_HI_mass_width}. Among these 26, 11 are bound satellites of the MW and M31 (names annotated in the Figure), and 6 of the 11 are close satellites within the hosts' projected $R_{200}$ (filled diamonds; see \citealt{zhu_census_2023}). The velocity widths of the Local Group dwarfs are adopted from \cite{karachentsev_updated_2013} and updated for individual galaxies where targeted deep \HIspace observations are available (
see Table 2 of \citealt{zhu_census_2023}). Despite the small numbers, there is a significant spread in the Local Group gaseous dwarf galaxies' $M_{\HI}-W_{50}$ distribution. The classical dwarf irregulars in the Local Group at $M_{\HI} \geq 10^{6} M_{\odot}$ agree well with the distribution of our satellite sample. The smallest gaseous dwarf galaxies in the Local Group ($M_{\HI} < 10^{6} M_{\odot}$), on the other hand, have masses below the typical sensitivity limits of our surveys and lack direct counterparts in our satellite sample. A few Local Group dwarfs have unusually low $M_{\HI}$ at fixed $W_{50}$, likely due to gas depletion; we return to these systems in the atomic gas fraction comparison below (\S \ref{subsec:sat_gas_star_properties}). The Local Group ``field" dwarfs (open diamond symbols without names) are also indistinguishable in the mass-kinematics plane from the satellites, similar to our extragalactic sample.

Our sample reaches \HIspace masses within a factor of two of Leo T, but still misses the lowest-mass Local Group-like gaseous dwarfs, particularly around the ALFALFA and WALLABY hosts. 
At the detected masses, the close satellite candidates and the nearby unbound dwarfs are qualitatively similar in the $M_{\HI}-W_{50}$ plane.

\subsection{Stellar masses and atomic gas fractions}\label{subsec:sat_gas_star_properties}

This section presents the relation between the stellar and atomic gas masses for the gaseous dwarf satellites. Stellar masses are derived for 110 out of the 127 total gaseous dwarf galaxies in our sample (Table \ref{table:sats}), and the remaining 17 lack data in one or more bands in Legacy DR10 (\S \ref{subsec:method_stellar_mass}). We first present the gas-and-stellar mass relation of our dwarf sample ($M_{\HI}-M_{\star}$; \S \ref{subsec:mgas_mstar}), characterizing the effects of survey \HIspace sensitivity and satellite-host separations. We then explore the corresponding atomic gas fractions ($f_{\HI} = M_{\HI}/M_{\star}$; \S \ref{subsec:gas_fraction}), comparing our satellite sample with \HI-bearing dwarf galaxy samples from the literature.

\subsubsection{Stellar and gas masses of dwarf satellites}\label{subsec:mgas_mstar}

\begin{figure*}
    \centering
    \includegraphics[width=1.0\linewidth]{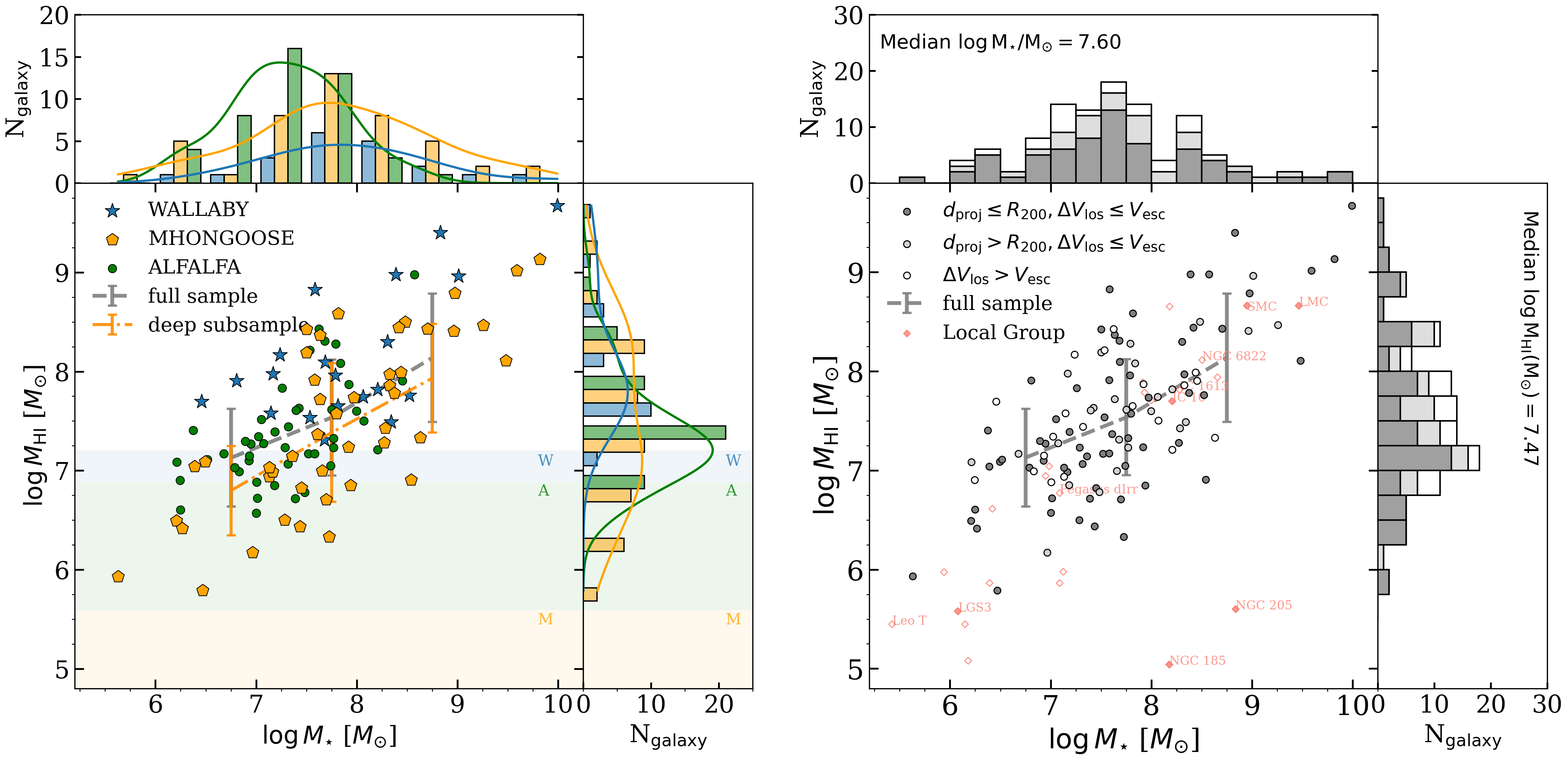}
    \caption{The \HIspace versus stellar mass distribution of the gaseous dwarf sample (Table \ref{table:sats}). Left: symbols distinguish the three \HIspace surveys, with average \HIspace sensitivities shaded as in Figure \ref{fig:sat_HI_mass_width}. Error bars show the binned means and standard deviations for the full sample (gray dashed lines) and the deep subsample (orange dashed-dotted lines). Right: shadings indicate the position-velocity separation from the host (\S \ref{subsec:method_sat_environment}): dark gray for satellite candidates within $R_{200}$, light gray for objects outside $R_{200}$ but bound, and open symbols for unbound nearby galaxies. The Local Group gas-rich population are shown in light red.}
    \label{fig:sat_mgas_mstar}
\end{figure*}

Figure \ref{fig:sat_mgas_mstar} shows the gas-to-stellar mass distribution of the dwarf galaxies. The \HIspace masses generally increase with stellar masses but show a large scatter across the full $M_{\star}$ range. The left panel distinguishes the three \HIspace surveys, displaying a gradient with \HIspace sensitivity (colored shadings): deeper surveys detect lower average $M_{\HI}$ values. To quantify this effect, we report $\log M_{\HI} - \log M_{\star}$ statistics for both the full sample (gray dashed lines) and the deeper half ($\log M_{\HI, lim}/M_{\odot} \leq 6.7$; orange dashed-dotted line), which includes all MHONGOOSE galaxies and satellites around the two deepest ALFALFA host systems (Figure \ref{fig:dwarf_mhi_sensitivity_compre} and Table \ref{table:hosts}). Binned results at $\log M_{\star}/M_{\odot}=[6.75, 7.75, 8.75]$ give $\log M_{\HI}/M_{\odot~\rm (all)}=[7.13 \pm 0.49, 7.54 \pm 0.59, 8.14 \pm 0.65]$ for the full sample and $\log M_{\HI}/M_{\odot~\rm (deep)} = [6.80 \pm 0.45, 7.38 \pm 0.70, 7.93 \pm 0.55]$ for the deep subsample. The $\sim0.2-0.3$ dex lower averages in the deep subsample confirm that in the \HIspace flux-limited regime, \HIspace sensitivity strongly affects the observed average $M_{\HI}$ values.

The right panel of Figure \ref{fig:sat_mgas_mstar} shows the same data as the left, now shaded by satellite-host separation (defined in \S \ref{subsec:method_sat_environment}). Among the dwarf galaxies with stellar mass measurements, about 58\% are close satellite candidates within the host projected $R_{200}$ (dark gray), 24\% are satellite candidates outside $R_{200}$ (light gray), and 18\% are unbound (open symbols). These three groups show no clear distinction in their $M_{\HI} - M_{\star}$ distribution (central-right panel). We note that the mass of the host galaxy also affects the satellite environment, which will be examined in the next section (\S \ref{subsec:sat_deplete_host_mass}).

The stellar masses span $\log M_{\star}/M_{\odot} = 7.60_{-1.28}^{+1.40}$ (median and $5-95\%$ range; upper-right panel of Figure \ref{fig:sat_mgas_mstar}). Low-mass satellites ($M_{\star} \approx 10^{6.5} M_{\odot}$) are rare, partially due to \HIspace sensitivities (see left panel) and likely also because they are more susceptible to environmental gas loss. Among the 30 MHONGOOSE dwarf galaxies with host sensitivities deeper than $M_{\HI} \approx 10^{6} M_{\odot}$, only one reaches $M_{\star,\rm sat} < 10^{6} M_{\odot}$: MKT J045726.5-532423.6, a satellite candidate of a Magellanic-mass host NGC 1705. Massive satellites above $M_{\star} \approx 10^{9} M_{\odot}$ are also rare: although they are more likely to retain their gas at $z=0$ and (unless extremely gas-depleted) are well above our \HIspace sensitivity limits, their number density per host is intrinsically low (e.g., \citealt{tollerud_small-scale_2011,robotham_galaxy_2012}). The most massive case, NGC 6215 ($M_{\star,\rm sat} \approx 10^{9.99} M_{\odot}$), is an interacting dwarf companion of NGC 6221 and has an unusually low $W_{50}$ for its gas and stellar mass (Table \ref{table:sats}; see also \citealt{koribalski_neutral_2004}), likely due to interactions with the host.

The right panel of Figure \ref{fig:sat_mgas_mstar} includes the 24 \HI-bearing dwarf galaxies\footnote{Out of the 26 \HI-bearing dwarfs from \cite{putman_gas_2021}, two objects, HIZSS3A and HIZSS3B, are in the zone of avoidance and do not have stellar mass measurements. We excluded them in the stellar mass comparison here.} in the Local Group. Stellar masses are adopted from the Local Volume Database\footnote{\url{https://github.com/apace7/local_volume_database}} \citep{pace_local_2025}. At comparable masses, the Local Group dwarf galaxies agree well with our satellite sample: nearly all systems above our sensitivity limits (left panel of Figure \ref{fig:sat_mgas_mstar}) fall within the $\pm 1\sigma$ range from our sample averages (gray error bars). As in our sample, Local Group dwarf galaxies in different environments --- satellites within the MW's or M31's projected $R_{200}$ (filled red diamonds), satellites outside of $R_{200}$ (open diamonds with labels), and the ``field" dwarfs unbound to either MW or M31 (unlabeled open diamonds) --- show no clear separation in the $M_{\HI}-M_{\star}$ plane. A few remaining Local Group dwarf galaxies have very low \HIspace masses ($M_{\HI} < 10^{5.6} M_{\odot}$) that are below our typical sensitivity limits. While they have no direct counterparts in our data, we can compare with their atomic gas fractions below.

The $M_{\HI}-M_{\star}$ relation shows that the stellar and gas masses are positively correlated, though with substantial scatter. Independent of environment, lower-mass galaxies typically have more gas than stars compared with massive galaxies (see the Introduction). To better quantify gas richness across the dwarf galaxy population and compare with literature samples spanning different environments, we next examine the atomic gas fraction ($f_{\HI}=M_{\HI}/M_{\star}$) as a function of stellar mass.

\subsubsection{Atomic gas fractions}\label{subsec:gas_fraction}

\begin{figure}
    \centering
    \includegraphics[width=1.0\linewidth]{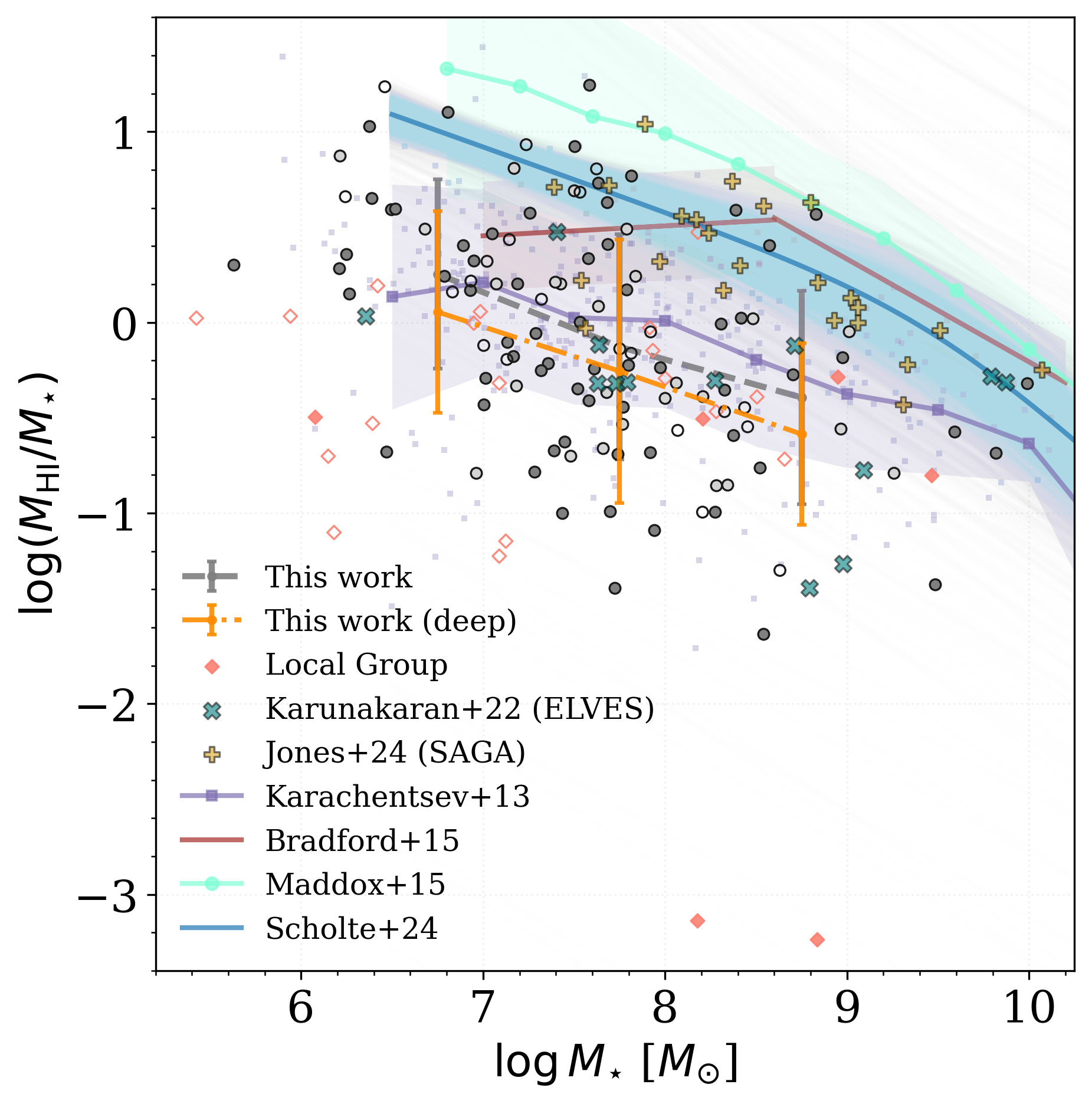}
    \caption{Atomic gas fraction ($\log M_{\HI}/M_{\star}$) versus stellar mass for dwarf galaxies, comparing our local volume satellite sample with literature results. Our sample (circles and error bars) and Local Group gaseous dwarfs (light-red diamonds) follow the same style as Figure \ref{fig:sat_mgas_mstar}, where symbol shading/fill indicates whether each dwarf is within $R_{200}$, bound beyond $R_{200}$, or unbound in projected phase space. Additional satellite comparison samples are shown from \cite{karunakaran_h_2022} (ELVES; crosses) and \cite{jones_gas_2024} (SAGA; pluses), including only \HIspace detections. Lines and shaded regions show the mean and $1\sigma$ scatter from field or mixed-environment comparison samples: \cite{karachentsev_updated_2013} (purple; individual galaxies in light squares), \cite{bradford_study_2015} (brown), \cite{maddox_variation_2015} (aqua), and \cite{scholte_atomic_2024} (blue; \HIspace flux-limited); see \S \ref{subsec:sat_gas_star_properties}.}
    \label{fig:sat_mgas_mstar_ratio}
\end{figure}

Figure \ref{fig:sat_mgas_mstar_ratio} presents the atomic gas fraction ($f_{\HI}$) for our satellite sample and literature comparison samples. Our data are shown with the same symbol style as in Figure \ref{fig:sat_mgas_mstar}, and we plot $\log f_{\HI}$ versus $\log M_{\star}$ with binned statistics for the full sample (gray dashed line) and the deep subsample (orange dashed-dotted line). For consistency with this \HI-selected sample, we compare only with \HIspace detections in the literature samples (excluding upper limits or gas-poor objects).

The field-dominated or mixed-environment comparison samples are shown as lines and shaded $1\sigma$ ranges. Nearby galaxies from \cite{karachentsev_updated_2013}\footnote{Updated data from \url{https://www.sao.ru/lv/lvgdb/tables.php}} include 314 dwarf galaxies at $2<D<11$ Mpc, roughly half satellites and half field systems. 
\cite{bradford_study_2015} includes 148 isolated dwarf galaxies at $M_{\star} = 10^{7}-10^{9.5} M_{\odot}$ out to $D \approx 150$ Mpc, where $M_{\rm gas}$ values are converted back to \HIspace masses via $M_{\HI} \approx M_{\rm gas}/1.4$. The \cite{maddox_variation_2015} and \cite{scholte_atomic_2024} samples contain $\sim$9000 (40\% of ALFALFA; \citealt{haynes_arecibo_2011}) and $\sim$24000 (100\% of ALFALFA; \citealt{haynes_arecibo_2018}) \HI-selected galaxies out to $z \approx 0.06$, respectively. 
These samples consistently show that $f_{\HI}$ decreases with increasing $M_{\star}$, but the observed averages depend strongly on \HIspace sensitivity. Because $M_{\HI,lim} \propto \sigma_{\rm survey} \cdot D^{2}$ (Equation \ref{eqn:mhi_sensitivity_vflux}; $\sigma_{\rm survey}$ is comparable here because these comparison samples all use ALFALFA data), distant, more flux-limited surveys (e.g., \citealt{maddox_variation_2015,scholte_atomic_2024}) are effectively less sensitive and yield higher average $f_{\HI}$ than nearby, more sensitive samples (e.g., \citealt{karachentsev_updated_2013}). 

Our satellite sample follows a similar decreasing $f_{\HI}-M_{\star}$ trend but lies systematically below these comparison samples. Our ALFALFA and WALLABY subsamples have comparable sensitivity limits to \cite{karachentsev_updated_2013} and also show nearly identical binned averages ($\log f_{\HI} \approx  [0.3, -0.1, -0.2]$ at the $M_{\star}$ bins of Figure \ref{fig:sat_mgas_mstar}), suggesting that at matched \HIspace sensitivity, detected satellites and field dwarf galaxies can have similar $f_{\HI}$ distributions (consistent with the finding of \citealt{bradford_study_2015}). The lower $f_{\HI}$ averages in our full sample and deep subsample arise from the inclusion of the deeper MHONGOOSE observations, which can recover lower $M_{\HI}$ at fixed $M_{\star}$ (Figure \ref{fig:dwarf_mhi_sensitivity_compre}).

We also compare with \HIspace follow-up observations from two larger satellite studies: ELVES satellites from \cite{karunakaran_h_2022} (crosses; 14 \HIspace detections out of 66) and SAGA satellites from \cite{jones_gas_2024} (pluses; 23 detections out of 46). The \HI-detected satellites in \cite{karunakaran_h_2022} broadly agree with our full-sample averages, while those from \cite{jones_gas_2024} lie systematically higher by $\sim 0.7$ dex. This offset is consistent with differences in \HIspace sensitivity, indicating that the low gas fractions in our sample relative to existing literature cannot be interpreted as a purely environmental effect. Matched-depth \HIspace observations of field and satellite samples are needed to isolate the environmental contribution.

The Local Group population (light-red symbols in Figure \ref{fig:sat_mgas_mstar_ratio}) is broadly consistent with our satellite sample (also see Figure \ref{fig:sat_mgas_mstar}). The lowest-mass systems ($M_{\star} \leq 10^{6.5}~M_{\odot}$) have $f_{\HI} \approx 0.1-1$; although many fall below our \HIspace sensitivity limit (e.g., Leo T), their gas fractions are consistent with the lowest-mass satellites in our sample (Figure \ref{fig:sat_mgas_mstar_ratio}). Two M31 dwarf ellipticals (dEs), NGC 185 and NGC 205, are extreme gas-poor outliers ($f_{\HI} < 10^{-3}$) with truncated \HIspace relative to optical emission \citep{young_neutral_1997}, possibly due to past interactions with M31. Our sample also contains one notably gas-poor lenticular galaxy, NGC 1581 ($\log M_{\star}/M_{\odot} \approx 9.5$ and $f_{\HI} \approx 10^{-1.4}$), whose diffuse \HIspace emission suggests past interaction with its host NGC 1566 \citep{maccagni_mhongoose_2024}, possibly driving the morphological transformation in the satellite.

\subsection{Satellite gas depletion versus host mass}\label{subsec:sat_deplete_host_mass}

In the previous section, we presented the atomic gas fraction ($f_{\HI}$; Figure \ref{fig:sat_mgas_mstar_ratio}) of our dwarf galaxy sample, which does not show a clear trend with the satellite-host separation scaled by host mass (right panel of Figure \ref{fig:sat_mgas_mstar}). In this section, we examine the dependence of the satellite gas depletion on their host masses.

\begin{figure}[!htb]
    \centering
    \includegraphics[width=1.0\linewidth]{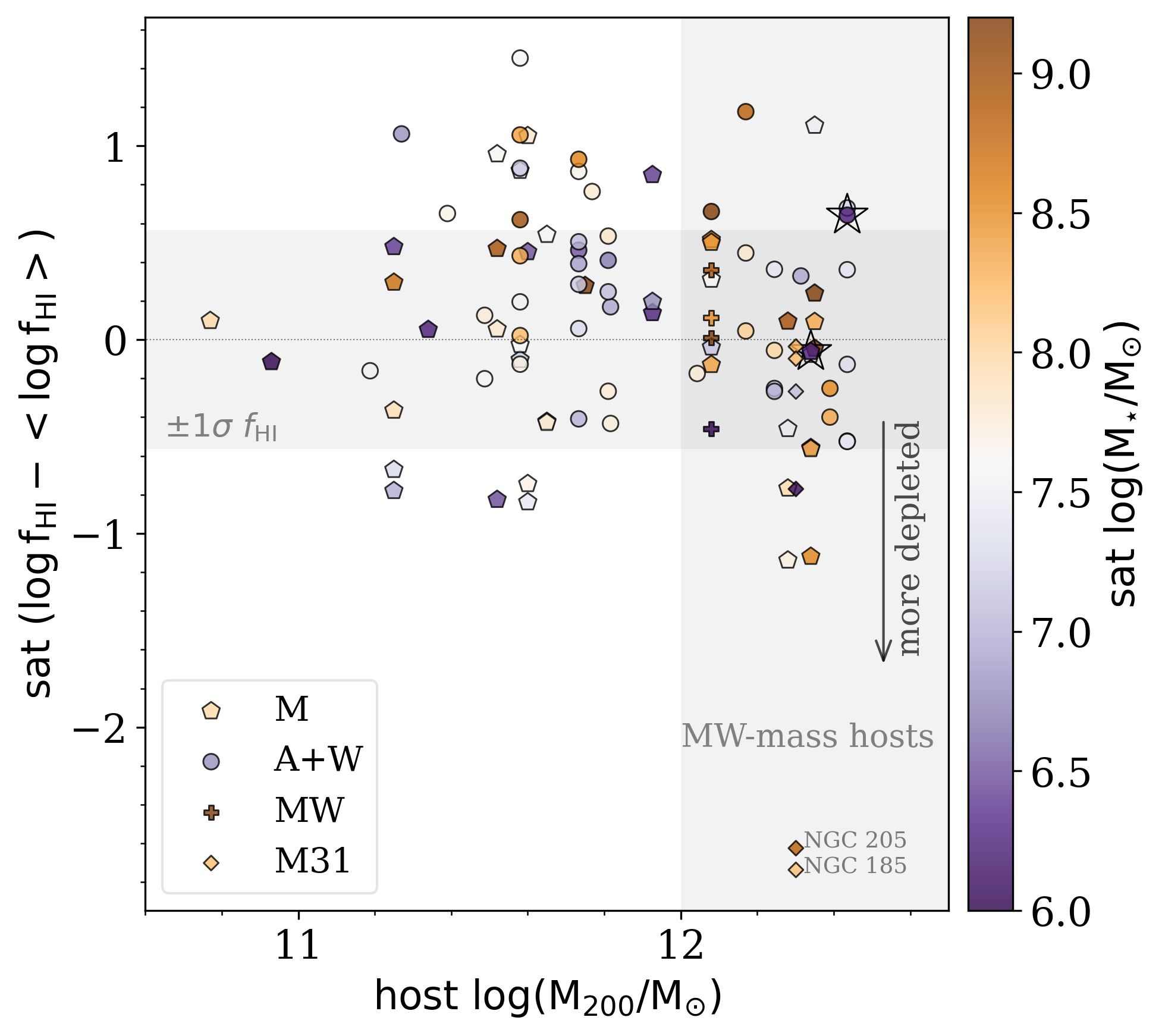}
    \caption{Satellite gas depletion versus host halo mass, colored by satellite stellar mass. Gas depletion is defined as a satellite's atomic gas fraction ($f_{\HI}=M_{\HI}/M_{\star}$) relative to the expected fraction at a given stellar mass, $<f_{\HI}>$ (Equation \ref{eqn:hi_depletion}). MHONGOOSE satellites are shown as pentagons, ALFALFA and WALLABY as circles, and MW/M31 gaseous satellites from \cite{putman_gas_2021} as crosses/diamonds. Gray shaded regions mark the $\pm 1\sigma$ scatter in $<\log f_{\HI}>$ (horizontal) and the MW-mass hosts ($\log M_{200}/M_{\odot} \geq 12$; vertical).  Low-mass satellite candidates of MW-mass hosts are highlighted with stars.}
    \label{fig:sat_mhi_delta_fhi_host_mhalo}
\end{figure}

Metrics of gas depletion (e.g., \citealt{cortese_dawes_2021}) need to account for the intrinsic stellar mass dependence of the atomic gas fraction (Figure \ref{fig:sat_mgas_mstar_ratio}). Here, we define gas depletion ($\Delta \log f_{\HI}$) as the difference between the measured $\log f_{\HI}$ of a dwarf galaxy and the expected average fraction ($<\log f_{\HI}>$) at a given stellar mass,
\begin{equation}\label{eqn:hi_depletion}
    \Delta \log f_{\HI} = 
    \log (M_{\HI}/M_{\star}) - <\log (M_{\HI}/M_{\star})>
\end{equation}
where the expected fraction, $<\log f_{\HI}>$, is derived from our deep sensitivity subsample (orange error bars; Figure \ref{fig:sat_mgas_mstar_ratio}). We parameterize the average $\log f_{\HI}-\log M_{\star}$ dependence using linear regression, obtaining $<\log f_{\HI}> = -0.32 \log M_{\star} + 2.22$ and a scatter of $\sigma \approx 0.56$ dex. In the absence of a field sample at comparable sensitivity, we use our deep subsample (mostly MHONGOOSE galaxies) as the baseline for gas fractions because it is least biased by \HIspace incompleteness. If using a field sample from the literature (Figure \ref{fig:sat_mgas_mstar_ratio}) instead, $<\log f_{\HI}>$ will adopt a similar slope but a higher intercept due to missing the most gas-poor dwarfs. Importantly, $\Delta \log f_{\HI}$ represents the \textit{relative} degree of gas depletion within our satellite sample; its baseline may differ for future field samples at similar sensitivities.

Figure \ref{fig:sat_mhi_delta_fhi_host_mhalo} shows the degree of gas depletion ($\Delta \log f_{\HI}$; Equation \ref{eqn:hi_depletion}) of the gaseous satellite candidates versus their host halo masses. Dwarf centrals in our sample ($M_{200} \leq 10^{11} M_{\odot}$) rarely contain gaseous satellites (see Section \ref{sec:host_sat_abundance} below) and are not included in the figure. ALFALFA and WALLABY galaxies (circles) are biased gas-rich relative to the MHONGOOSE galaxies (pentagons, including three ALFALFA galaxies with deep sensitivities; see \S \ref{subsec:sat_gas_star_properties}) due to the sensitivity differences. 

In lower-mass satellites ($\log M_{\star,\rm sat}/M_{\odot} \leq 7.6$; Figure \ref{fig:sat_mhi_delta_fhi_host_mhalo}), gas depletion shows no significant correlation with host mass. But for intermediate-to-massive satellites ($\log M_{\star,\rm sat}/M_{\odot} > 7.6$), $\Delta \log f_{\HI}$ decreases with increasing host halo mass (Pearson $r=-0.3$, $p=0.04$; more negative $\Delta \log f_{\HI}$ indicates higher depletion). The most depleted candidates in our sample ($\Delta \log f_{\HI} \approx -1$) are massive satellites found around massive spirals ($\log M_{200}/M_{\odot} \approx 12.3$; NGC 1566 and NGC 1371), similar to but not as extreme as M31's highly depleted dEs discussed above. Massive satellites around lower-mass hosts show no signs of gas depletion, often appearing gas-rich ($\Delta \log f_{\HI} > 0.5$). Around the MW-mass hosts ($\log M_{200}/M_{\odot} > 12$; gray vertical band), the fact that only massive satellites appear strongly depleted may reflect their longer quenching timescales and ability to retain part of the ISM under environmental stripping (e.g., \citealt{zhu_too_2026}). Less massive satellites experiencing similar environmental effects may have already been stripped and are therefore less likely to be observed in our \HI-selected sample.

Among the nine lowest-mass satellite candidates in our sample with $M_{\star} \leq 10^{6.5} M_{\odot}$, only two are found around MW-mass hosts (marked as open stars in Figure \ref{fig:sat_mhi_delta_fhi_host_mhalo})\footnote{Occurrence rates of these low-mass satellites are generally limited by \HIspace sensitivities.}. These two detections are of particular interest not for their degree of gas depletion, but for the fact that they retain any gas at all. Simulations predict that the ISM in satellites with $M_{\star} \leq 10^{6.5} M_{\odot}$ should be completely stripped within $\lesssim 1$ Gyr after a typical pericentric passage in a MW-mass halo (see \citealt{rodriguez-cardoso_agora_2025,zhu_too_2026} and references therein). These two ``survived" galaxies appear similar to Leo T (MW) and LGS 3 (M31) --- the two lowest-mass Local Group satellites that retain their \HIspace gas, while all other satellites of the MW and M31 at $M_{\star} \leq 10^{6.5} M_{\odot}$ are gas poor \citep{putman_gas_2021}. We will later show that, unlike Leo T and LGS 3, these two galaxies are likely interlopers (Section \ref{sec:gas_loss_discuss}).

\subsection{The abundance of gaseous satellites}\label{sec:host_sat_abundance}

In this section, we examine the abundance of gaseous satellites ($N_{\HI~sat}$) per host galaxy. We focus on the satellite candidates within the host's projected $R_{200}$, where contamination is minimized and spatial coverage is most complete (Section \ref{subsec:method_sat_environment}). Satellite counts are limited by survey sensitivity limits, characterized for each host galaxy in Section \ref{subsec:method_sensitivity}. We present these limits along with the satellite abundance results throughout the section.

\begin{figure*}[!htb]
    \centering
    \includegraphics[width=1.0\linewidth]{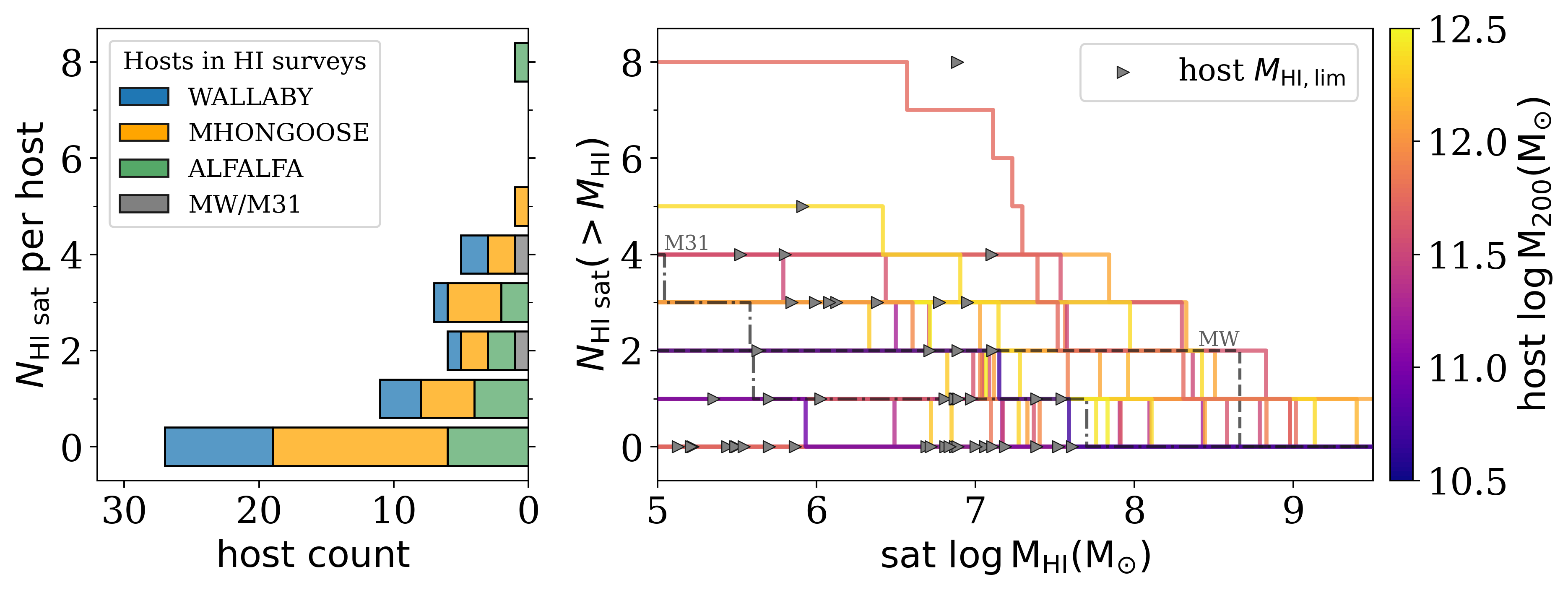}
    \caption{The abundance of \HI-bearing satellites per host. Left: distribution of close satellite abundance (within projected $R_{200}$) per host, colored by survey. This histogram combines a range of host masses; we show the $N_{\HI~\rm sat}-M_{\rm host}$ relation separately in Figure \ref{fig:host_mass_nsat_rvir_cut}. Right: the \HI-mass function of the gaseous satellites per host, defined as the number of gaseous satellites above a given mass, colored by the host halo mass. Right triangles mark the host sensitivity limit $M_{\HI,lim}$ for typical dwarfs ($W_{50}=25$ km/s). The MW and M31 are shown in gray for reference.}
    \label{fig:host_nsat_hist_and_sat_himf}
\end{figure*}

The left panel of Figure \ref{fig:host_nsat_hist_and_sat_himf} shows the distribution of gaseous satellite counts across the 56 host galaxies. Overall, the abundance of gaseous satellites is low: about half (30/56) have none within $R_{200}$, and almost all (54/56) have $N_{\HI~\rm sat} \leq 4$. For comparison, the MW ($N_{\HI~\rm sat}=2$; LMC and SMC) and M31 ($N_{\HI~\rm sat}=4$, NGC 205, NGC 185, IC 10, and LGS 3) are shown in gray\footnote{The MW and M31 population are based on satellite \textit{projected} distances within $R_{200}$, following mock Local Group observations from \cite{zhu_census_2023}.}. Our host sample spans over two dex in halo mass (Table \ref{table:hosts}), from dwarf centrals to M31-mass spirals. We explore the dependence of gaseous satellite abundance on host mass later in this section (Figure \ref{fig:host_mass_nsat_rvir_cut}).

The right panel shows the cumulative \HIspace mass function (HIMF) of satellites per host galaxy. Each colored line represents one host galaxy and counts the cumulative number of its satellites above a given \HIspace mass on the x-axis. For example, the MW's (gray dashed line) two Magellanic Clouds both have $M_{\HI} \approx 10^{8.7} M_{\odot}$, where the step-like line rises by two. The total count at the lowest-mass end ($M_{\HI} \rightarrow 10^{5} M_{\odot}$) matches $N_{\HI~sat}$ in the left panel. Sensitivity limits for each host ($M_{\HI,lim}$; Table \ref{table:hosts}) are marked with triangles, below which the dwarf galaxy mass distributions are typically truncated (except for two ALFALFA hosts with slightly deeper detections).

For the 17 hosts with multiple satellites ($N_{\HI~\rm sat} \geq 2$), the \HIspace mass distribution spans a wide range. The median and $10\%-90\%$ spread between the most and least massive satellites of the same host is $\Delta \log M_{\HI, (max-min)} = 1.3 ^{+0.9}_{-1.0}$ dex, consistent with the overall broad $M_{\HI}$ distribution seen in our full satellite sample (Section \ref{subsec:sat_hi_properties}) and with the relatively flat HIMF found in galaxy group environments (\citealt{jones_h_2020}; also see \citealt{pisano_h_2011} for Local Group-like environments).

\subsubsection{Gaseous satellite count versus host mass}

\begin{figure}[!htb]
    \centering
    \includegraphics[width=1.0\linewidth]{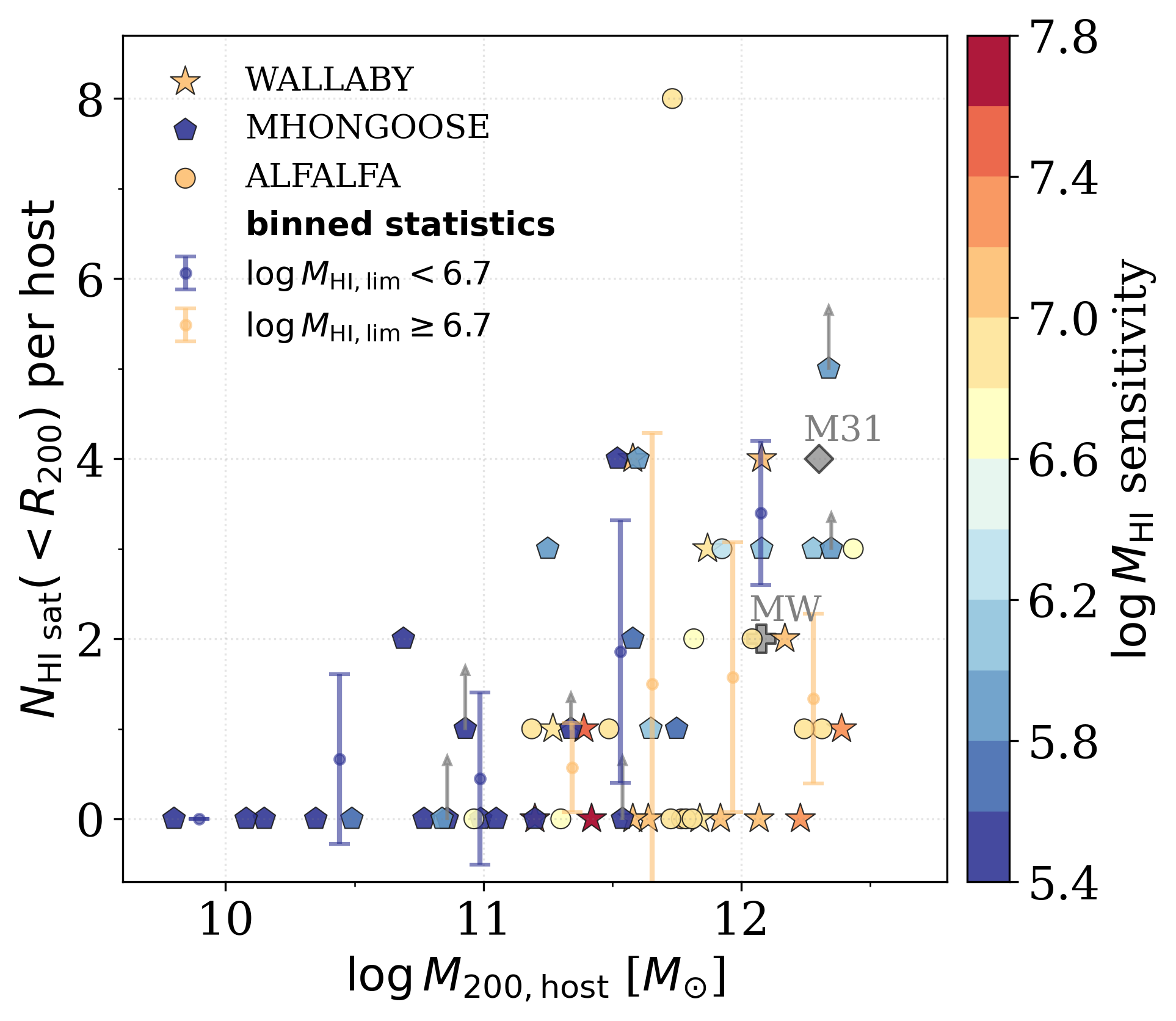}
    \caption{The number of \HI-bearing satellite candidates within projected $R_{200}$ versus the host halo mass. Points show individual hosts, colored by their \HIspace sensitivity limits. Hosts with partial halo coverage (Table \ref{table:hosts}) are marked by upper arrows (four with $R_{\rm cover}/R_{200}=70-90\%$ in full-size arrows, three with $90-100\%$ in half-size arrows). Error bars show binned means and $1\sigma$ scatter for two groups of our sample separated by median sensitivity: the deeper half with $\log M_{\HI,lim}< 6.7$ (in blue) and the rest with $\log M_{\HI,lim} \geq 6.7$ (in yellow).}
    \label{fig:host_mass_nsat_rvir_cut}
\end{figure}

Figure \ref{fig:host_mass_nsat_rvir_cut} shows the number of \HI-bearing satellites within the host virial radius ($N_{\HI~\rm sat}(<R_{200})$) versus the host halo mass ($M_{\rm 200,host}$), colored by the \HIspace sensitivity limit ($\log M_{\HI,lim}$). We find a positive correlation (Pearson $r=0.42$, $p=0.001$) that more massive hosts tend to have more gaseous satellites. The lowest-mass dwarf hosts ($M_{200} < 10^{11} M_{\odot}$) are mostly MHONGOOSE galaxies (Figures \ref{fig:host_m200_halo_rcover} and \ref{fig:dwarf_mhi_sensitivity_compre}) with excellent mass sensitivities ($M_{\HI,lim} \approx 10^{5.4} M_{\odot}$ or Leo T-level; Figure \ref{fig:host_mass_nsat_rvir_cut}). Yet, most (12/14) dwarf centrals have no gaseous satellites detected within $R_{200}$. The two exceptions are (i) UGCA 320 (J1303-17b; \citealt{de_blok_mhongoose_2024}), which has two companions at very close velocity separations (both at $\delta V_{\odot}\approx -16$ \kms), and (ii) NGC 1705 (J0454–53; \citealt{de_blok_mhongoose_2024}), which has an incomplete halo coverage ($R_{\rm cover}/R_{200} \approx 0.8$) and one very low-mass satellite within coverage, detected in the deep release MHONGOOSE data (Section \ref{subsec:method_source_finding}). Above $M_{200} \geq 10^{11} M_{\odot}$, over 60\% (29/45) host galaxies have one or more gaseous satellites detected. At spiral masses ($M_{200} \geq 10^{11.5} M_{\odot}$), the relation between satellite count and host mass becomes weaker (Pearson $r=0.22$, $p=0.19$), with large host-to-host scatter likely driven by observational sensitivity effects combined with intrinsic variations.

To assess sensitivity effects, we split the sample at the median sensitivity ($\log M_{\HI,lim}/M_{\odot} \approx 6.7$) when presenting the binned statistics in Figure \ref{fig:host_mass_nsat_rvir_cut}. The deeper half (28/56 hosts, mostly MHONGOOSE; blue) shows a stronger correlation between increasing $N_{\HI~sat}$ and increasing $M_{200,\rm host}$, which extends to MW-mass and higher. At $M_{200} \approx 10^{12} M_{\odot}$, the deep subsample reaches an average $N_{\HI~sat} = 3.4 \pm 0.8$, while the shallow subsample (WALLABY+ALFALFA; yellow) flattens to $N_{\HI ~sat} = 1.5 \pm 1.3$ (Figure \ref{fig:host_mass_nsat_rvir_cut}). The WALLABY and ALFALFA subsample thus likely misses $\sim 1-2$ low-mass satellites ($M_{\HI,\rm sat} \approx 10^{5.5}-10^{7} M_{\odot}$) per host.

Despite the narrow range (typical $N_{\HI~sat} = 0-4$), there is a non-negligible scatter in satellite counts among the spiral hosts ($\sigma_{N_{\rm sat}} \approx 1-2$). Two ``outlier'' hosts, NGC 3486 ($N_{\HI~sat}=8$) and NGC 1566 ($N_{\HI~sat}\geq 5$), drive much of this spread. NGC 3486 is an intermediate mass spiral where all eight companions reside at very close velocity separations ($|\Delta V_{\rm los}| \lesssim 30$ km/s). NGC 1566 is a massive grand-design spiral ($\log M_{200} = 12.34$; Table \ref{table:hosts}) where only $72\%$ of the virial area ($R_{\rm cover}/R_{200} \approx 85\%$) is covered in MHONGOOSE, thus the five satellite count is a lower limit.

\subsubsection{Local environment effects}\label{subsec:sat_abundance_local_env}

\begin{figure}[!htb]
    \centering
    \includegraphics[width=1.0\linewidth]{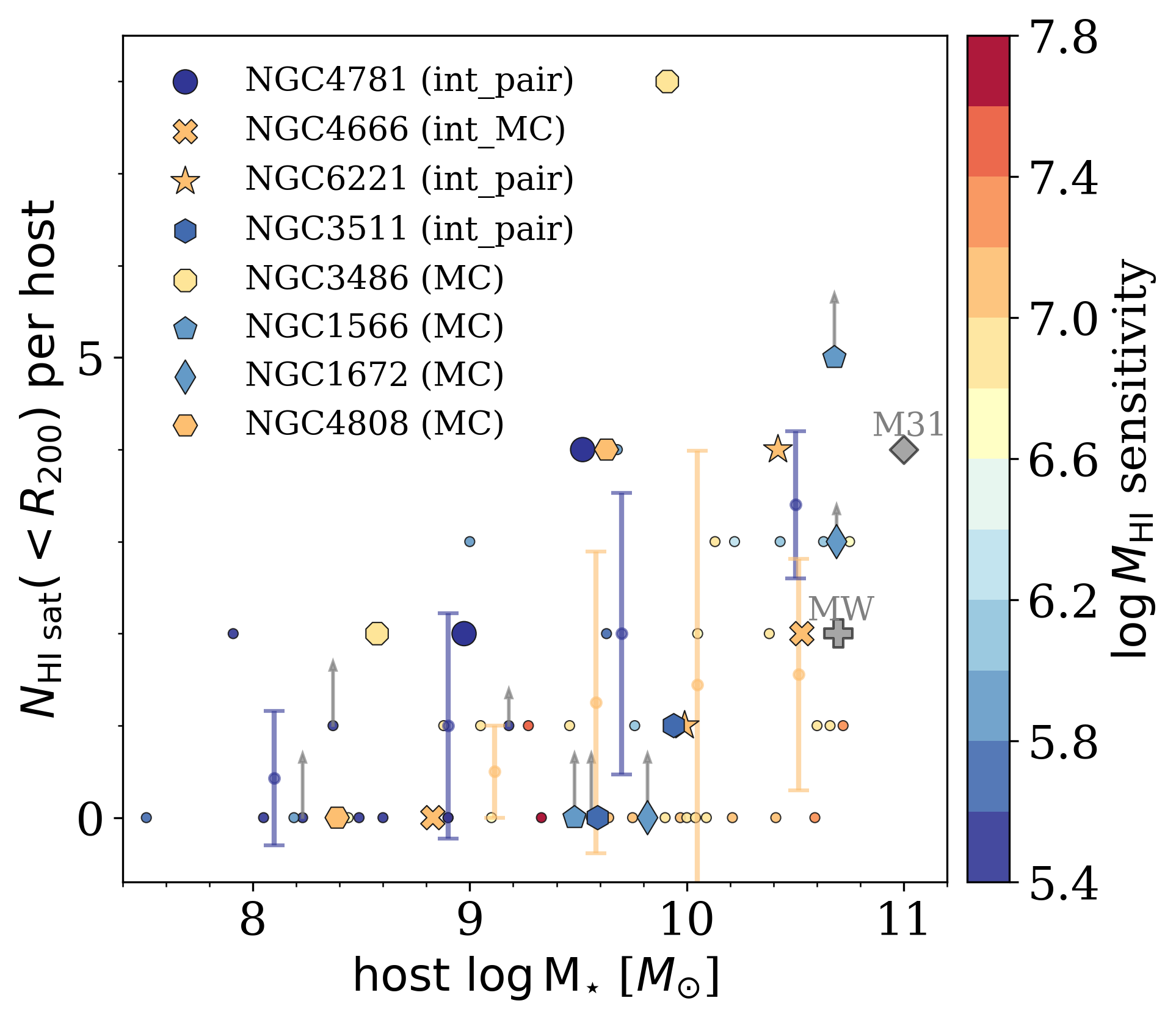}
    \caption{The number of \HI-bearing satellite candidates versus host stellar mass for hosts in interacting pairs (``int") or with Magellanic Cloud-like massive satellites (``MC"). Matching symbols identify each primary-secondary system; the lower-mass point counts the satellites associated with the secondary. More isolated hosts (small circles) and the binned statistics of the full sample are shown for comparison (as in Figure \ref{fig:host_mass_nsat_rvir_cut}; host $M_{*}-M_{200}$ conversions are listed in Table \ref{table:hosts}).}
    \label{fig:host_lmstar_nsat_sat_of_sat}
\end{figure}

We further test how a host galaxy's local environment, such as close pairs of Magellanic Cloud (MC)-like massive companions, affects the abundance of gaseous satellites (Figure \ref{fig:host_lmstar_nsat_sat_of_sat}). Eight host systems meet these criteria: four demonstrate signs of tidal interactions in \HIspace with a close secondary, and four have non-interacting MC-like satellites. In Figure \ref{fig:host_lmstar_nsat_sat_of_sat}, $N_{\HI~sat}$ is plotted against the host $M_{\star}$ (more reliable than $M_{200}$ for secondaries in interacting systems). The same symbols mark each host and its secondary, where $N_{\HI~sat}$ of the secondary counts the \HI-bearing ``satellite of satellite", i.e., dwarf galaxies that are closer to the secondary in position-velocity space than to the host. This satellite-of-satellite classification is approximate as only projected distances and line-of-sight velocities are available. Satellites of satellites are also counted in the primary host's abundance as long as they satisfy the definition in \S \ref{subsec:method_sat_environment}.

Seven out of these eight hosts (with interacting secondaries or massive satellites) have more gaseous satellites than average for their mass and sensitivity. Excluding the pair secondaries or MC-like companions reduces most counts by $\sim 0-2$, bringing them back to the sample mean. NGC 3486 remains an outlier ($N_{\HI~\rm sat}=5$) even after excluding its MC-like satellite UGC 6126 and its two gas-rich satellites. The satellite-of-satellite counts, however, should be regarded as lower limits, as they are more affected by spatial coverage and sensitivity incompleteness.

Overall, gaseous satellite abundance within $R_{200}$ increases with host mass but remains low, with $N_{\HI~\rm sat}=0-2$ per dwarf host and $N_{\HI~\rm sat}=0-5$ per spiral host (with one outlier at $N_{\HI~ sat}=8$). Host-to-host scatter at fixed mass is non-negligible, which can reflect environmental effects like the presence of massive satellites or close-interacting pairs. The MW and M31 abundances lie well within this range, consistent with the broader extragalactic sample.

\section{Comparison with optical studies: the abundance of dwarf satellites}\label{subsec:discussion_literature_compare}

In this section, we compare our \HIspace sample with deep optical surveys of dwarf satellites around nearby host galaxies: the ELVES survey of spiral hosts \citep{carlsten_exploration_2022}, the SAGA survey of Milky Way-analog hosts \citep{mao_saga_2024}, and the ELVES-Dwarfs survey of isolated Magellanic-mass dwarf hosts \citep{li_elves-dwarf_2025}. We first define how we select galaxies from each survey to perform a fair comparison (\S \ref{sect:def}), then compare the abundance of gas-rich satellites from our \HIspace sample with the star-forming satellites from the optical surveys, and estimate the total (gas-rich/star-forming and quenched) satellite populations (\S \ref{sect:compare_and_f_q}).

\subsection{Gaseous versus star-forming: defining the comparison samples}\label{sect:def}

We select star-forming dwarf satellites from the optical surveys to compare with our \HIspace sample. Star formation requires the presence of a cold gas reservoir. At the lowest star formation rates (SFR) traceable by \Haspace ($\rm SFR \lesssim 10^{-4} M_{\odot}/yr$, produced by a single O star), the expected \HIspace mass based on the star-forming main sequence is $M_{\HI} \approx 10^{7} M_{\odot}$ \citep{mcgaugh_star-forming_2017}. Typical star-forming dwarf galaxies should therefore be detectable in \HIspace given the sensitivities of our surveys (Figure \ref{fig:dwarf_mhi_sensitivity_compre}). Conversely, \HI-bearing galaxies at $M_{\HI} \geq 10^{7.5} M_{\odot}$ are almost ubiquitously detected in \Haspace \citep{meurer_survey_2006,van_sistine_alfalfa_2016}.

However, the equivalence can break down at low masses. Dwarf galaxies exhibit time variability or ``burstiness" in star formation histories \citep{mcquinn_nature_2010,mcquinn_nature_2010-1,weisz_acs_2011,weisz_modeling_2011,iyer_diversity_2020}. Gas-rich systems may appear quenched when observed during quiescent episodes of their star-forming cycles \citep{mintz_taking_2025}. Furthermore, different indicators for star formation, like in SAGA (\Haspace and NUV emission) versus ELVES (morphology and color), can yield a $20-40\%$ difference in the ultimate quenched fractions \citep{karunakaran_quenched_2023,geha_saga_2024}.

Thus, gas-rich and star-forming classifications are largely interchangeable for massive dwarfs (e.g., $M_{\HI} \geq 10^{7.5} M_{\odot}$), but may diverge at lower masses because of tracer selection, tracer sensitivities, and intrinsic burstiness. As a consistency check, we cross-matched our ALFALFA sample with the ELVES survey in our previous work \citep{zhu_census_2023}, and found all eight \HI-rich dwarf galaxies were indeed classified as late-type (star-forming) in ELVES. We therefore assume that the sample-wise misclassification rate is low, and treat the \HI-bearing satellites in our sample as comparable to the star-forming population in the optical surveys.

\begin{figure}[!htb]
    \centering
    \includegraphics[width=1.0\linewidth]{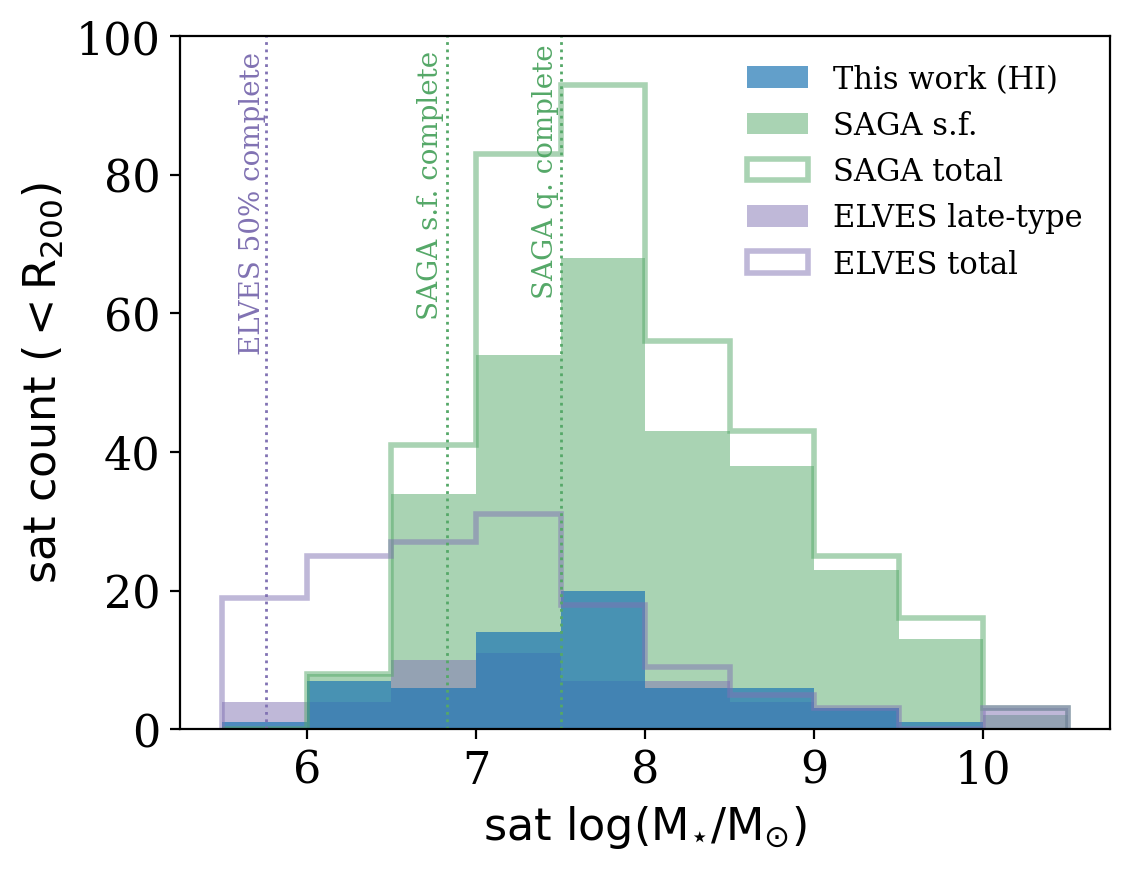}
    \caption{Overview of the dwarf satellites within the host projected $R_{200}$ from our \HIspace sample and from optical surveys: SAGA (green; \citealt{mao_saga_2024}), ELVES (purple; \citealt{carlsten_exploration_2022}), see \S \ref{sect:def}. Solid histograms count the star-forming (gas-rich) satellites from the optical (\HI) surveys, while step histograms count the total (star-forming and quenched) satellites from the optical surveys. Dotted lines mark the $\sim$50\% completeness limits in dwarf stellar mass: $M_{\star,\rm lim} \approx 10^{7.5} M_{\odot}$ for SAGA quenched, $M_{\star,\rm lim} \approx 10^{6.75} M_{\odot}$ for SAGA star-forming \citep{mao_saga_2024}; $M_{\star,\rm lim} \approx 10^{5.76} M_{\odot}$ for ELVES \citep{carlsten_exploration_2022,danieli_elves_2023}.}
    \label{fig:sat_hist_compare_surveys}
\end{figure}

We selected satellites within the hosts' projected $R_{200}$ from SAGA and ELVES, recalculating the host $M_{200}, R_{200}$ using \cite{behroozi_universemachine_2019} for consistency with our methodology (\S \ref{subsec:host_sample}). The $R_{200}$ selection slightly changes the total satellite counts (as reported below) from the published values in \cite{carlsten_exploration_2022,mao_saga_2024} but does not affect any scientific results. Figure \ref{fig:sat_hist_compare_surveys} summarizes the comparison samples:

\begin{itemize}
    \item \textbf{SAGA} (green; \citealt{mao_saga_2024}): 283 star-forming out of 368 total satellite candidates within the projected $R_{200}$ of 101 MW-like spirals.
    \item \textbf{ELVES} (purple; \citealt{carlsten_exploration_2022}): 55 late-type (star-forming) out of 143 total satellite candidates around 21 spiral hosts (four early-type hosts are excluded here). 
    
    \item \textbf{Our HI sample} (blue): 64 satellites with stellar mass estimations out of 73 total \HI-bearing satellite candidates within $R_{200}$.
\end{itemize}

The SAGA survey captures a larger sample and is sensitive to the brighter dwarfs; the ELVES survey in the local volume captures a smaller and deeper sample, including many more quenched galaxies at lower masses. At comparable satellite masses ($M_{\star,\rm sat} \geq 10^{7.5} M_{\odot}$), the two surveys agree in the satellite quenched fractions. Our sample is subject to \HIspace sensitivity limits (\S \ref{subsec:method_sensitivity}) rather than optical magnitude (effectively, stellar mass) limits. Assuming a typical gas-to-stellar mass ratio of $M_{\HI}/M_{\star} \approx 1$ (Figure \ref{fig:sat_mgas_mstar_ratio}), our \textit{effective} stellar sensitivities are $M_{\star,\rm lim} \approx M_{\HI,lim}\sim 10^{7} M_{\odot}$ for ALFALFA/WALLABY and $\lesssim 10^{6} M_{\odot}$ for MHONGOOSE hosts (Table \ref{table:hosts}). In this comparison (Figure \ref{fig:sat_hist_compare_surveys}), all gas-rich/star-forming satellite populations are uniformly sensitive to the classical dwarf irregulars down to $M_{\star} \approx 10^{7} M_{\odot}$; the ELVES survey and our MHONGOOSE subsample are additionally sensitive to lower-mass dwarf galaxies down to $M_{\star} \approx 10^{6} M_{\odot}$.

\subsection{Satellite abundance and quenched fractions}
\label{sect:compare_and_f_q}

\begin{figure*}[!htb]
    \centering
    \includegraphics[width=1.0\linewidth]{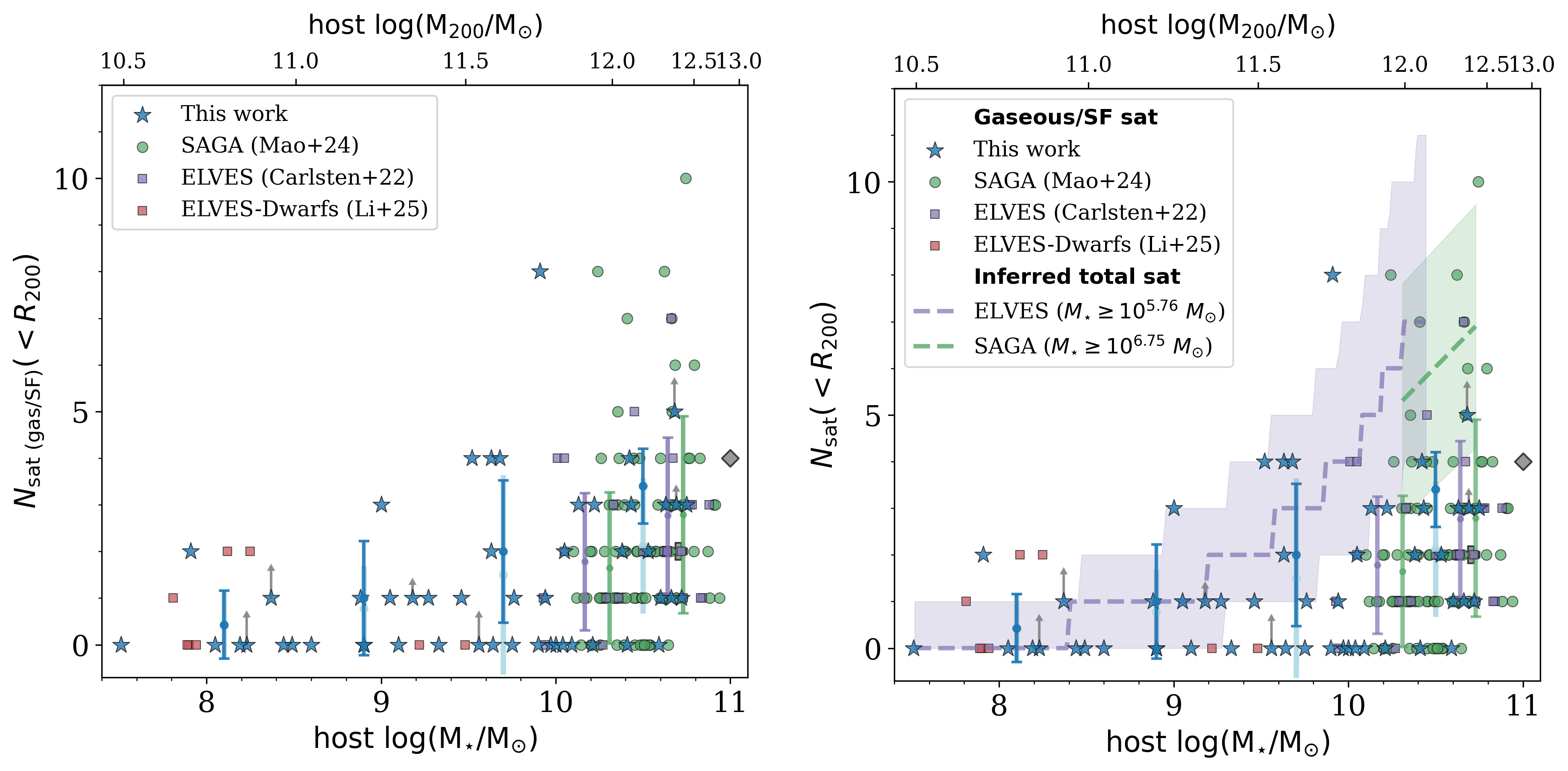}
    \caption{The number of dwarf satellites within the host's projected $R_{200}$ versus host mass, comparing multiple surveys (\S \ref{subsec:discussion_literature_compare}). Left: Gas-rich satellites from this work (blue; arrows for hosts with partial coverage as in Figure \ref{fig:host_mass_nsat_rvir_cut}) versus star-forming satellites from SAGA (green; \citealt{mao_saga_2024}), ELVES (purple; \citealt{carlsten_exploration_2022}), and ELVES-Dwarfs (red; \citealt{li_elves-dwarf_2025}). Points represent individual hosts, and error bars show binned averages and $1\sigma$ scatter. For our \HIspace sample, darker and lighter error bars represent the deeper subsample (as in Figure \ref{fig:host_lmstar_nsat_sat_of_sat}) and full sample, respectively. The Milky Way (plus symbol) and M31 (diamond symbol) are shown for reference. Host halo masses from \cite{behroozi_universemachine_2019} are given on the top x-axis. Right: the gaseous/star-forming satellites as the left panel, with dashed lines and shaded regions showing the mean and $\pm 1\sigma$ range of the \textit{total} (star-forming + quenched) satellite counts from the optical surveys. For SAGA (green), the total counts are the observed, incompleteness-corrected averages (Table C1 in \citealt{mao_saga_2024}). For ELVES (purple), these are semi-analytic predictions from \cite{danieli_elves_2023} extended from spirals down to Magellanic-mass dwarf hosts (see \citealt{li_elves-dwarf_2025}).}
    \label{fig:host_mass_nsat_literature_compare}
\end{figure*}

Figure \ref{fig:host_mass_nsat_literature_compare} compares the abundances of gas-rich and star-forming dwarf satellites (left) and the total number of satellites (right) across the surveys (samples defined in \S \ref{sect:def}). Overall, spiral hosts from various surveys show consistent satellite abundances.

We selected the host stellar mass bins in Figure \ref{fig:host_mass_nsat_literature_compare} based on the covered mass range of each sample. For the SAGA survey of MW-mass spirals, the binned averages and $1\sigma$ scatter of star-forming satellites are $N_{\rm sat~(SF)} = [1.6\pm1.6, 2.8\pm2.1]$ across $\log M_{\star}/M_{\odot} = [10.31, 10.73]$. The ELVES survey (excluding the four early-type hosts) includes slightly lower-mass spirals, yielding $N_{\rm sat~(SF)} = [1.8\pm1.5, 2.8\pm1.7]$ at $\log M_{\star}/M_{\odot} = [10.17, 10.64]$. Our \HIspace sample also includes dwarf hosts, so we adopt wider bins $\log M_{\star}/M_{\odot} = [8.1, 8.9, 9.7, 10.5]$. The deep subsample (dark blue; mostly MHONGOOSE galaxies) shows $N_{\rm sat~(gas)} = [0.4\pm0.7, 1 \pm 1.2, 2 \pm 1.5, 3.4 \pm 0.8]$, while the full sample (lighter blue; ALFALFA+WALLABY) averages $\sim 1.3$ fewer gas-rich satellites at the highest mass bin due to sensitivity. Across all spiral galaxies at $\log M_{\star}/M_{\odot} \geq 10$, the three samples agree on $N_{\rm sat~(gas/SF)} = 0-5$, averaging at $2-3$ per host. Less than 10\% of the spiral hosts have five or more star-forming satellites (9 out of 101 in SAGA; 2 out of 21 in ELVES).

Within the spiral regime, both SAGA and ELVES show an increasing $N_{\rm sat~(SF)}$ with host mass, consistent with our deep subsample. Beyond $M_{\star} \geq 10^{11} M_{\odot}$ (or $M_{200} \geq 10^{13} M_{\odot}$; see the top x-axis), gas-rich/star-forming satellite counts will likely decrease, as more massive galaxies tend to be early-type hosts (lenticular or elliptical), for which the satellites are mostly quiescent and gas-poor (see, e.g., \citealt{trentham_dwarf_2009,crnojevic_faint_2019,carlsten_exploration_2022}). The gas-rich satellite abundances of the MW and M31 fit well within the overall range of $N_{\rm sat~(gas/SF)} = 0-5$, and also within the $1\sigma$ scatter from the SAGA and ELVES averages\footnote{For the MW and M31, we adopt halo masses from dynamical measurements  \citep{tamm_stellar_2012,cautun_milky_2020}, which are significantly lower than SMHM-based estimates \citep{mcgaugh_dark_2021}.}.

At lower host masses, we can compare the MHONGOOSE dwarf centrals with the ELVES-Dwarfs survey (red; \citealt{li_elves-dwarf_2025}). ELVES-Dwarfs catalogs satellites around eight isolated Magellanic-mass dwarf centrals at $4 < D < 10$ Mpc. Similar to the ELVES survey, satellite star formation is characterized by late-type morphology and color. Both our \HIspace sample and ELVES-Dwarfs show $N_{\rm sat~(gas/SF)} = 0-2$ for host masses $10^{7.5}M_{\odot} < M_{\star} < 10^{9.5} M_{\odot}$. There is no clear trend with host mass due to the small sample size. Below $M_{\star} < 10^{7.5} M_{\odot}$ ($M_{200} < 10^{10.5} M_{\odot}$; truncated here but see Figure \ref{fig:host_mass_nsat_rvir_cut}), we detect no gas-rich satellites within the projected $R_{200}$ of any MHONGOOSE hosts.

The right panel of Figure \ref{fig:host_mass_nsat_literature_compare} shows the total satellite populations ($N_{\rm sat,total}$; star-forming and quenched) from the optical surveys. We first examine the ELVES/ELVES-Dwarfs samples (purple dashed line and shaded region), which span a wide range of host masses. For dwarf hosts ($10^{7.5} M_{\odot} < M_{\star,\rm host} < 10^{9.5} M_{\odot}$, the total number of satellites above the ELVES sensitivity limit (50\% complete at $M_{\star, \rm sat} = 10^{5.76} M_{\odot}$; \citealt{danieli_elves_2023}) is $N_{\rm sat,total} = 0-3$, which is comparable with their gaseous/star-forming satellite number. This suggests that classical dwarf satellites of isolated Magellanic-mass hosts are few in number (also found in \citealt{hunter_identifying_2025,hunter_id-mage_2026}: $N_{\rm sat,total}=1.0\pm 0.3$ per SMC-mass host and $1.7\pm0.7$ per LMC-mass host) and mostly star-forming. At $M_{\star,\rm host} = 10^{10.5} M_{\odot}$, the ELVES-inferred total satellite number increases to $N_{\rm sat,total}=7^{+4}_{-3}$. Given that $N_{\rm sat~(gas/SF)} \approx 2-3$, we can infer global quenched fractions of $f_{q} = N_{\rm sat,quenched}/N_{\rm sat,total} \approx 60-70\%$. For the SAGA survey, the incompleteness-corrected total satellite numbers ($M_{\star, \rm sat} \geq 10^{6.75} M_{\odot}$) across the host mass bins are $N_{\rm sat,total} = [5.3\pm 2.5, 6.9\pm2.6]$, which gives quenched fractions of $f_{q} \approx 40-70\%$.

The low abundance of gas-rich/star-forming satellites relative to quenched counterparts suggests that environmental quenching is efficient in MW-mass hosts ($\log M_{200}/M_{\odot} > 12$). Based on the SAGA and ELVES results, we expect an average of $\sim 4-5$ quenched satellites per host (down to $M_{\star,\rm sat} \approx 10^{5.76}-10^{6.76} M_{\odot}$) around the MW-mass hosts in our sample. This sample of $\sim 100$ quenched dwarfs can be discovered in current and future optical and infrared surveys such as \textit{Roman, Euclid, Rubin} LSST, with the \HIspace mass upper limits already characterized by our data ($M_{\HI,\rm lim}$ in Table \ref{table:hosts}). 

The observed host-to-host scatter in satellite abundance arises from several factors, including differences in host mass \citep{carlsten_exploration_2022,li_elves-dwarf_2025,hunter_identifying_2025} and in local environment, particularly the existence of a massive companion (\citealt{mao_saga_2024}; also see \S \ref{subsec:sat_abundance_local_env}). The role of the large-scale environment, such as galaxy group associations, is more complex for the gaseous satellites. Spiral galaxies in denser environments can host a higher number of satellites overall \citep{mutlu-pakdil_faint_2024}, although these satellites may experience additional gas loss from RPS by an intragroup medium \citep{brown_cold_2017,kleiner_meerkat_2021,putman_gas_2021}.

Finally, we note that lower-mass satellites are more abundant \citep{nadler_milky_2020} and also more susceptible to environmental gas loss due to their lower gravitational restoring forces \citep{sales_baryonic_2022}. Below the sensitivity limits of current extragalactic optical surveys ($M_{\star,\rm sat} \lesssim 10^{6} M_{\odot}$), there likely exists a large population of predominantly quenched low-mass satellites. Gaseous or star-forming satellites are typically the most massive (thus easier to detect) dwarf galaxies around a host \citep{font_quenching_2022}. As a result, the global quenched fractions estimated here ($f_{q} \approx 40-70\%$) are likely lower limits; accounting for the undetected low-mass population may increase it to MW/M31 levels ($>90\%$; \citealt{wetzel_rapid_2015}).

\section{Comparison with gas stripping theory}\label{sec:gas_loss_discuss}

The satellite abundance presented in the previous section suggests that environmental gas loss is effective in MW-mass hosts. Around these MW-mass hosts, we also identified intermediate-to-massive satellite candidates ($M_{\star} \gtrsim 10^{7.6}~M_{\odot}$) that are gas-depleted (see Section \ref{subsec:sat_deplete_host_mass}). At these stellar masses, stellar feedback alone is unlikely to drive significant gas loss (e.g., \citealt{jones_pavo_2025}) and environmental gas-removal mechanisms are likely required.

In this section, we examine our satellite sample in the context of ram pressure stripping (RPS) theory, the leading quenching mechanism for dwarf satellites in MW-like hosts (see the Introduction). The strength of ram pressure is set by $P_{\rm ram}=\rho_{\rm host} V_{\rm sat}^{2}$ \citep{gunn_infall_1972}, where the stripping medium ($\rho_{\rm host}$) is the circumgalactic medium (CGM) of the host galaxy \citep{tumlinson_circumgalactic_2017}. While RPS likely also occurs in lower-mass hosts, the CGM densities of dwarf centrals are lower and more uncertain \citep{christensen_-n-out_2016,hafen_origins_2019,zheng_comprehensive_2024,piacitelli_marvelous_2025}. We therefore focus on MW-mass host environments ($\log M_{200,\rm host}/M_{\odot} > 12$; identified in \S \ref{subsec:sat_deplete_host_mass}), where RPS is likely more effective and the CGM properties are better characterized. 

We construct a simple model to estimate the instantaneous ram pressure at the present-day projected phase space position ($d_{\rm proj} - \Delta V_{\rm los}$; \S \ref{subsec:method_sat_environment}), adopting a MW-like CGM profile \citep{zhu_when_2024} and applying an averaged de-projection of satellite velocities ($V_{\rm sat}$) assuming randomized viewing angles and spherical symmetry:

\begin{equation}\label{eqn:pram_in_phase_space}
\begin{split}
    P_{\rm ram} (r) & = \rho_{\rm host}(r) \cdot V_{\rm sat}(r)^{2} \\
                    & \approx \rho_{\rm host}(r=\sqrt{3/2} \cdot d_{\rm proj}) \cdot (\sqrt{3} \cdot \Delta V_{\rm los} )^{2}
\end{split}
\end{equation}

\begin{figure}[!htb]
    \centering
    \includegraphics[width=1.0\linewidth]{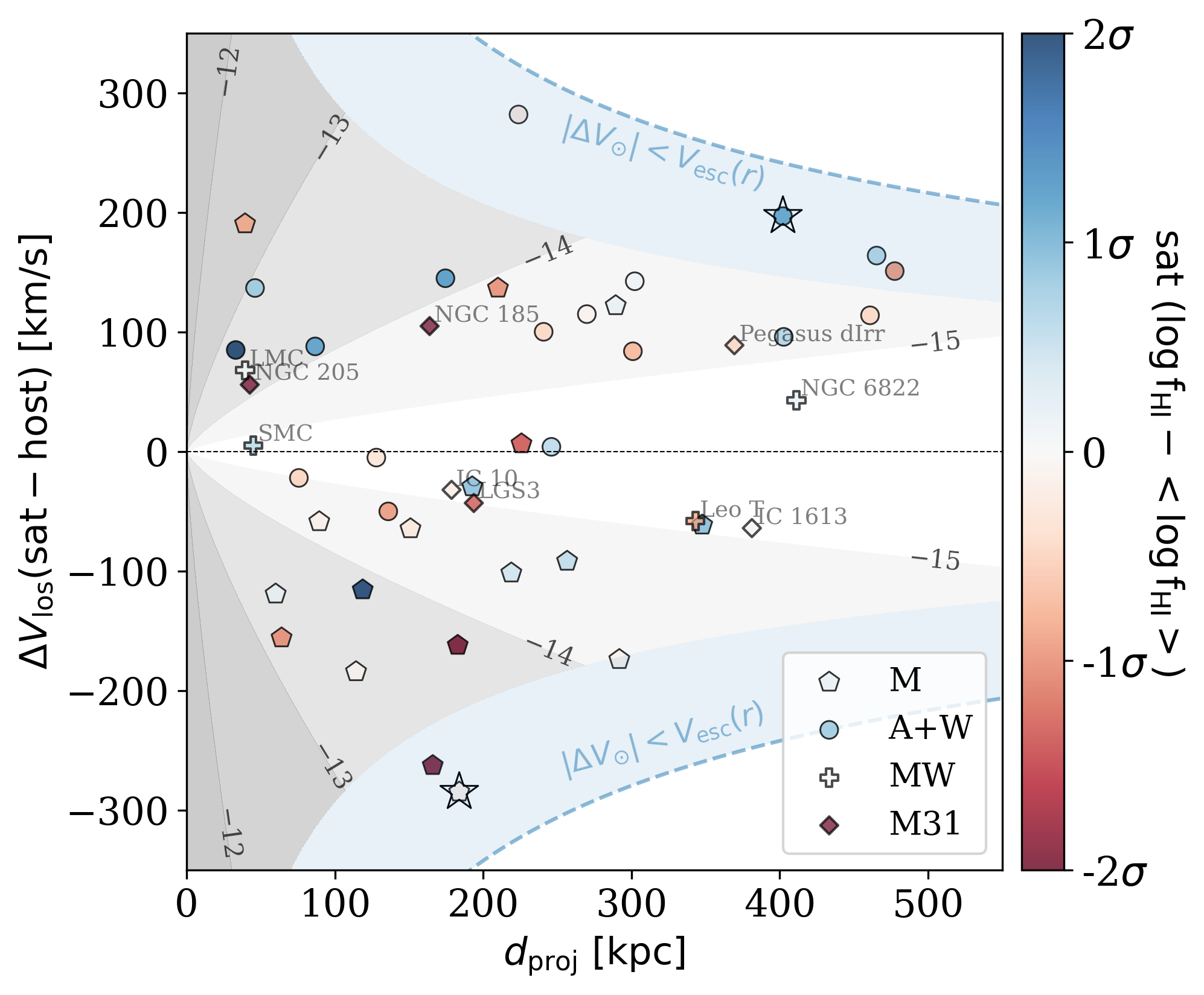}
    \caption{Satellite-host spatial ($d_{\rm proj}$) and velocity ($\Delta V_{\rm los}$) separations (similar to Figure \ref{fig:sat_kinematic_distribution}) for the satellite candidates of Milky Way-mass hosts ($\log M_{200}/M_{\odot} > 12$; see Figure \ref{fig:sat_mhi_delta_fhi_host_mhalo}). Colors of the data points represent the degree of satellite gas depletion (Equation \ref{eqn:hi_depletion}; see \S \ref{subsec:sat_deplete_host_mass}). Light-blue shaded regions mark the escape velocity curves (Equation \ref{eqn:grav_bound_theory}; see \S \ref{subsec:method_sat_environment}) for the full range of host halo masses shown here ($\log M_{200}/M_{\odot} \in [12, 12.44]$). Gray-scale shadings show contours of estimated ram pressure strengths (annotating $\log P_{\rm ram}$ values in $\rm dyne/cm^{2}$) based on a Milky Way-like CGM model and simple de-projections (Equation \ref{eqn:pram_in_phase_space}; see \S \ref{sec:gas_loss_discuss}). Two low-mass ``outlier" satellites (also see Figure \ref{fig:sat_mhi_delta_fhi_host_mhalo}) are marked in open stars. Names of the Milky Way's and M31's gaseous satellites are annotated.}
    \label{fig:massive_sat_fhi_pv_ram_pressure}
\end{figure}

The resulting $P_{\rm ram}$ across the projected $(d_{\rm proj}, \Delta V_{\rm los})$ phase space is shown as grayscale shadings in Figure \ref{fig:massive_sat_fhi_pv_ram_pressure}. 
As expected, ram pressure increases with closer distance (higher host CGM densities) and higher satellite velocities, ranging from $P_{\rm ram} > 10^{-12}~ \rm dyne/cm^{2}$ (darkest shading) to $P_{\rm ram} < 10^{-15}~ \rm dyne/cm^{2}$ (lightest shading near zero velocity) in this MW-like halo model.

Figure \ref{fig:massive_sat_fhi_pv_ram_pressure} shows the satellite candidates of MW-mass hosts in the projected phase space, color-coded by gas depletion. The color map is scaled to $\pm 2\sigma$ where $\sigma \approx 0.56$ dex is the scatter in our gas fraction fit (\S \ref{subsec:sat_deplete_host_mass}). We find no clear correlation between gas depletion and satellite-host separations. This is expected in part because our \HI-selected sample contains only surviving gas-bearing satellites, which by definition have not been fully stripped by the present day. The degree of gas depletion is determined by the maximum ram pressure experienced over a satellite's orbital history, which cannot be recovered from projected phase-space positions at $z=0$ alone.

The two low-mass ($M_{\star} \leq 10^{6.5} M_{\odot}$) satellite candidates (open stars in Figure \ref{fig:massive_sat_fhi_pv_ram_pressure}; also see Figure \ref{fig:sat_mhi_delta_fhi_host_mhalo}) 
lie very close to the $V_{\rm esc}$ boundary and are likely foreground/background interlopers \citep{oman_satellite_2016,rhee_phase-space_2017}, for which environmental gas loss is likely negligible. At $M_{\star} \approx 10^{6} M_{\odot}$, they also exceed the typical mass threshold for quenching by reionization ($M_{\star} \leq 10^{5} M_{\odot}$; \citealt{brown_quenching_2014,weisz_star_2014-1}) and are expected to remain gas-bearing if they are field galaxies.

Two other familiar cases of low-mass dwarf satellites that remain gas-bearing are Leo T of the MW and LGS 3 of M31. Both fall in the phase space of large separation and low velocity, which produces a very low instantaneous ram pressure ($P_{\rm ram} \lesssim 10^{-15} \rm ~dyne/cm^{2}$; Figure \ref{fig:massive_sat_fhi_pv_ram_pressure}). If $P_{\rm ram}$ 
remained comparably low throughout their orbital histories, 
stripping would be highly incomplete \citep{emerick_gas_2016}, consistent with their survival as the lowest-mass gaseous satellites of the MW and M31, while all other satellites at similar stellar masses are gas-poor \citep{putman_gas_2021}.

The most depleted satellite candidates (darker red points), including M31's NGC 185, 
are found at intermediate projected distances ($d_{\rm proj} \approx 200$ kpc; 3D distances are higher), consistent with ``backsplash" galaxies that have completed at least one pericentric passage and are moving outward \citep{rhee_phase-space_2017,borrow_there_2023}. Previous work has found that a significant fraction of quenched satellites are backsplash galaxies at large distances \citep{simpson_quenching_2018,bhattacharyya_environmental_2024,benavides_environmental_2025}. In contrast, few galaxies are detected in the first-infall phase space (small $d_{\rm proj}$, large $\Delta V$; similar to the LMC), 
as satellites have the highest velocities and are least likely to be detected there. The data point next to LMC is NGC 4668, an interacting Magellanic analog of NGC 4666 \citep{walter_superwind_2004}. Neither NGC 4668 nor the LMC appear gas-depleted: at their high masses ($M_{\star} > 10^{9} M_{\odot}$), RPS of the cold gas is highly incomplete even at perigalacticon (e.g., \citealt{salem_ram_2015}).

Projected phase space location provides a useful statistical view of the orbital stage, such as first infall versus backsplash, but it does not uniquely determine the orbital or gas stripping history of individual satellites. This limitation is reflected in the lack of correlation between satellite gas depletion and present-day phase-space position, or equivalently, instantaneous ram pressure. Several observational and theoretical ingredients are needed to improve existing constraints. These include accurate dwarf galaxy distances to help distinguish true satellites from interlopers \citep{mcquinn_accurate_2017,cohen_dragonfly_2018,carlsten_using_2019}, improved statistics on 3D satellite velocity distributions from local observations \citep{pace_proper_2022} and high-resolution simulations \citep{wetzel_orbits_2011,rocha_infall_2012,nadler_symphony_2023,santos-santos_unabridged_2025}, deeper \HIspace observations of field dwarf galaxies to assess the baseline \HI-to-stellar mass fractions \citep{teich_shield_2016,jones_pavo_2025}, and high-resolution hydrodynamical simulations that constrain the effectiveness of gas loss in dwarf satellite galaxies \citep{mayer_environmental_2010,emerick_gas_2016,zhu_its_2024,zhu_too_2026}.

\section{Summary and Conclusions}\label{sec:summary}

In this work, we studied gas-bearing dwarf galaxies around late-type host galaxies (Milky Way-mass and lower; Table \ref{table:hosts}), using \HIspace data from WALLABY and MHONGOOSE. We combine these data with the ALFALFA data in \cite{zhu_census_2023} to form a sample of 56 host galaxies within 30 Mpc. We find 127 gaseous dwarfs and use deep optical imaging from Legacy DR10 \citep{dey_2019AJ....157..168D} for 110 of them to derive baryonic properties ($\log (M_{\HI}/M_{\odot}) = 5.8-9.7, \log (M_{\star}/M_{\odot}) = 5.6-10.0$). We identify the dwarf galaxies to be either satellite galaxies within the host projected $R_{200}$, possible satellites beyond $R_{200}$, or unbound.

The proximity of the host galaxies and new survey data enable the detection of \HIspace masses comparable to Local Group (LG) dwarf galaxies (Table \ref{table:sats}). We developed an injection test (Appendix \ref{appendix:inject}) in conjunction with the \HIspace source finder SoFiA-2 \citep{westmeier_sofia_2021} to assess the \HIspace sensitivity limits for each host (\S \ref{subsec:method_sensitivity}). The typical \HIspace mass sensitivity at 10 Mpc is $M_{\HI,lim} = 10^{6.7} M_{\odot}$ for WALLABY and $M_{\HI,lim} = 10^{5.4} M_{\odot}$ for MHONGOOSE (Figure \ref{fig:dwarf_mhi_sensitivity_compre}).
    
Our main results are summarized as follows:

\begin{itemize}[nolistsep]
    
    \item The median$\pm34\%$ \HIspace mass range of our dwarf galaxies is $\log (M_{\HI}/M_{\odot}) = 7.56_{-0.52}^{+0.81}$ for WALLABY and ALFALFA and $\log (M_{\HI}/M_{\odot}) = 7.35_{-0.85}^{+1.08}$ for MHONGOOSE, with a median line-width of $W_{50} = 40_{-18}^{+40} ~\rm km~s^{-1}$ (Figure \ref{fig:sat_HI_mass_width}).

    \item The satellite stellar and \HIspace mass distributions span a similar range, correlation, and scatter as the Local Group population (Figure \ref{fig:sat_mgas_mstar}). Our sample shows lower average gas fractions ($M_{\HI}/M_{\star}$) than field samples outside the Local Group (Figure \ref{fig:sat_mgas_mstar_ratio}). This result also indicates that \HIspace sensitivity of the data is key to consider when comparing gas fractions across samples.

    \item Satellites are more gas-depleted around more massive, Milky Way-like host galaxies (Figure \ref{fig:sat_mhi_delta_fhi_host_mhalo}). In particular, we identify a small group of highly depleted massive satellites ($M_{\star} \geq 10^{7.6} M_{\odot}$) around the Milky Way-mass hosts.

    \item The abundance of \HI-bearing satellites is low overall and increases with host mass: dwarf hosts ($\log M_{200}/M_{\odot} \leq 11$) have $0-2$ within their projected halos, while Milky Way-mass spirals ($\log M_{200}/M_{\odot} \geq 12$) have $0-5$, averaging at $N_{\HI~sat} \approx 3$ (Figure \ref{fig:host_mass_nsat_rvir_cut}). Hosts with massive companions tend to host more gaseous satellites (Figure \ref{fig:host_lmstar_nsat_sat_of_sat}).
    
    \item For Milky Way-mass spirals, the number of \HI-bearing satellites agrees with the Milky Way and M31 (\S \ref{sec:host_sat_abundance}) as well as the star-forming satellite abundance from deep optical surveys (Figure \ref{fig:host_mass_nsat_literature_compare}). The inferred satellite quenched fraction for classical dwarf masses is $f_{q} = 40-70\%$ (\S \ref{subsec:discussion_literature_compare}).
    
    \item We estimated ram pressure strengths in Milky Way-mass host halos across satellite phase space (Figure \ref{fig:massive_sat_fhi_pv_ram_pressure}), providing a statistical view of possible first-infall, backsplash, and low-mass interloper populations (\S \ref{sec:gas_loss_discuss}).
    
\end{itemize}

The low abundance of \HI-bearing satellites is a key result of this work, showing that effective environmental quenching around spiral galaxies is not unique to the Milky Way and M31. Future deep \HIspace observations of field galaxies, such as with MeerKAT or local WALLABY observations, will help calibrate gas fractions in low-density environments and refine gas-depletion measurements ($\Delta \log f_{\HI}$; \S \ref{subsec:sat_deplete_host_mass}). Improved line-width corrections and kinematic modeling of a resolved subsample will enable studies of Tully-Fisher scaling relations \citep{tully_new_1977,mcgaugh_baryonic_2000,lelli_small_2016}, as demonstrated by recent Apertif observations of relatively massive dwarfs \citep{siljeg_photometry_2024}. High-resolution simulations will further quantify the roles of different gas loss mechanisms in low-mass satellite galaxies (\S \ref{sec:gas_loss_discuss}).

\vspace{5mm}

{
\textit{Acknowledgments}. We thank the referee for their insightful and constructive feedback. JZ thanks Enrico Di Teodoro for help with the pyBBarolo software, Kelly M. Hess for publicizing the \texttt{SoFiA-image-pipeline} software, Viraj Pandya for conversations on stellar mass-halo mass relations, Rory Smith for conversations on phase space satellite-vs-interloper fractions, Jiaxuan Li for sharing summary statistics from the ELVES-Dwarfs survey, and Yao-Yuan Mao for conversations on the SAGA data. The authors thank Stephanie Tonnesen, Greg Bryan, and Marla Geha for helpful conversations.  
MEP and JZ acknowledge support from NASA through grant HST-AR-17562 from STScI. STScI is operated by the Association of Universities for Research in Astronomy, Inc., under NASA contract NAS5-26555.  Support for this work was also provided by the NSF through award NRAO CU24-0734. FMM carried out part of the research activities described in this paper with contribution of the Next Generation EU funds within the National Recovery and Resilience Plan (PNRR), Mission 4 - Education and Research, Component 2 - From Research to Business (M4C2), Investment Line 3.1 - Strengthening and creation of Research Infrastructures, Project IR0000034 – ``STILES - Strengthening the Italian Leadership in ELT and SKA."

WALLABY is an Australian SKA Pathfinder (ASKAP) survey, which uses data obtained from Inyarrimanha Ilgari Bundara, the CSIRO Murchison Radio-astronomy Observatory. We acknowledge the Wajarri Yamaji People as the Traditional Owners and native title holders of the Observatory site. CSIRO’s ASKAP radio telescope is part of the Australia Telescope National Facility (https://ror.org/05qajvd42). Operation of ASKAP is funded by the Australian Government with support from the National Collaborative Research Infrastructure Strategy. ASKAP uses the resources of the Pawsey Supercomputing Research Centre. Establishment of ASKAP, Inyarrimanha Ilgari Bundara, the CSIRO Murchison Radio-astronomy Observatory and the Pawsey Supercomputing Research Centre are initiatives of the Australian Government, with support from the Government of Western Australia and the Science and Industry Endowment Fund. WALLABY is supported by the Australian SKA Regional Centre (AusSRC), Australia’s portion of the international SKA Regional Centre Network (SRCNet), funded by the Australian Government through the Department of Industry, Science, and Resources (DISR; grant SKARC000001). AusSRC is an equal collaboration between CSIRO – Australia’s national science agency, Curtin University, the Pawsey Supercomputing Research Centre, and the University of Western Australia.

The MeerKAT telescope is operated by the South African Radio Astronomy Observatory, which is a facility
of the National Research Foundation, an agency of the Department of Science and Innovation.
This work has received funding from the European Research Council (ERC) under the European Union’s Horizon 2020 research and innovation programme (grant agreement No. 882793 "MeerGas").}


\software{SoFiA-2 \citep{serra_sofia_2015,westmeier_sofia_2021}, pySersic \citep{pasha_2023JOSS....8.5703P}, BBarolo \citep{di_teodoro_3d_2015}, NumPy \citep{harris_array_2020}, Astropy \citep{astropy_collaboration_astropy_2013,astropy_collaboration_astropy_2018,astropy_collaboration_astropy_2022}}



\appendix

\section{Catalog of \HI-bearing Dwarf Galaxies}\label{appendix:dwarfs}

This appendix presents the properties of the \HIspace sources identified in Sections \ref{subsec:method_source_finding} and \ref{subsec:method_sat_environment}. Table \ref{table:sats} lists all 127 dwarf galaxies within the host $R_{\rm cover}$ and $\pm 300$ km/s range, drawn from the WALLABY, MHONGOOSE, and ALFALFA surveys. Descriptions of the table columns are provided below.

\vspace{-3mm}
\begin{itemize}[leftmargin=*,labelwidth=2.5em, itemsep=0em, align=left]
    \item[(1)] Dwarf galaxy name. New detections follow WALLABY and MeerKAT (MKT) naming conventions.
    \item[(2)] Host galaxy name (see Table \ref{table:hosts}).
    \item[(3-4)] Right ascension and declination of the \HIspace centroid in $\deg$ (J2000); optical counterparts may have minor spatial offsets (sub-arcmin level).
    \item[(5)] Heliocentric (line-of-sight) velocity.
    \item[(6)] Projected separation between the dwarf galaxy and its host, evaluated at the host distance ($D$; Table \ref{table:hosts}).
    \item[(7)] Line-of-sight velocity offset from the host.
    \item[(8)]  Environment flag ($\rm s_{\rm flag}$) as characterized in Section \ref{subsec:method_sat_environment}:
    \begin{itemize}[nolistsep,topsep=0pt]
        \item $\rm s_{\rm flag}$=1: satellite candidates within the host's projected $R_{200}$
        \item $\rm s_{\rm flag}$=2: satellite candidate outside the host $R_{200}$
        \item $\rm s_{\rm flag}$=3: unbound field galaxies.
    \end{itemize}
    \item[(9)] \HIspace line-width ($W_{50}$; velocity width at $50\%$ of the peak flux density), uncorrected for inclination (see Section \ref{subsec:sat_hi_properties}).
    \item[(10)] \HIspace mass, scaled to the host distance. WALLABY fluxes have been corrected following \cite{westmeier_wallaby_2022,murugeshan_wallaby_2024} to account for missing flux relative to single-dish data. MHONGOOSE fluxes are measured from full-depth data cubes and corrected for primary beam attenuation.
    \item[(11)] Stellar mass, scaled to the host distance (see Section \ref{subsec:method_stellar_mass}). Missing values correspond to galaxies lacking coverage in one or more bands in Legacy DR10 unless otherwise noted.
    \item[(12-13)] Absolute $g$- and $r$-band magnitudes from Legacy DR10, scaled to the host distance.
    \item[(14-15)] Apparent axis ratio of the optical counterpart ($b/a$) and optical half-light radius ($R_{\rm opt.}$), derived from Bayesian Sérsic profile fits (Section \ref{subsec:method_stellar_mass}).
\end{itemize}

\startlongtable
\centerwidetable 
\begin{deluxetable*}{llccccccccccccc}\label{table:sats}
\tabletypesize{\scriptsize}
\tablecaption{Properties of the \HI-containing dwarf galaxies around the host galaxies.} 
\tablewidth{0pt}
\tablehead{\colhead{Name} & \colhead{Host} & \colhead{RA$_{(\HI)}$} & \colhead{Dec$_{(\HI)}$} & \colhead{$V_{\rm los}$} & \colhead{$d_{\rm proj}$} & \colhead{$\Delta V_{\rm los}$} & \colhead{s$_{\rm flag}$} & \colhead{$W_{50}$} & \colhead{$\log M_{\HI}$} & \colhead{$\log M_{*}$} & \colhead{$M_{g}$} & \colhead{$M_{r}$} & \colhead{$b/a$} & \colhead{$R_{\rm opt.}$}\\
\colhead{} & \colhead{} & \colhead{(deg)}  & \colhead{(deg)} & \colhead{(km/s)}  & \colhead{(kpc)} &
\colhead{(km/s)} & \colhead{} & \colhead{(km/s)} & \colhead{($M_{\odot}$)}  & \colhead{($M_{\odot}$)}  & \colhead{(mag)} & \colhead{(mag)} & \colhead{} & \colhead{(kpc)}}
\colnumbers
\startdata
WALLABY J131929-200819 & ESO576-G037 & 199.8714 & -20.1388 & 1949 & 167 & 239 & 3 &  53 &  8.41 & \nodata & \nodata & \nodata & \nodata & \nodata \\
WALLABY J132030-202723 & ESO576-G037 & 200.1257 & -20.4564 & 1621 & 271 & -89 & 3 &  21 &  7.65 & 7.81 & -14.44 & -14.87 & 0.81 & 1.35 \\
WALLABY J132118-193418 & ESO576-G037 & 200.3256 & -19.5718 & 1681 & 146 & -30 & 2 &  69 &  8.52 & \nodata & \nodata & \nodata & \nodata & \nodata \\
WALLABY J132202-201335 & ESO576-G037 & 200.5119 & -20.2263 & 1880 & 235 & 169 & 3 &  94 &  8.77 & \nodata & \nodata & \nodata & \nodata & \nodata \\
WALLABY J123903-003850$^{a,d}$ & NGC4592 & 189.7641 & -0.6475 & 1125 & 26 & 54 & 1 &  19 &  7.91 & 6.81 & -12.42 & -12.65 & 0.48 & 0.80 \\
WALLABY J124119+013253$^{a}$ & UGC07841 & 190.3306 & 1.5482 & 1674 & 64 & -20 & 1 &  29 &  8.09 & 7.68 & -14.50 & -14.82 & 0.81 & 1.23 \\
WALLABY J124046-013557$^{a}$ & NGC4629 & 190.1934 & -1.5993 & 1250 & 247 & 142 & 3 &  41 &  8.17 & 7.24 & -13.62 & -13.85 & 0.67 & 0.56 \\
WALLABY J124009-002103 & NGC4632 & 190.0383 & -0.3511 & 1701 & 180 & -19 & 2 &  25 &  7.82 & 8.21 & -15.11 & -15.65 & 0.50 & 1.14 \\
WALLABY J124035+004620$^{a}$ & NGC4632 & 190.1475 & 0.7722 & 1574 & 270 & -140 & 3 &  44 &  7.58 & 7.14 & -13.01 & -13.33 & 0.57 & 0.46 \\
WALLABY J125215+042728 & NGC4808 & 193.0642 & 4.4578 & 700 & 245 & -57 & 2 &  63 &  7.98 & 7.17 & -13.25 & -13.52 & 0.38 & 0.90 \\
WALLABY J125313+042746 & NGC4808 & 193.3064 & 4.4628 & 724 & 181 & -33 & 2 &  62 &  8.96 & 9.01 & -17.85 & -18.24 & 0.72 & 0.66 \\
WALLABY J125339+040434 & NGC4808 & 193.4163 & 4.0762 & 886 & 158 & 129 & 2 &  36 &  7.31 & 7.68 & -13.82 & -14.32 & 0.70 & 0.51 \\
WALLABY J125343+040920 & NGC4808 & 193.4301 & 4.1557 & 762 & 147 & 6 & 1 &  20 &  7.54 & 7.53 & -13.91 & -14.27 & 0.72 & 0.30 \\
WALLABY J125549+040049 & NGC4808 & 193.9563 & 4.0138 & 716 & 78 & -41 & 1 &  118 &  8.98 & 8.39 & -16.24 & -16.61 & 0.55 & 1.54 \\
WALLABY J125604+034843 & NGC4808 & 194.0193 & 3.8121 & 618 & 135 & -138 & 1 &  89 &  8.83 & 7.58 & -14.45 & -14.70 & 0.63 & 1.00 \\
WALLABY J125656+040352 & NGC4808 & 194.2369 & 4.0646 & 830 & 101 & 73 & 1 &  80 &  8.30 & 8.31 & -15.98 & -16.36 & 0.49 & 0.97 \\
WALLABY J125718+045929 & NGC4808 & 194.3261 & 4.9915 & 883 & 214 & 126 & 3 &  20 &  7.70 & 6.46 & -11.86 & -11.98 & 0.84 & 0.24 \\
WALLABY J131139-165121 & NGC4984 & 197.9148 & -16.8560 & 1284 & 312 & 34 & 2 &  24 &  6.98 & \nodata & \nodata & \nodata & \nodata & \nodata \\
WALLABY J131204-161647$^{a}$ & NGC4984 & 198.0171 & -16.2798 & 1367 & 224 & 117 & 2 &  10 &  6.84 & \nodata & \nodata & \nodata & \nodata & \nodata \\
WALLABY J165437-602755 & ESO138-G010 & 253.6549 & -60.4654 & 1046 & 137 & -97 & 1 &  24 &  7.58 & \nodata & \nodata & \nodata & \nodata & \nodata \\
WALLABY J165804-605308 & ESO138-G010 & 254.5175 & -60.8856 & 1035 & 156 & -108 & 1 &  48 &  8.83 & \nodata & \nodata & \nodata & \nodata & \nodata \\
WALLABY J170248-602617 & ESO138-G010 & 255.7022 & -60.4383 & 1078 & 119 & -65 & 1 &  27 &  7.57 & \nodata & \nodata & \nodata & \nodata & \nodata \\
WALLABY J131431-162247 & NGC5054 & 198.6302 & -16.3798 & 1474 & 193 & -263 & 3 &  77 &  8.42 & \nodata & \nodata & \nodata & \nodata & \nodata \\
WALLABY J165105-585918 & NGC6221 & 252.7742 & -58.9884 & 1560 & 87 & 88 & 1 &  56 &  9.67 & 9.99$^{f}$ & \nodata & \nodata & \nodata & \nodata \\
WALLABY J165131-590247$^{a}$ & NGC6221 & 252.8818 & -59.0466 & 1590 & 65 & 118 & 1 &  48 &  7.84 & \nodata & \nodata & \nodata & \nodata & \nodata \\
WALLABY J165348-590530 & NGC6221 & 253.4524 & -59.0917 & 1645 & 49 & 169 & 1 &  45 &  8.51 & \nodata & \nodata & \nodata & \nodata & \nodata \\
WALLABY J165429-592515 & NGC6221 & 253.6210 & -59.4210 & 1509 & 80 & 33 & 1 &  62 &  8.33 & \nodata & \nodata & \nodata & \nodata & \nodata \\
WALLABY J124319-012805$^{a}$ & NGC4666 & 190.8303 & -1.4682 & 1677 & 302 & 143 & 2 &  52 &  7.74 & 8.06 & -15.17 & -15.58 & 0.38 & 0.85 \\
WALLABY J124531-003203 & NGC4666 & 191.3824 & -0.5343 & 1615 & 33 & 85 & 1 &  140 &  9.40 & 8.83$^{f}$ & \nodata & \nodata & \nodata & \nodata \\
WALLABY J124548-002600 & NGC4666 & 191.4504 & -0.4335 & 1671 & 46 & 137 & 1 &  47 &  7.96 & 7.79 & -14.52 & -14.91 & 0.55 & 0.62 \\
WALLABY J132726-173924 & NGC5170 & 201.8599 & -17.6567 & 1615 & 241 & 100 & 1 &  58 &  7.76 & 8.52 & -16.23 & -16.70 & 0.41 & 1.32 \\
WALLABY J132831-171324 & NGC5170 & 202.1331 & -17.2236 & 1599 & 301 & 84 & 2 &  11 &  7.49 & 8.34 & -15.50 & -16.04 & 0.94 & 1.21 \\
LEDA 46885 & KK98-195 & 201.1488 & -30.9718 & 484 & 84 & -87 & 3 &  18 &  6.94$^{e}$ & 7.13 & -12.74 & -13.12 & 0.70 & 0.40 \\
UGCA 319 & UGCA320 & 195.5604 & -17.2383 & 724 & 32 & -16 & 1 &  24 &  7.59 & \nodata & \nodata & \nodata & \nodata & \nodata \\
LEDA 886203 & UGCA320 & 196.2663 & -17.2567 & 724 & 48 & -16 & 1 &  27 &  7.15 & \nodata & \nodata & \nodata & \nodata & \nodata \\
LCRS B125209.0-112324 & UGCA307 & 193.6912 & -11.6610 & 858 & 95 & -66 & 1 &  23 &  7.74 & 7.97 & -14.96 & -15.36 & 0.55 & 0.76 \\
MKT J045726.5-532423.6$^{b}$ & NGC1705 & 74.3604 & -53.4066 & 666 & 48 & 34 & 1 &  17 &  5.93 & 5.63 & -9.29 & -9.49 & 0.35 & 0.16 \\
MKT J054422.3-514523.5$^{b}$ & NGC2101 & 86.0929 & -51.7565 & 1235 & 131 & 48 & 2 &  30 &  6.17 & 6.96 & -11.64 & -12.20 & 0.46 & 0.43 \\
ESO 204-G034 & NGC2101 & 86.1729 & -51.9635 & 1227 & 83 & 41 & 1 &  97 &  8.43 & 8.70 & -16.21 & -16.82 & 0.80 & 1.65 \\
MKT J054449.0-522431.8$^{b}$ & NGC2101 & 86.2042 & -52.4088 & 1082 & 115 & -105 & 1 &  14 &  6.50 & 7.28 & -12.80 & -13.28 & 0.64 & 0.36 \\
LEDA 454649 & NGC2101 & 86.8779 & -51.5960 & 1182 & 23 & 19 & 1 &  52 &  7.24 & 7.92 & -14.52 & -15.00 & 0.25 & 0.86 \\
MKT J054753.5-515635 & NGC2101 & 86.9729 & -51.9433 & 1172 & 78 & -15 & 1 &  16 &  7.04 & 6.39 & -11.07 & -11.35 & 0.50 & 0.64 \\
MKT J045920.2-252959 & NGC1744 & 74.8342 & -25.4999 & 692 & 88 & -48 & 1 &  30 &  6.49 & 6.21 & -10.99 & -11.15 & 0.90 & 0.13 \\
MKT J125412.6-102440.9$^{b}$ & NGC4781 & 193.5525 & -10.4114 & 1338 & 22 & 82 & 1 &  16 &  5.79 & 6.47 & -10.52 & -11.01 & 0.48 & 0.27 \\
NGC 4790 & NGC4781 & 193.7146 & -10.2454 & 1354 & 56 & 98 & 1 &  182 &  8.79 & 8.97 & -17.59 & -18.02 & 0.70 & 0.92 \\
2MASX J12551835-1103580 & NGC4781 & 193.8258 & -11.0673 & 1105 & 118 & -151 & 1 &  80 &  7.57 & 7.80 & -14.62 & -14.99 & 0.61 & 0.31 \\
UGCA 308 & NGC4781 & 193.8796 & -10.3944 & 1317 & 56 & 62 & 1 &  55 &  8.37 & 7.63 & -14.25 & -14.60 & 0.56 & 0.88 \\
LEDA 457253 & IC1954 & 52.5471 & -51.3540 & 1102 & 131 & 45 & 1 &  42 &  7.37 & 7.61 & -14.14 & -14.49 & 0.67 & 0.33 \\
MKT J033131.9-520400 & IC1954 & 52.8829 & -52.0667 & 1014 & 36 & -43 & 1 &  17 &  6.99 & 7.16 & -12.43 & -12.92 & 0.77 & 0.85 \\
ESO 200-G045 & IC1954 & 53.7533 & -51.4542 & 1023 & 157 & -34 & 2 &  42 &  8.19 & 7.50 & -13.97 & -14.29 & 0.88 & 0.95 \\
WISEA J051522.96-370331.3 & ESO362-G011 & 78.8462 & -37.0586 & 1361 & 70 & 24 & 1 &  25 &  6.71 & 7.70 & -13.79 & -14.31 & 0.56 & 0.62 \\
LEDA 630916 & ESO362-G011 & 79.1863 & -36.9247 & 1329 & 49 & -8 & 1 &  18 &  7.09 & 6.50 & -11.47 & -11.73 & 0.87 & 0.65 \\
MKT J051704.6-372940.9$^{b}$ & ESO362-G011 & 79.2696 & -37.4947 & 1418 & 109 & 81 & 1 &  34 &  6.44 & 7.44 & -13.16 & -13.65 & 0.78 & 0.30 \\
ESO 362-G016 & ESO362-G011 & 79.8283 & -37.1089 & 1335 & 145 & -2 & 1 &  76 &  8.58 & 7.81 & -15.16 & -15.39 & 0.56 & 1.33 \\
ESO 440-G014 & UGCA250 & 177.5142 & -28.6716 & 1856 & 262 & 160 & 3 &  40 &  7.86 & 8.33 & -15.89 & -16.30 & 0.87 & 0.69 \\
6dFGS gJ115153.3-281047 & UGCA250 & 177.9721 & -28.1790 & 1424 & 177 & -272 & 3 &  56 &  7.33 & 8.63 & -16.20 & -16.76 & 0.56 & 1.34 \\
MKT J115221.7-280126.9$^{b}$ & UGCA250 & 178.0904 & -28.0241 & 1637 & 203 & -59 & 2 &  26 &  7.00 & 7.66 & -13.86 & -14.33 & 0.50 & 0.97 \\
ESO 440-G023 & UGCA250 & 178.1387 & -29.1223 & 1866 & 211 & 170 & 3 &  85 &  7.99 & 8.44 & -15.79 & -16.31 & 0.35 & 1.51 \\
LEDA 746393 & UGCA250 & 178.2937 & -28.0579 & 1833 & 175 & 137 & 2 &  80 &  7.43 & 8.28 & -14.81 & -15.49 & 0.49 & 1.09 \\
LEDA 737259 & UGCA250 & 178.5617 & -28.8056 & 1764 & 110 & 68 & 1 &  26 &  7.91 & 7.58 & -14.37 & -14.65 & 0.53 & 0.57 \\
MKT J115513.5-280114.5$^{b}$ & UGCA250 & 178.8063 & -28.0207 & 1951 & 235 & 255 & 3 &  22 &  6.50 & \nodata & \nodata & \nodata & \nodata & \nodata \\
NGC 3513 & NGC3511 & 165.9421 & -23.2421 & 1191 & 43 & 92 & 1 &  80 &  9.02 & 9.59 & -17.87 & -18.34 & 0.85 & 1.59 \\
LEDA 715942 & NGC0289 & 12.9667 & -30.5461 & 1528 & 257 & -92 & 2 &  56 &  7.72 & 7.63 & -13.99 & -14.40 & 0.30 & 1.23 \\
2dFGRS TGS370Z277 & NGC0289 & 13.1071 & -30.9071 & 1436 & 114 & -184 & 1 &  22 &  7.03 & 7.13 & -12.60 & -13.02 & 0.90 & 0.37 \\
ESO 411-G026 & NGC0289 & 13.1883 & -31.7190 & 1591 & 193 & -29 & 1 &  102 &  8.44 & 8.42 & -16.20 & -16.60 & 0.58 & 1.79 \\
2dFGRS TGS369Z051 & NGC0289 & 13.3754 & -30.8414 & 1556 & 151 & -64 & 1 &  36 &  7.78 & 8.37 & -14.98 & -15.68 & 0.85 & 0.84 \\
ESO 411-G031 & NGC0289 & 13.9271 & -30.5399 & 1559 & 348 & -61 & 2 &  63 &  8.50$^{e}$ & 8.48 & -16.07 & -16.55 & 0.86 & 1.63 \\
MKT J033305.0-250432.8$^{b}$ & NGC1371 & 53.2713 & -25.0758 & 1294 & 183 & -162 & 1 &  20 &  6.33 & 7.72 & -14.06 & -14.52 & 0.65 & 1.00 \\
MKT J033442.4-245150.3$^{b}$ & NGC1371 & 53.6767 & -24.8640 & 1647 & 39 & 191 & 1 &  22 &  6.82 & 7.45 & -13.37 & -13.82 & 0.88 & 0.78 \\
ESO 482-G011 & NGC1371 & 54.0721 & -25.6052 & 1578 & 289 & 122 & 2 &  136 &  8.41 & 8.96 & -17.12 & -17.68 & 0.33 & 2.01 \\
MKT J033654.4-251839.4$^{b}$ & NGC1371 & 54.2271 & -25.3109 & 1463 & 226 & 7 & 1 &  12 &  6.85 & 7.94 & -14.53 & -15.03 & 0.65 & 0.79 \\
MKT J041616.8-545338.0$^{b}$ & NGC1566 & 64.0700 & -54.8939 & 1233 & 166 & -263 & 1 &  70 &  6.91 & 8.54 & -15.81 & -16.41 & 0.37 & 1.18 \\
LSBG F157-052 & NGC1566 & 65.1117 & -54.7403 & 1340 & 64 & -156 & 1 &  61 &  7.28 & 8.27 & -15.20 & -15.76 & 0.40 & 1.14 \\
LEDA 414611 & NGC1566 & 65.5037 & -54.9115 & 1437 & 89 & -59 & 1 &  31 &  7.14 & 7.36 & -13.25 & -13.66 & 0.90 & 0.37 \\
MKT J042327.1-551618.6$^{b}$ & NGC1566 & 65.8629 & -55.2718 & 1211 & 184 & -285 & 1 &  10 &  6.42 & 6.26 & -10.61 & -10.92 & 0.57 & 0.31 \\
NGC 1581 & NGC1566 & 66.1850 & -54.9420 & 1633 & 210 & 137 & 1 &  208 &  8.11 & 9.48 & -18.18 & -18.84 & 0.42 & 1.42 \\
ESO 118-G034 & NGC1672 & 70.0692 & -58.7437 & 1153 & 292 & -174 & 2 &  51 &  8.47$^{e}$ & 9.26 & -18.16 & -18.65 & 0.98 & 0.83 \\
ESO 118-G042 & NGC1672 & 71.3650 & -59.0726 & 1208 & 60 & -119 & 1 &  45 &  7.97 & 8.33 & -15.69 & -16.16 & 0.54 & 1.30 \\
ESO 119-G003 & NGC1672 & 72.0308 & -59.4152 & 1211 & 119 & -116 & 1 &  65 &  8.43 & 7.50 & -14.49 & -14.68 & 0.29 & 1.86 \\
NGC 1688 & NGC1672 & 72.0975 & -59.7973 & 1226 & 219 & -102 & 1 &  149 &  9.13 & 9.82 & -18.45 & -18.96 & 0.59 & 2.54 \\
AGC 749315 & NGC3274 & 157.2837 & 26.9042 & 645 & 181 & 104 & 3 &  31 &  6.90 & 6.24 & -11.20 & -11.34 & 0.80 & 0.12 \\
AGC 731457 & NGC3274 & 157.9950 & 28.0303 & 454 & 64 & -87 & 1 &  36 &  7.17 & 7.52 & -13.52 & -13.97 & 0.76 & 0.37 \\
AGC 225760 & NGC4517 & 187.2737 & 0.1011 & 1198 & 132 & 71 & 1 &  25 &  7.17 & 7.58 & -13.62 & -14.09 & 0.85 & 0.57 \\
AGC 221570 & NGC4517 & 187.7658 & 1.6844 & 1104 & 236 & -23 & 2 &  39 &  7.61 & 7.75 & -13.91 & -14.43 & 0.71 & 0.56 \\
UGC 7300 & NGC4274 & 184.1854 & 28.7311 & 1209 & 187 & 292 & 3 &  74 &  8.43 & 7.62 & -14.22 & -14.56 & 0.52 & 1.03 \\
AGC 747848 & NGC4274 & 184.2337 & 30.5533 & 1145 & 189 & 228 & 3 &  34 &  7.34 & 7.02 & -12.71 & -13.02 & 0.41 & 0.50 \\
AGC 220408 & NGC4274 & 185.1442 & 30.7975 & 1101 & 200 & 184 & 3 &  38 &  7.50 & 8.07 & -14.88 & -15.38 & 0.86 & 0.53 \\
UGC 7438 & NGC4274 & 185.5817 & 30.0711 & 694 & 118 & -223 & 3 &  101 &  7.21 & 8.21 & -14.72 & -15.36 & 0.34 & 0.91 \\
AGC 724774 & NGC4274 & 185.8438 & 30.5658 & 679 & 204 & -238 & 3 &  53 &  7.44 & 7.32 & -13.33 & -13.69 & 0.45 & 0.41 \\
AGC 208569 & NGC3486 & 164.6950 & 29.2200 & 697 & 73 & 20 & 1 &  81 &  7.11 & 6.52 & -11.78 & -11.96 & 0.75 & 0.25 \\
AGC 722731 & NGC3486 & 164.9325 & 28.6067 & 698 & 67 & 21 & 1 &  12 &  6.57 & 7.00 & -11.98 & -12.48 & 0.83 & 0.21 \\
AGC 212945 & NGC3486 & 165.1646 & 29.7078 & 673 & 125 & -4 & 1 &  44 &  7.52 & 7.05 & -12.71 & -13.04 & 0.44 & 0.77 \\
AGC 215232 & NGC3486 & 165.3096 & 29.8489 & 693 & 151 & 16 & 1 &  23 &  7.23 & 7.29 & -13.33 & -13.66 & 0.45 & 0.44 \\
UGC 6102 & NGC3486 & 165.4492 & 28.6886 & 697 & 71 & 20 & 1 &  70 &  8.31 & 7.68 & -14.59 & -14.87 & 0.62 & 0.80 \\
UGC 6126 & NGC3486 & 165.9367 & 28.8881 & 704 & 125 & 27 & 1 &  184 &  8.98 & 8.57 & -16.81 & -17.16 & 0.24 & 1.60 \\
AGC 219369 & NGC3486 & 165.9633 & 28.6858 & 667 & 137 & -10 & 1 &  22 &  7.30 & 6.89 & -11.71 & -12.19 & 0.81 & 0.77 \\
AGC 210027 & NGC3486 & 166.2454 & 29.1392 & 647 & 172 & -30 & 1 &  44 &  7.39 & 7.19 & -14.50 & -14.44 & 0.71 & 0.25 \\
UGC 1924 & NGC925 & 36.9571 & 31.7306 & 595 & 294 & 43 & 2 &  111 &  8.28 & 7.79 & -14.52 & -14.91 & 0.22 & 0.87 \\
AGC 732129 & NGC4559 & 186.8521 & 27.8372 & 1051 & 295 & 238 & 3 &  29 &  6.99 & 6.83 & -12.10 & -12.43 & 0.47 & 0.27 \\
AGC 749238$^{c}$ & NGC4559 & 187.3763 & 27.6169 & 974 & 229 & 161 & 3 &  106 &  7.05 & \nodata & \nodata & \nodata & \nodata & \nodata \\
AGC 724906 & NGC4559 & 187.7358 & 26.5122 & 927 & 284 & 114 & 2 &  22 &  7.17 & 6.68 & -12.57 & -12.66 & 0.96 & 0.30 \\
UGC 7673 & NGC4559 & 187.9950 & 29.7139 & 643 & 305 & -170 & 3 &  70 &  8.22 & 7.53 & -14.12 & -14.42 & 0.69 & 0.88 \\
AGC 229111 & NGC4559 & 188.4046 & 29.4378 & 1001 & 244 & 188 & 3 &  31 &  7.15 & 6.93 & -12.27 & -12.63 & 0.71 & 0.25 \\
AGC 220847 & NGC4559 & 189.3092 & 29.6270 & 795 & 263 & -18 & 2 &  44 &  7.23 & 7.77 & -14.28 & -14.71 & 0.56 & 0.37 \\
AGC 724993 & NGC4559 & 189.6300 & 29.0522 & 756 & 191 & -57 & 2 &  37 &  7.27 & 7.07 & -13.09 & -13.33 & 0.87 & 0.34 \\
AGC 220984 & NGC4559 & 191.1587 & 28.4731 & 945 & 307 & 132 & 2 &  50 &  8.08 & 7.84 & -15.04 & -15.32 & 0.55 & 0.50 \\
AGC 202256 & NGC3593 & 168.6892 & 12.6503 & 630 & 26 & -1 & 1 &  42 &  7.10 & 6.93 & -12.21 & -12.58 & 0.60 & 0.23 \\
AGC 210220 & NGC3593 & 169.2529 & 13.0975 & 588 & 101 & -43 & 1 &  25 &  7.05 & 7.74 & -13.99 & -14.48 & 0.82 & 0.54 \\
AGC 742601 & NGC4826 & 192.4008 & 21.9181 & 539 & 130 & 131 & 1 &  27 &  6.60 & 6.25 & -10.59 & -10.89 & 0.70 & 0.13 \\
AGC 229386 & NGC4826 & 192.9408 & 21.7353 & 582 & 90 & 174 & 1 &  25 &  7.03 & 6.79 & -12.22 & -12.48 & 0.46 & 0.37 \\
UGC 8030 & NGC4826 & 193.6204 & 26.3025 & 628 & 360 & 220 & 3 &  53 &  6.88 & 7.00 & -12.33 & -12.73 & 0.43 & 0.49 \\
AGC 221120 & NGC4826 & 193.9242 & 19.2097 & 419 & 189 & 11 & 1 &  22 &  7.40 & 6.38 & -11.73 & -11.82 & 0.43 & 0.60 \\
UGC 5761 & NGC3351 & 159.0967 & 12.7119 & 604 & 342 & -173 & 3 &  112 &  7.91 & 8.45 & -15.48 & -16.10 & 0.82 & 0.85 \\
AGC 200532 & NGC3351 & 160.5033 & 12.3325 & 772 & 128 & -5 & 1 &  36 &  7.33 & 7.77 & -14.21 & -14.66 & 0.41 & 0.71 \\
AGC 201970$^{c}$ & NGC3351 & 161.7246 & 12.7392 & 636 & 205 & -141 & 1 &  38 &  7.12 & \nodata & \nodata & \nodata & \nodata & \nodata \\
AGC 200512 & NGC3368 & 159.9825 & 13.9044 & 1007 & 461 & 114 & 2 &  21 &  6.85 & 7.18 & -12.69 & -13.12 & 0.57 & 0.76 \\
UGC 5812 & NGC3368 & 160.2367 & 12.4725 & 1008 & 270 & 115 & 2 &  56 &  7.60 & 8.00 & -14.73 & -15.21 & 0.38 & 1.10 \\
AGC 202024 & NGC3368 & 161.2508 & 11.9128 & 871 & 76 & -22 & 1 &  24 &  6.72 & 7.01 & -12.47 & -12.84 & 0.53 & 0.49 \\
UGC 6014 & NGC3368 & 163.4246 & 9.7292 & 1133 & 467 & 240 & 3 &  94 &  7.87 & 7.92 & -14.92 & -15.29 & 0.48 & 0.78 \\
AGC 202035 & NGC3368 & 164.0617 & 12.0108 & 989 & 403 & 96 & 2 &  30 &  7.63 & 7.43 & -13.46 & -13.86 & 0.65 & 0.67 \\
AGC 191706 & NGC2903 & 142.5554 & 19.9919 & 561 & 246 & 4 & 1 &  23 &  7.27 & 6.95 & -12.55 & -12.84 & 0.89 & 0.43 \\
AGC 215282 & NGC3627 & 168.6129 & 15.5342 & 867 & 478 & 151 & 2 &  27 &  6.78 & 7.48 & -13.57 & -13.98 & 0.51 & 0.34 \\
AGC 215286 & NGC3627 & 169.7971 & 14.3275 & 998 & 224 & 282 & 1 &  28 &  7.07 & 7.32 & -12.75 & -13.27 & 0.70 & 0.67 \\
AGC 202257 & NGC3627 & 169.8104 & 11.9531 & 861 & 175 & 145 & 1 &  51 &  7.83 & 7.26 & -13.41 & -13.70 & 0.49 & 0.72 \\
AGC 213440 & NGC3627 & 170.9083 & 12.8956 & 666 & 136 & -50 & 1 &  22 &  6.72 & 7.39 & -13.09 & -13.57 & 0.54 & 0.46 \\
AGC 215296 & NGC3627 & 171.7292 & 14.8281 & 913 & 402 & 197 & 2 &  44 &  7.08 & 6.21 & -11.24 & -11.34 & 0.45 & 0.18 \\
AGC 212837 & NGC3627 & 172.7242 & 14.1453 & 880 & 465 & 164 & 2 &  22 &  7.61 & 7.40 & -13.36 & -13.77 & 0.84 & 0.65 \\
\enddata
\tablecomments{ 
Summary of the \HI-bearing dwarf galaxies. See Appendix \ref{appendix:dwarfs} for a description of the table columns.\\
$^{a}$ New detections from the WALLABY pilot data.\\
$^{b}$ New detections from the MHONGOOSE full-depth data. We list the \HIspace names instead of the candidate optical counterparts to avoid source confusion. Only sources within $\pm 300$ \kms~of the host's systemic velocity are included (\S \ref{subsec:method_sat_environment}).\\
$^{c}$ These two ALFALFA objects have Legacy DR10 coverage, but their optical counterparts are uncertain (potentially \HIspace clouds).\\
$^{d}$ The \HIspace shows clear tidal disturbance in the direction of the host galaxy.\\
$^{e}$ Reliable primary-beam attenuation corrections cannot be applied for these three galaxies near the MeerKAT primary beam edge. We adopted literature fluxes to derive their \HIspace masses: LEDA 46885 \citep{karachentsev_updated_2013}, ESO 411-G031 \citep{courtois_update_2015}, and ESO 118-G034 \citep{meyer_hipass_2004}. \\
$^{f}$ These two massive galaxies lack Legacy DR10 coverage but have recent stellar mass measurements available in the literature. We adopt $M_{\star}$ from the z0MGS catalog \citep{leroy_z_2019}. 
}
\end{deluxetable*}

\section{Injection Test for Sensitivity Assessment}\label{appendix:inject}

In this section, we describe our \HIspace flux injection pipeline for assessing the sensitivity of a data cube. The pipeline includes two main steps, (i) \textit{Modeling} (\S \ref{subsubsec:model_in_sensitivity_test}): we model the three-dimensional \HIspace emission of gaseous dwarf galaxies across a range of masses; (ii) \textit{Injection and source recovery} (\S \ref{subsubsec:injection_in_sensitivity_test}): we inject, that is, add the fluxes of the modeled dwarfs into the host \HIspace cubes, and rerun source finding using SoFiA-2 on the injected cubes (\S \ref{subsec:method_source_finding}). The recovery rate, defined as the ratio of detected sources to total injected sources as a function of $M_{\HI}$, is then used to quantify the sensitivity limit of the host field, as presented in Section \ref{subsec:method_sensitivity}.

\subsection{Modeling \HIspace in dwarf galaxies: from radial profiles to three-dimensional cubes}\label{subsubsec:model_in_sensitivity_test}

We model a dwarf galaxy's \HIspace distribution following the observed column density profiles of nearby dwarf galaxies. From the literature, we select low-mass gaseous dwarfs that have spatially resolved and inclination-corrected radial profiles, including targeted deep imaging of Leo T \citep{adams_deep_2018}, as well as low-mass dwarfs with $M_{\HI} \leq 10^{7} M_{\odot}$ from the LITTLE THINGS survey \citep{hunter_little_2012,hunter_relationships_2021}. Based on these observed galaxies, we calibrate a simple model that outputs the \HIspace column density profile, $\Sigma_{\HI}(R)$, given an input $M_{\HI}$.

\begin{figure*}
    \centering
    \includegraphics[width=1.0\linewidth]{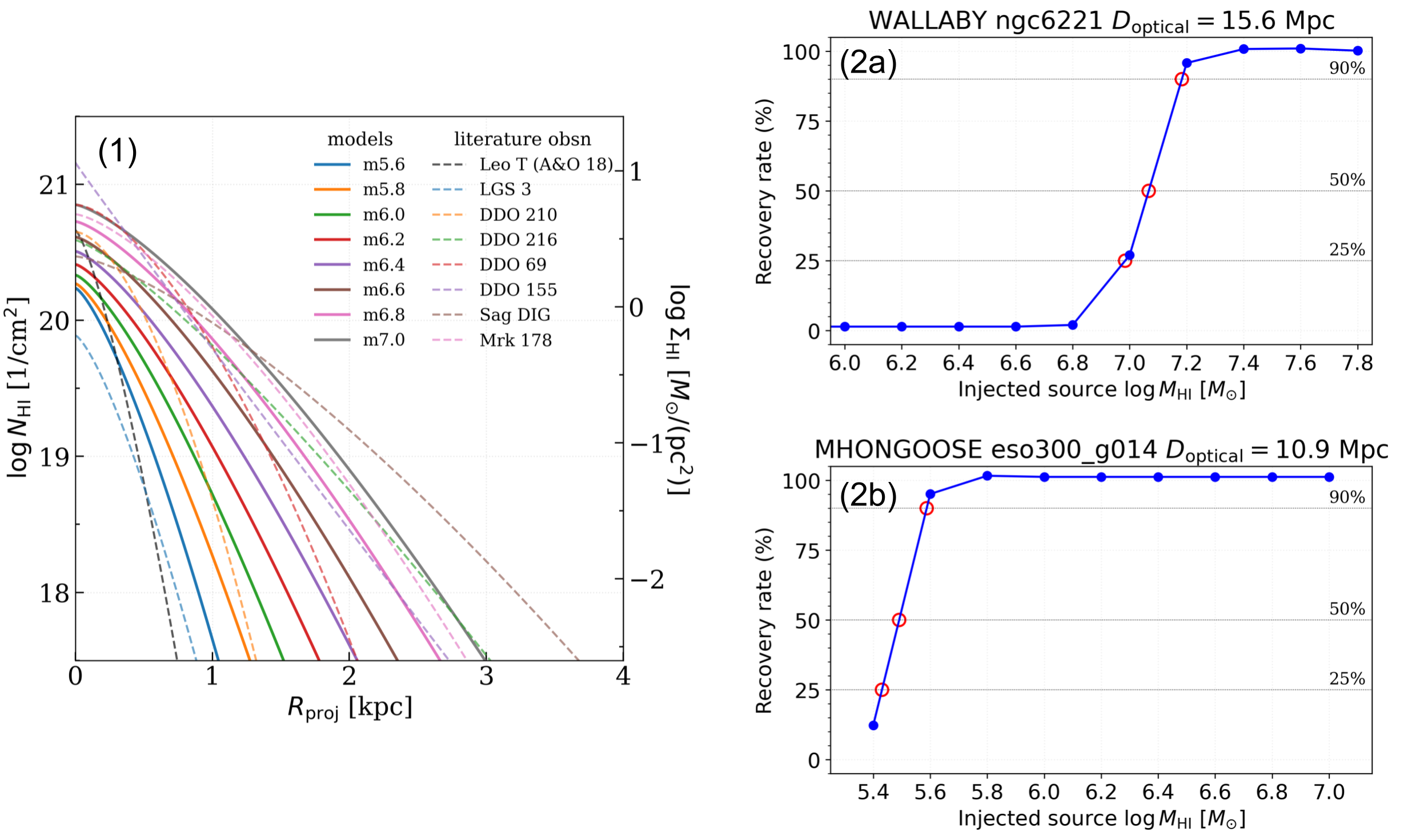}
    \caption{Key ingredients of our injection method for assessing the \HIspace sensitivity. Panel (1). Dwarf \HIspace column density profiles as a function of radius. The model dwarfs (Equation \ref{eqn:sersic_model_dwarfs}) are shown in solid lines, while the observed low-mass dwarfs from literature ($M_{\HI} < 10^{7} M_{\odot}$; from which the model parameters are derived) are shown in dashed lines \citep{adams_deep_2018,hunter_relationships_2021}. Panels (2a) and (2b). The recovery rate as a function of injected dwarf mass. The WALLABY and MHONGOOSE example hosts shown here are at the median distance among each survey (Table \ref{table:hosts}). The red circles mark the $M_{\HI}$ detectability thresholds at $25, 50, 90\%$ recovery rate.}
    \label{fig:injection_pipeline}
\end{figure*}

We adopt the simple S\'ersic-profile parametrization for the \HIspace column density profile of a galaxy, following \cite{hunter_relationships_2021},
\begin{equation}\label{eqn:sersic_model_dwarfs}
    \log \Sigma_{\HI}(R) = \log \Sigma_{\HI}^{0} - 0.434 (R/R_{0,\HI})^{1/n_{\HI}}
\end{equation}
where $\Sigma_{\HI}^{0}$ is the central density, $R_{0,\HI}$ is the scale radius, and $n_{\HI}$ the S\'ersic index. We derive an average radial profile that scales only with the galaxy's $M_{\HI}$, and model the density profiles for a uniform grid of $M_{\HI}$ to bracket the transition between recovery rates 0 to 100\%. Based on partial correlation analyses, the S\'ersic index is least correlated with $M_{\HI}$, and the scale radius is most correlated with $M_{\HI}$. We therefore adopt the simplifications that the S\'ersic index is a constant ($n_{\HI} = 0.743$), and the scale radius is a linear function of $\log M_{\HI}$ ($R_{0,\HI} = 0.278 \log M_{\HI} - 1.292$), and fit these values from the observed profiles from literature. We then solve for the central density $\Sigma_{\HI}^{0}$ given a target $M_{\HI}$ following Equation \ref{eqn:sersic_model_dwarfs}. Panel (1) of Figure \ref{fig:injection_pipeline} summarizes the column density profiles of the observed galaxies and our modeled galaxies.

We use the \texttt{SimulatedGalaxyCube} method in the \texttt{pyBBarolo} software \citep{di_teodoro_3d_2015} to model three-dimensional (3D) cubes of low-mass dwarf galaxies from the column density profiles (Figure \ref{fig:injection_pipeline}, panel 1). \texttt{pyBBarolo} takes the input of gas column densities as a function of angular radius to create 3D cubes. For a model galaxy at a given mass, the column density profile is independent of distance, but the angular size does scale with the distance $D$ to the observer, $R \propto 1/D$. Low-mass sources at larger distances inevitably become unresolved (i.e., on-sky sizes comparable to or smaller than the beam size). We need to correct for the beam dilution effect, where the observed intensity becomes lower than the underlying physical intensity for unresolved sources. To do so, we calculate the theoretical integrated flux of a galaxy (see, e.g., \citealt{meyer_tracing_2017} for a review), $S_{V, \rm theory} = M_{\HI}/(2.356\times 10^{5} \cdot D^{2})$, and multiply the intensities in all pixels of the 3D model cube by $S_{V, \rm theory}/S_{V, \rm model}$. This step ensures that both the analytical radial profile (Equation \ref{eqn:sersic_model_dwarfs}) and the correct total fluxes are preserved at any distance $D$, regardless of whether the source is resolved.

The column density profile is the most important input to \texttt{pyBBarolo}, as it determines the model galaxy's \HIspace mass. A set of additional input parameters can be specified. We use the WALLABY and MHONGOOSE survey-specific information (pixel scale, velocity channel size, and beam size) to set the grid sizes for these 3D model cubes. We choose to keep most other variables constant across masses for simplicity, setting them to reasonable values for low-mass dwarfs: the maximum rotation velocity $V_{\rm max}=17.67$ km/s, the dispersion velocity $V_{\rm disp}=5$ km/s, and the inclination angle $i=45 \deg$. Collectively, the inclination angle and maximum rotation velocity set the velocity width via $W_{50} = 2 \sin(i) \cdot V_{\rm max} = 25$ km/s, which is approximately the average $W_{50}$ for Local Group gaseous dwarfs \citep{putman_gas_2021}.

\subsection{Injection and source recovery}\label{subsubsec:injection_in_sensitivity_test}

To determine how sensitive a host galaxy cube is to a typical dwarf galaxy of a certain $M_{\HI,test}$, we inject (i.e., add the fluxes of) many copies of the $M_{\HI,test}$ model cube (\S \ref{subsubsec:model_in_sensitivity_test}) to a uniform 3D grid within the host cube. The model dwarf is scaled to the distance of the host (Table \ref{table:hosts}). The grid spacing is selected such that injected models are separable by the source finder, while the total injected number ($N_{\rm inject,tot}$) is sufficient for a robust recovery rate calculation. Given the sizes of our WALLABY and MHONGOOSE host cubes (\S \ref{subsec:method_source_finding}), $N_{\rm inject,tot}$ is typically a few hundred. Then, we rerun source finding using SoFiA-2 as described in \S \ref{subsec:method_source_finding} and report the recovered number of sources, $N_{\rm recover}$, at this mass. Because of the large number of injected sources, $N_{\rm inject,tot}$ is typically much greater than the original number of real detections or false positives (\S \ref{subsec:method_source_finding}), most of the SoFiA-2 detections in this step are indeed injected model dwarfs. The recovery rate is thus given by $N_{\rm recover}/N_{\rm inject,tot}$. We repeat this step iteratively for a grid of $M_{\HI,test}$ until the entire mass range is scanned, where the recovery rate varies from $0\%$ to $100\%$. 

The empirical $M_{\HI}$ thresholds for the host galaxies are a key result of this injection test, and are summarized in Section \ref{subsec:method_sensitivity}. Here in Figure \ref{fig:injection_pipeline}, we show the example recovery rates as a function of injected $M_{\HI,test}$, panel (2a) for a WALLABY host galaxy (NGC 6221), and panel (2b) for a MHONGOOSE host galaxy (ESO300-G014). Each galaxy is selected to be at the median distance of its survey's subsample (Table \ref{table:hosts}), so the result is representative. We calculate the $M_{\HI}$ thresholds at $25, 50, 90\%$ recovery rates by linearly interpolating the recovery rate results, as annotated by the circles in Figure \ref{fig:injection_pipeline}. The lowest mass model in our injection mass grid is set to be $\log M_{\HI} = 5.4$ based on the smallest observed gas-rich dwarf galaxies (LGS 3, Leo T-like mass; see, e.g., \citealt{hunter_little_2012,adams_deep_2018}). For some MHONGOOSE galaxies within 10 Mpc, the recovery rate is $>50\%$ at the lowest model mass, indicating that $\log M_{\HI}/M_{\odot} = 5.4$ is an upper limit (Table \ref{table:hosts}) and the true sensitivities are deeper.

\bibliography{yasmeen-refs,gas_stripping_sf_thesis}{}

@article{adams_deep_2018,
  title = {Deep Neutral Hydrogen Observations of {{Leo T}} with the {{Westerbork Synthesis Radio Telescope}}},
  author = {Adams, Elizabeth A. K. and Oosterloo, Tom A.},
  year = 2018,
  month = apr,
  journal = {A\&A},
  volume = {612},
  pages = {A26},
  publisher = {EDP},
  doi = {10.1051/0004-6361/201732017}
}

@article{akins_quenching_2021,
  title = {Quenching {{Timescales}} of {{Dwarf Satellites}} around {{Milky Way-mass Hosts}}},
  author = {Akins, Hollis B. and Christensen, Charlotte R. and Brooks, Alyson M. and Munshi, Ferah and Applebaum, Elaad and Engelhardt, Anna and Chamberland, Lucas},
  year = 2021,
  month = mar,
  journal = {ApJ},
  volume = {909},
  pages = {139},
  publisher = {IOP},
  doi = {10.3847/1538-4357/abe2ab}
}

@article{astropy_collaboration_astropy_2013,
  title = {Astropy: {{A}} Community {{Python}} Package for Astronomy},
  shorttitle = {Astropy},
  author = {{Astropy Collaboration} and Robitaille, Thomas P. and Tollerud, Erik J. and Greenfield, Perry and Droettboom, Michael and Bray, Erik and Aldcroft, Tom and Davis, Matt and Ginsburg, Adam and {Price-Whelan}, Adrian M. and Kerzendorf, Wolfgang E. and Conley, Alexander and Crighton, Neil and Barbary, Kyle and Muna, Demitri and Ferguson, Henry and Grollier, Fr{\'e}d{\'e}ric and Parikh, Madhura M. and Nair, Prasanth H. and Unther, Hans M. and Deil, Christoph and Woillez, Julien and Conseil, Simon and Kramer, Roban and Turner, James E. H. and Singer, Leo and Fox, Ryan and Weaver, Benjamin A. and Zabalza, Victor and Edwards, Zachary I. and Azalee Bostroem, K. and Burke, D. J. and Casey, Andrew R. and Crawford, Steven M. and Dencheva, Nadia and Ely, Justin and Jenness, Tim and Labrie, Kathleen and Lim, Pey Lian and Pierfederici, Francesco and Pontzen, Andrew and Ptak, Andy and Refsdal, Brian and Servillat, Mathieu and Streicher, Ole},
  year = 2013,
  month = oct,
  journal = {A\&A},
  volume = {558},
  pages = {A33},
  doi = {10.1051/0004-6361/201322068}
}

@article{astropy_collaboration_astropy_2018,
  title = {The {{Astropy Project}}: {{Building}} an {{Open-science Project}} and {{Status}} of the v2.0 {{Core Package}}},
  shorttitle = {The {{Astropy Project}}},
  author = {{Astropy Collaboration} and {Price-Whelan}, A. M. and Sip{\H o}cz, B. M. and G{\"u}nther, H. M. and Lim, P. L. and Crawford, S. M. and Conseil, S. and Shupe, D. L. and Craig, M. W. and Dencheva, N. and Ginsburg, A. and VanderPlas, J. T. and Bradley, L. D. and {P{\'e}rez-Su{\'a}rez}, D. and {de Val-Borro}, M. and Aldcroft, T. L. and Cruz, K. L. and Robitaille, T. P. and Tollerud, E. J. and Ardelean, C. and Babej, T. and Bach, Y. P. and Bachetti, M. and Bakanov, A. V. and Bamford, S. P. and Barentsen, G. and Barmby, P. and Baumbach, A. and Berry, K. L. and Biscani, F. and Boquien, M. and Bostroem, K. A. and Bouma, L. G. and Brammer, G. B. and Bray, E. M. and Breytenbach, H. and Buddelmeijer, H. and Burke, D. J. and Calderone, G. and Cano Rodr{\'i}guez, J. L. and Cara, M. and Cardoso, J. V. M. and Cheedella, S. and Copin, Y. and Corrales, L. and Crichton, D. and D'Avella, D. and Deil, C. and Depagne, {\'E}. and Dietrich, J. P. and Donath, A. and Droettboom, M. and Earl, N. and Erben, T. and Fabbro, S. and Ferreira, L. A. and Finethy, T. and Fox, R. T. and Garrison, L. H. and Gibbons, S. L. J. and Goldstein, D. A. and Gommers, R. and Greco, J. P. and Greenfield, P. and Groener, A. M. and Grollier, F. and Hagen, A. and Hirst, P. and Homeier, D. and Horton, A. J. and Hosseinzadeh, G. and Hu, L. and Hunkeler, J. S. and Ivezi{\'c}, {\v Z}. and Jain, A. and Jenness, T. and Kanarek, G. and Kendrew, S. and Kern, N. S. and Kerzendorf, W. E. and Khvalko, A. and King, J. and Kirkby, D. and Kulkarni, A. M. and Kumar, A. and Lee, A. and Lenz, D. and Littlefair, S. P. and Ma, Z. and Macleod, D. M. and Mastropietro, M. and McCully, C. and Montagnac, S. and Morris, B. M. and Mueller, M. and Mumford, S. J. and Muna, D. and Murphy, N. A. and Nelson, S. and Nguyen, G. H. and Ninan, J. P. and N{\"o}the, M. and Ogaz, S. and Oh, S. and Parejko, J. K. and Parley, N. and Pascual, S. and Patil, R. and Patil, A. A. and Plunkett, A. L. and Prochaska, J. X. and Rastogi, T. and Reddy Janga, V. and Sabater, J. and Sakurikar, P. and Seifert, M. and Sherbert, L. E. and {Sherwood-Taylor}, H. and Shih, A. Y. and Sick, J. and Silbiger, M. T. and Singanamalla, S. and Singer, L. P. and Sladen, P. H. and Sooley, K. A. and Sornarajah, S. and Streicher, O. and Teuben, P. and Thomas, S. W. and Tremblay, G. R. and Turner, J. E. H. and Terr{\'o}n, V. and {van Kerkwijk}, M. H. and {de la Vega}, A. and Watkins, L. L. and Weaver, B. A. and Whitmore, J. B. and Woillez, J. and Zabalza, V. and {Astropy Contributors}},
  year = 2018,
  month = sep,
  journal = {AJ},
  volume = {156},
  pages = {123},
  doi = {10.3847/1538-3881/aabc4f}
}

@article{astropy_collaboration_astropy_2022,
  title = {The {{Astropy Project}}: {{Sustaining}} and {{Growing}} a {{Community-oriented Open-source Project}} and the {{Latest Major Release}} (v5.0) of the {{Core Package}}},
  shorttitle = {The {{Astropy Project}}},
  author = {{Astropy Collaboration} and {Price-Whelan}, Adrian M. and Lim, Pey Lian and Earl, Nicholas and Starkman, Nathaniel and Bradley, Larry and Shupe, David L. and Patil, Aarya A. and Corrales, Lia and Brasseur, C. E. and N{\"o}the, Maximilian and Donath, Axel and Tollerud, Erik and Morris, Brett M. and Ginsburg, Adam and Vaher, Eero and Weaver, Benjamin A. and Tocknell, James and Jamieson, William and {van Kerkwijk}, Marten H. and Robitaille, Thomas P. and Merry, Bruce and Bachetti, Matteo and G{\"u}nther, H. Moritz and Aldcroft, Thomas L. and {Alvarado-Montes}, Jaime A. and Archibald, Anne M. and B{\'o}di, Attila and Bapat, Shreyas and Barentsen, Geert and Baz{\'a}n, Juanjo and Biswas, Manish and Boquien, M{\'e}d{\'e}ric and Burke, D. J. and Cara, Daria and Cara, Mihai and Conroy, Kyle E. and Conseil, Simon and Craig, Matthew W. and Cross, Robert M. and Cruz, Kelle L. and D'Eugenio, Francesco and Dencheva, Nadia and Devillepoix, Hadrien A. R. and Dietrich, J{\"o}rg P. and Eigenbrot, Arthur Davis and Erben, Thomas and Ferreira, Leonardo and {Foreman-Mackey}, Daniel and Fox, Ryan and Freij, Nabil and Garg, Suyog and Geda, Robel and Glattly, Lauren and Gondhalekar, Yash and Gordon, Karl D. and Grant, David and Greenfield, Perry and Groener, Austen M. and Guest, Steve and Gurovich, Sebastian and Handberg, Rasmus and Hart, Akeem and {Hatfield-Dodds}, Zac and Homeier, Derek and Hosseinzadeh, Griffin and Jenness, Tim and Jones, Craig K. and Joseph, Prajwel and Kalmbach, J. Bryce and Karamehmetoglu, Emir and Ka{\l}uszy{\'n}ski, Miko{\l}aj and Kelley, Michael S. P. and Kern, Nicholas and Kerzendorf, Wolfgang E. and Koch, Eric W. and Kulumani, Shankar and Lee, Antony and Ly, Chun and Ma, Zhiyuan and MacBride, Conor and Maljaars, Jakob M. and Muna, Demitri and Murphy, N. A. and Norman, Henrik and O'Steen, Richard and Oman, Kyle A. and Pacifici, Camilla and Pascual, Sergio and {Pascual-Granado}, J. and Patil, Rohit R. and Perren, Gabriel I. and Pickering, Timothy E. and Rastogi, Tanuj and Roulston, Benjamin R. and Ryan, Daniel F. and Rykoff, Eli S. and Sabater, Jose and Sakurikar, Parikshit and Salgado, Jes{\'u}s and Sanghi, Aniket and Saunders, Nicholas and Savchenko, Volodymyr and Schwardt, Ludwig and {Seifert-Eckert}, Michael and Shih, Albert Y. and Jain, Anany Shrey and Shukla, Gyanendra and Sick, Jonathan and Simpson, Chris and Singanamalla, Sudheesh and Singer, Leo P. and Singhal, Jaladh and Sinha, Manodeep and Sip{\H o}cz, Brigitta M. and Spitler, Lee R. and Stansby, David and Streicher, Ole and {\v S}umak, Jani and Swinbank, John D. and Taranu, Dan S. and Tewary, Nikita and Tremblay, Grant R. and {de Val-Borro}, Miguel and Van Kooten, Samuel J. and Vasovi{\'c}, Zlatan and Verma, Shresth and {de Miranda Cardoso}, Jos{\'e} Vin{\'i}cius and Williams, Peter K. G. and Wilson, Tom J. and Winkel, Benjamin and {Wood-Vasey}, W. M. and Xue, Rui and Yoachim, Peter and Zhang, Chen and Zonca, Andrea and {Astropy Project Contributors}},
  year = 2022,
  month = aug,
  journal = {ApJ},
  volume = {935},
  pages = {167},
  doi = {10.3847/1538-4357/ac7c74}
}

@article{behroozi_universemachine_2019,
  title = {{{UNIVERSEMACHINE}}: {{The}} Correlation between Galaxy Growth and Dark Matter Halo Assembly from z = 0-10},
  shorttitle = {{{UNIVERSEMACHINE}}},
  author = {Behroozi, Peter and Wechsler, Risa H. and Hearin, Andrew P. and Conroy, Charlie},
  year = 2019,
  month = sep,
  journal = {MNRAS},
  volume = {488},
  pages = {3143--3194},
  doi = {10.1093/mnras/stz1182}
}

@article{benavides_environmental_2025,
  title = {The {{Environmental Quenching Mechanisms}} of {{Field Dwarf Galaxies}}},
  author = {Benavides, Jos{\'e} A. and Navarro, Julio F. and Sales, Laura V. and P{\'e}rez, Isabel and Bidaran, Bahar},
  year = 2025,
  month = may,
  journal = {ApJ},
  volume = {985},
  number = {1},
  pages = {86},
  publisher = {The American Astronomical Society},
  doi = {10.3847/1538-4357/adced0},
  langid = {english}
}

@article{bennet_m101_2019,
  title = {The {{M101 Satellite Luminosity Function}} and the {{Halo}}--{{Halo Scatter}} among {{Local Volume Hosts}}},
  author = {Bennet, P. and Sand, D. J. and Crnojevi{\'c}, D. and Spekkens, K. and Karunakaran, A. and Zaritsky, D. and {Mutlu-Pakdil}, B.},
  year = 2019,
  month = nov,
  journal = {ApJ},
  volume = {885},
  number = {2},
  pages = {153},
  publisher = {The American Astronomical Society},
  doi = {10.3847/1538-4357/ab46ab},
  langid = {english}
}

@article{bhattacharyya_environmental_2024,
  title = {Environmental {{Quenching}} of {{Low-surface-brightness Galaxies Near Hosts}} from {{Large Magellanic Cloud}} to {{Milky Way Mass Scales}}},
  author = {Bhattacharyya, J. and Peter, A. H. G. and Martini, P. and {Mutlu-Pakdil}, B. and {Drlica-Wagner}, A. and Pace, A. B. and Strigari, L. E. and Cheng, T.-Y. and Roberts, D. and Tanoglidis, D. and Aguena, M. and Alves, O. and {Andrade-Oliveira}, F. and Bacon, D. and Brooks, D. and Carnero Rosell, A. and Carretero, J. and {da Costa}, L. N. and Pereira, M. E. S. and Davis, T. M. and Desai, S. and Doel, P. and Ferrero, I. and Frieman, J. and {Garc{\'i}a-Bellido}, J. and Giannini, G. and Gruen, D. and Gruendl, R. A. and Hinton, S. R. and Hollowood, D. L. and Honscheid, K. and James, D. J. and Kuehn, K. and Marshall, J. L. and {Mena-Fern{\'a}ndez}, J. and Miquel, R. and Palmese, A. and Pieres, A. and Plazas Malag{\'o}n, A. A. and Sanchez, E. and Santiago, B. and Schubnell, M. and {Sevilla-Noarbe}, I. and Smith, M. and Suchyta, E. and Swanson, M. E. C. and Tarle, G. and Vincenzi, M. and Walker, A. R. and Weaverdyck, N. and Wiseman, P. and Collaboration), (Dark Energy Survey},
  year = 2024,
  month = nov,
  journal = {ApJ},
  volume = {975},
  number = {2},
  pages = {244},
  publisher = {The American Astronomical Society},
  doi = {10.3847/1538-4357/ad79fe},
  langid = {english}
}

@article{binggeli_abundance_1990,
  title = {The Abundance and Morphological Segregation of Dwarf Galaxies in the Field},
  author = {Binggeli, B. and Tarenghi, M. and Sandage, A.},
  year = 1990,
  month = feb,
  journal = {A\&A},
  volume = {228},
  pages = {42--60},
  publisher = {EDP}
}

@article{blumenthal_formation_1984,
  title = {Formation of Galaxies and Large-Scale Structure with Cold Dark Matter},
  author = {Blumenthal, George R. and Faber, S. M. and Primack, Joel R. and Rees, Martin J.},
  year = 1984,
  month = oct,
  journal = {Nature},
  volume = {311},
  number = {5986},
  pages = {517--525},
  publisher = {Nature Publishing Group},
  doi = {10.1038/311517a0},
  copyright = {1984 Springer Nature Limited},
  langid = {english}
}

@article{borrow_there_2023,
  title = {There and Back Again: {{Understanding}} the Critical Properties of Backsplash Galaxies},
  shorttitle = {There and Back Again},
  author = {Borrow, Josh and Vogelsberger, Mark and O'Neil, Stephanie and McDonald, Michael A. and Smith, Aaron},
  year = 2023,
  month = mar,
  journal = {MNRAS},
  volume = {520},
  pages = {649--667},
  doi = {10.1093/mnras/stad045}
}

@article{bradford_study_2015,
  title = {A {{Study}} in {{Blue}}: {{The Baryon Content}} of {{Isolated Low-mass Galaxies}}},
  shorttitle = {A {{Study}} in {{Blue}}},
  author = {Bradford, Jeremy D. and Geha, Marla C. and Blanton, Michael R.},
  year = 2015,
  month = aug,
  journal = {ApJ},
  volume = {809},
  pages = {146},
  publisher = {IOP},
  doi = {10.1088/0004-637X/809/2/146}
}

@article{brown_cold_2017,
  title = {Cold Gas Stripping in Satellite Galaxies: From Pairs to Clusters},
  shorttitle = {Cold Gas Stripping in Satellite Galaxies},
  author = {Brown, Toby and Catinella, Barbara and Cortese, Luca and Lagos, Claudia del P. and Dav{\'e}, Romeel and Kilborn, Virginia and Haynes, Martha P. and Giovanelli, Riccardo and Rafieferantsoa, Mika},
  year = 2017,
  month = apr,
  journal = {MNRAS},
  volume = {466},
  pages = {1275--1289},
  doi = {10.1093/mnras/stw2991}
}

@article{brown_quenching_2014,
  title = {{{THE QUENCHING OF THE ULTRA-FAINT DWARF GALAXIES IN THE REIONIZATION ERA}}*},
  author = {Brown, Thomas M. and Tumlinson, Jason and Geha, Marla and Simon, Joshua D. and Vargas, Luis C. and VandenBerg, Don A. and Kirby, Evan N. and Kalirai, Jason S. and Avila, Roberto J. and Gennaro, Mario and Ferguson, Henry C. and Mu{\~n}oz, Ricardo R. and Guhathakurta, Puragra and Renzini, Alvio},
  year = 2014,
  month = nov,
  journal = {ApJ},
  volume = {796},
  number = {2},
  pages = {91},
  publisher = {The American Astronomical Society},
  doi = {10.1088/0004-637X/796/2/91},
  langid = {english}
}

@article{bryan_statistical_1998,
  title = {Statistical {{Properties}} of {{X-Ray Clusters}}: {{Analytic}} and {{Numerical Comparisons}}},
  shorttitle = {Statistical {{Properties}} of {{X-Ray Clusters}}},
  author = {Bryan, Greg L. and Norman, Michael L.},
  year = 1998,
  month = mar,
  journal = {ApJ},
  volume = {495},
  pages = {80--99},
  publisher = {IOP},
  doi = {10.1086/305262}
}

@article{bullock_small-scale_2017,
  title = {Small-{{Scale Challenges}} to the {{$\Lambda$CDM Paradigm}}},
  author = {Bullock, James S. and {Boylan-Kolchin}, Michael},
  year = 2017,
  month = aug,
  journal = {ARA\&A},
  volume = {55},
  pages = {343--387},
  doi = {10.1146/annurev-astro-091916-055313}
}

@article{carlsten_exploration_2022,
  title = {The {{Exploration}} of {{Local VolumE Satellites}} ({{ELVES}}) {{Survey}}: {{A Nearly Volume-limited Sample}} of {{Nearby Dwarf Satellite Systems}}},
  shorttitle = {The {{Exploration}} of {{Local VolumE Satellites}} ({{ELVES}}) {{Survey}}},
  author = {Carlsten, Scott G. and Greene, Jenny E. and Beaton, Rachael L. and Danieli, Shany and Greco, Johnny P.},
  year = 2022,
  month = jul,
  journal = {ApJ},
  volume = {933},
  pages = {47},
  doi = {10.3847/1538-4357/ac6fd7}
}

@article{carlsten_structures_2021,
  title = {Structures of {{Dwarf Satellites}} of {{Milky Way-like Galaxies}}: {{Morphology}}, {{Scaling Relations}}, and {{Intrinsic Shapes}}},
  shorttitle = {Structures of {{Dwarf Satellites}} of {{Milky Way-like Galaxies}}},
  author = {Carlsten, Scott G. and Greene, Jenny E. and Greco, Johnny P. and Beaton, Rachael L. and {Kado-Fong}, Erin},
  year = 2021,
  month = dec,
  journal = {ApJ},
  volume = {922},
  pages = {267},
  doi = {10.3847/1538-4357/ac2581}
}

@article{carlsten_using_2019,
  title = {Using {{Surface Brightness Fluctuations}} to {{Study Nearby Satellite Galaxy Systems}}: {{Calibration}} and {{Methodology}}},
  shorttitle = {Using {{Surface Brightness Fluctuations}} to {{Study Nearby Satellite Galaxy Systems}}},
  author = {Carlsten, Scott G. and Beaton, Rachael L. and Greco, Johnny P. and Greene, Jenny E.},
  year = 2019,
  month = jun,
  journal = {ApJ},
  volume = {879},
  number = {1},
  pages = {13},
  publisher = {The American Astronomical Society},
  doi = {10.3847/1538-4357/ab22c1},
  langid = {english}
}

@article{cautun_milky_2020,
  title = {The {{Milky Way}} Total Mass Profile as Inferred from {{Gaia DR2}}},
  author = {Cautun, Marius and {Benitez-Llambay}, Alejandro and Deason, Alis J. and Frenk, Carlos S. and Fattahi, Azadeh and G{\'o}mez, Facundo A. and Grand, Robert J. J. and Oman, Kyle A. and Navarro, Julio F. and Simpson, Christine M.},
  year = 2020,
  month = may,
  journal = {MNRAS},
  volume = {494},
  number = {3},
  eprint = {1911.04557},
  primaryclass = {astro-ph},
  pages = {4291--4313},
  doi = {10.1093/mnras/staa1017},
  archiveprefix = {arXiv}
}

@article{christensen_-n-out_2016,
  title = {{{IN-N-OUT}}: {{THE GAS CYCLE FROM DWARFS TO SPIRAL GALAXIES}}},
  shorttitle = {{{IN-N-OUT}}},
  author = {Christensen, Charlotte R. and Dav{\'e}, Romeel and Governato, Fabio and Pontzen, Andrew and Brooks, Alyson and Munshi, Ferah and Quinn, Thomas and Wadsley, James},
  year = 2016,
  month = jun,
  journal = {ApJ},
  volume = {824},
  number = {1},
  pages = {57},
  publisher = {The American Astronomical Society},
  doi = {10.3847/0004-637X/824/1/57},
  langid = {english}
}

@article{christensen_environment_2024,
  title = {Environment {{Matters}}: {{Predicted Differences}} in the {{Stellar Mass}}--{{Halo Mass Relation}} and {{History}} of {{Star Formation}} for {{Dwarf Galaxies}}},
  shorttitle = {Environment {{Matters}}},
  author = {Christensen, Charlotte R. and Brooks, Alyson M. and Munshi, Ferah and Riggs, Claire and Van Nest, Jordan and Akins, Hollis and Quinn, Thomas R. and Chamberland, Lucas},
  year = 2024,
  month = feb,
  journal = {ApJ},
  volume = {961},
  pages = {236},
  doi = {10.3847/1538-4357/ad0c5a}
}

@article{cohen_dragonfly_2018,
  title = {The {{Dragonfly Nearby Galaxies Survey}}. {{V}}. {{HST}}/{{ACS Observations}} of 23 {{Low Surface Brightness Objects}} in the {{Fields}} of {{NGC}} 1052, {{NGC}} 1084, {{M96}}, and {{NGC}} 4258},
  author = {Cohen, Yotam and van Dokkum, Pieter and Danieli, Shany and Romanowsky, Aaron J. and Abraham, Roberto and Merritt, Allison and Zhang, Jielai and Mowla, Lamiya and Kruijssen, J. M. Diederik and Conroy, Charlie and Wasserman, Asher},
  year = 2018,
  month = nov,
  journal = {ApJ},
  volume = {868},
  number = {2},
  pages = {96},
  publisher = {The American Astronomical Society},
  doi = {10.3847/1538-4357/aae7c8},
  langid = {english}
}

@article{cortese_dawes_2021,
  title = {The {{Dawes Review}} 9: {{The}} Role of Cold Gas Stripping on the Star Formation Quenching of Satellite Galaxies},
  shorttitle = {The {{Dawes Review}} 9},
  author = {Cortese, L. and Catinella, B. and Smith, R.},
  year = 2021,
  month = aug,
  journal = {PASA},
  volume = {38},
  pages = {e035},
  doi = {10.1017/pasa.2021.18}
}

@article{courtois_update_2015,
  title = {Update on {{H I}} Data Collection from {{Green Bank}}, {{Parkes}} and {{Arecibo}} Telescopes for the {{Cosmic Flows}} Project},
  author = {Courtois, H{\'e}l{\`e}ne M. and Tully, R. Brent},
  year = 2015,
  month = feb,
  journal = {MNRAS},
  volume = {447},
  pages = {1531--1534},
  publisher = {OUP},
  doi = {10.1093/mnras/stu2405}
}

@article{crnojevic_faint_2019,
  title = {The {{Faint End}} of the {{Centaurus A Satellite Luminosity Function}}},
  author = {Crnojevi{\'c}, D. and Sand, D. J. and Bennet, P. and Pasetto, S. and Spekkens, K. and Caldwell, N. and Guhathakurta, P. and McLeod, B. and Seth, A. and Simon, J. D. and Strader, J. and Toloba, E.},
  year = 2019,
  month = feb,
  journal = {ApJ},
  volume = {872},
  pages = {80},
  doi = {10.3847/1538-4357/aafbe7}
}

@article{danieli_elves_2023,
  title = {{{ELVES}}. {{IV}}. {{The Satellite Stellar-to-halo Mass Relation Beyond}} the {{Milky Way}}},
  author = {Danieli, Shany and Greene, Jenny E. and Carlsten, Scott and Jiang, Fangzhou and Beaton, Rachael and Goulding, Andy D.},
  year = 2023,
  month = oct,
  journal = {ApJ},
  volume = {956},
  pages = {6},
  publisher = {IOP},
  doi = {10.3847/1538-4357/acefbd}
}

@misc{davis_lbt_2024,
  title = {The {{LBT Satellites}} of {{Nearby Galaxies Survey}} ({{LBT-SONG}}): {{The Diffuse Satellite Population}} of {{Local Volume Hosts}}},
  shorttitle = {The {{LBT Satellites}} of {{Nearby Galaxies Survey}} ({{LBT-SONG}})},
  author = {Davis, A. Bianca and Garling, Christopher T. and Nierenberg, Anna M. and Peter, Annika H. G. and Sardone, Amy and Kochanek, Christopher S. and Leroy, Adam K. and Casey, Kirsten J. and Pogge, Richard W. and Roberts, Daniella M. and Sand, David J. and Greco, Johnny P.},
  year = 2024,
  month = sep,
  publisher = {arXiv},
  doi = {10.48550/arXiv.2409.03999}
}

@article{de_blok_mhongoose_2024,
  title = {{{MHONGOOSE}}: {{A MeerKAT}} Nearby Galaxy {{H I}} Survey},
  shorttitle = {{{MHONGOOSE}}},
  author = {{de Blok}, W. J. G. and Healy, J. and Maccagni, F. M. and Pisano, D. J. and Bosma, A. and English, J. and Jarrett, T. and Marasco, A. and Meurer, G. R. and Veronese, S. and Bigiel, F. and Chemin, L. and Fraternali, F. and Holwerda, B. W. and Kamphuis, P. and Kl{\"o}ckner, H. R. and Kleiner, D. and Leroy, A. K. and Mogotsi, M. and Oman, K. A. and Schinnerer, E. and {Verdes-Montenegro}, L. and Westmeier, T. and Wong, O. I. and Zabel, N. and Amram, P. and Carignan, C. and Combes, F. and Brinks, E. and Dettmar, R. J. and Gibson, B. K. and Jozsa, G. I. G. and Koribalski, B. S. and McGaugh, S. S. and Oosterloo, T. A. and Spekkens, K. and Schr{\"o}der, A. C. and Adams, E. A. K. and Athanassoula, E. and Bershady, M. A. and Beswick, R. J. and Blyth, S. and Elson, E. C. and Frank, B. S. and Heald, G. and Henning, P. A. and Kurapati, S. and Loubser, S. I. and Lucero, D. and Meyer, M. and Namumba, B. and Oh, S. -H. and Sardone, A. and Sheth, K. and Smith, M. W. L. and Sorgho, A. and Walter, F. and Williams, T. and Woudt, P. A. and Zijlstra, A.},
  year = 2024,
  month = aug,
  journal = {A\&A},
  volume = {688},
  pages = {A109},
  publisher = {EDP},
  doi = {10.1051/0004-6361/202348297}
}

@article{de_los_reyes_stellar_2025,
  title = {Stellar {{Mass Calibrations}} for {{Local Low-mass Galaxies}}},
  author = {{de los Reyes}, Mithi A. C. and Asali, Yasmeen and Wechsler, Risa H. and Geha, Marla and Mao, Yao-Yuan and {Kado-Fong}, Erin and Pucha, Ragadeepika and Grant, William and Gandhi, Pratik J. and Manwadkar, Viraj and Engelhardt, Anna and Munshi, Ferah and Wang, Yunchong},
  year = 2025,
  month = aug,
  journal = {ApJ},
  volume = {989},
  pages = {91},
  publisher = {IOP},
  doi = {10.3847/1538-4357/ade4c5}
}

@article{deg_wallaby_2022,
  title = {{{WALLABY Pilot Survey}}: {{Public}} Release of {{HI}} Kinematic Models for More than 100 Galaxies from Phase 1 of {{ASKAP}} Pilot Observations},
  shorttitle = {{{WALLABY Pilot Survey}}},
  author = {Deg, N. and Spekkens, K. and Westmeier, T. and Reynolds, T. N. and Venkataraman, P. and Goliath, S. and Shen, A. X. and Halloran, R. and Bosma, A. and Catinella, B. and {de Blok}, W. J. G. and D{\'e}nes, H. and DiTeodoro, E. M. and Elagali, A. and For, B. -Q. and Howlett, C. and J{\'o}zsa, G. I. G. and Kamphuis, P. and Kleiner, D. and Koribalski, B. and {Lee-Waddell}, K. and Lelli, F. and Lin, X. and Murugeshan, C. and Oh, S. and Rhee, J. and Scott, T. C. and {Staveley-Smith}, L. and {van der Hulst}, J. M. and {Verdes-Montenegro}, L. and Wang, J. and Wong, O. I.},
  year = 2022,
  month = nov,
  journal = {PASA},
  volume = {39},
  pages = {e059},
  doi = {10.1017/pasa.2022.43}
}

@article{di_teodoro_3d_2015,
  title = {{{3D BAROLO}}: A New {{3D}} Algorithm to Derive Rotation Curves of Galaxies},
  shorttitle = {{{3D BAROLO}}},
  author = {Di Teodoro, E. M. and Fraternali, F.},
  year = 2015,
  month = aug,
  journal = {MNRAS},
  volume = {451},
  pages = {3021--3033},
  publisher = {OUP},
  doi = {10.1093/mnras/stv1213}
}

@article{emerick_gas_2016,
  title = {Gas {{Loss}} by {{Ram Pressure Stripping}} and {{Internal Feedback From Low Mass Milky Way Satellites}}},
  author = {Emerick, Andrew and Low, Mordecai-Mark Mac and Grcevich, Jana and Gatto, Andrea},
  year = 2016,
  month = aug,
  journal = {ApJ},
  volume = {826},
  number = {2},
  eprint = {1605.02746},
  primaryclass = {astro-ph},
  pages = {148},
  doi = {10.3847/0004-637X/826/2/148},
  archiveprefix = {arXiv}
}

@article{engler_satellites_2023,
  title = {Satellites of {{Milky Way-}} and {{M31-like}} Galaxies with {{TNG50}}: Quenched Fractions, Gas Content, and Star Formation Histories},
  shorttitle = {Satellites of {{Milky Way-}} and {{M31-like}} Galaxies with {{TNG50}}},
  author = {Engler, Christoph and Pillepich, Annalisa and Joshi, Gandhali D. and Pasquali, Anna and Nelson, Dylan and Grebel, Eva K.},
  year = 2023,
  month = jul,
  journal = {MNRAS},
  volume = {522},
  pages = {5946--5972},
  doi = {10.1093/mnras/stad1357}
}

@article{fillingham_taking_2015,
  title = {Taking Care of Business in a Flash: Constraining the Time-Scale for Low-Mass Satellite Quenching with {{ELVIS}}},
  shorttitle = {Taking Care of Business in a Flash},
  author = {Fillingham, Sean P. and Cooper, Michael C. and Wheeler, Coral and {Garrison-Kimmel}, Shea and {Boylan-Kolchin}, Michael and Bullock, James S.},
  year = 2015,
  month = dec,
  journal = {MNRAS},
  volume = {454},
  pages = {2039--2049},
  doi = {10.1093/mnras/stv2058}
}

@article{font_quenching_2022,
  title = {Quenching of Satellite Galaxies of {{Milky Way}} Analogues: Reconciling Theory and Observations},
  shorttitle = {Quenching of Satellite Galaxies of {{Milky Way}} Analogues},
  author = {Font, Andreea S. and McCarthy, Ian G. and Belokurov, Vasily and Brown, Shaun T. and Stafford, Sam G.},
  year = 2022,
  month = mar,
  journal = {MNRAS},
  volume = {511},
  pages = {1544--1556},
  doi = {10.1093/mnras/stac183}
}

@article{garling_search_2021,
  title = {A Search for Satellite Galaxies of Nearby Star-Forming Galaxies with Resolved Stars in {{LBT-SONG}}},
  author = {Garling, Christopher T and Peter, Annika H G and Kochanek, Christopher S and Sand, David J and Crnojevi{\'c}, Denija},
  year = 2021,
  month = nov,
  journal = {MNRAS},
  volume = {507},
  number = {4},
  pages = {4764--4778},
  doi = {10.1093/mnras/stab2447}
}

@article{garrison-kimmel_organized_2017,
  title = {Organized Chaos: Scatter in the Relation between Stellar Mass and Halo Mass in Small Galaxies},
  shorttitle = {Organized Chaos},
  author = {{Garrison-Kimmel}, Shea and Bullock, James S. and {Boylan-Kolchin}, Michael and Bardwell, Emma},
  year = 2017,
  month = jan,
  journal = {MNRAS},
  volume = {464},
  number = {3},
  pages = {3108--3120},
  doi = {10.1093/mnras/stw2564}
}

@article{geha_baryon_2006,
  title = {The {{Baryon Content}} of {{Extremely Low Mass Dwarf Galaxies}}},
  author = {Geha, M. and Blanton, M. R. and Masjedi, M. and West, A. A.},
  year = 2006,
  month = dec,
  journal = {ApJ},
  volume = {653},
  pages = {240--254},
  doi = {10.1086/508604}
}

@article{geha_saga_2024,
  title = {The {{SAGA Survey}}. {{IV}}. {{The Star Formation Properties}} of 101 {{Satellite Systems}} around {{Milky Way}}--Mass {{Galaxies}}},
  author = {Geha, Marla and Mao, Yao-Yuan and Wechsler, Risa H. and Asali, Yasmeen and {Kado-Fong}, Erin and Kallivayalil, Nitya and Nadler, Ethan O. and Tollerud, Erik J. and Weiner, Benjamin and {de los Reyes}, Mithi A. C. and Wang, Yunchong and Wu, John F.},
  year = 2024,
  month = nov,
  journal = {ApJ},
  volume = {976},
  pages = {118},
  publisher = {IOP},
  doi = {10.3847/1538-4357/ad61e7}
}

@article{geha_stellar_2012,
  title = {A {{Stellar Mass Threshold}} for {{Quenching}} of {{Field Galaxies}}},
  author = {Geha, M. and Blanton, M. R. and Yan, R. and Tinker, J. L.},
  year = 2012,
  month = sep,
  journal = {ApJ},
  volume = {757},
  pages = {85},
  doi = {10.1088/0004-637X/757/1/85}
}

@article{giovanelli_arecibo_2005-1,
  title = {The {{Arecibo Legacy Fast ALFA Survey}}. {{I}}. {{Science Goals}}, {{Survey Design}}, and {{Strategy}}},
  author = {Giovanelli, Riccardo and Haynes, Martha P. and Kent, Brian R. and Perillat, Philip and Saintonge, Amelie and Brosch, Noah and Catinella, Barbara and Hoffman, G. Lyle and Stierwalt, Sabrina and Spekkens, Kristine and Lerner, Mikael S. and Masters, Karen L. and Momjian, Emmanuel and Rosenberg, Jessica L. and Springob, Christopher M. and Boselli, Alessandro and Charmandaris, Vassilis and Darling, Jeremy K. and Davies, Jonathan and Lambas, Diego Garcia and Gavazzi, Giuseppe and Giovanardi, Carlo and Hardy, Eduardo and Hunt, Leslie K. and Iovino, Angela and Karachentsev, Igor D. and Karachentseva, Valentina E. and Koopmann, Rebecca A. and Marinoni, Christian and Minchin, Robert and Muller, Erik and Putman, Mary and Pantoja, Carmen and Salzer, John J. and Scodeggio, Marco and Skillman, Evan and Solanes, Jose M. and Valotto, Carlos and van Driel, Wim and van Zee, Liese},
  year = 2005,
  month = dec,
  journal = {AJ},
  volume = {130},
  number = {6},
  pages = {2598},
  publisher = {IOP Publishing},
  doi = {10.1086/497431},
  langid = {english}
}

@article{gnedin_cosmological_2000,
  title = {Cosmological {{Reionization}} by {{Stellar Sources}}},
  author = {Gnedin, Nickolay Y.},
  year = 2000,
  month = jun,
  journal = {ApJ},
  volume = {535},
  pages = {530--554},
  doi = {10.1086/308876}
}

@article{grcevich_h_2009,
  title = {H {{I}} in {{Local Group Dwarf Galaxies}} and {{Stripping}} by the {{Galactic Halo}}},
  author = {Grcevich, Jana and Putman, Mary E.},
  year = 2009,
  month = may,
  journal = {ApJ},
  volume = {696},
  pages = {385--395},
  doi = {10.1088/0004-637X/696/1/385}
}

@article{gunn_infall_1972,
  title = {On the {{Infall}} of {{Matter Into Clusters}} of {{Galaxies}} and {{Some Effects}} on {{Their Evolution}}},
  author = {Gunn, James E. and Gott, III, J. Richard},
  year = 1972,
  month = aug,
  journal = {ApJ},
  volume = {176},
  pages = {1},
  doi = {10.1086/151605}
}

@article{hafen_origins_2019,
  title = {The Origins of the Circumgalactic Medium in the {{FIRE}} Simulations},
  author = {Hafen, Zachary and {Faucher-Gigu{\`e}re}, Claude-Andr{\'e} and {Angl{\'e}s-Alc{\'a}zar}, Daniel and Stern, Jonathan and Kere{\v s}, Du{\v s}an and Hummels, Cameron and Esmerian, Clarke and {Garrison-Kimmel}, Shea and {El-Badry}, Kareem and Wetzel, Andrew and Chan, T K and Hopkins, Philip F and Murray, Norman},
  year = 2019,
  month = sep,
  journal = {MNRAS},
  volume = {488},
  number = {1},
  pages = {1248--1272},
  doi = {10.1093/mnras/stz1773}
}

@article{harris_array_2020,
  title = {Array Programming with {{NumPy}}},
  author = {Harris, Charles R. and Millman, K. Jarrod and {van der Walt}, St{\'e}fan J. and Gommers, Ralf and Virtanen, Pauli and Cournapeau, David and Wieser, Eric and Taylor, Julian and Berg, Sebastian and Smith, Nathaniel J. and Kern, Robert and Picus, Matti and Hoyer, Stephan and {van Kerkwijk}, Marten H. and Brett, Matthew and Haldane, Allan and {del R{\'i}o}, Jaime Fern{\'a}ndez and Wiebe, Mark and Peterson, Pearu and {G{\'e}rard-Marchant}, Pierre and Sheppard, Kevin and Reddy, Tyler and Weckesser, Warren and Abbasi, Hameer and Gohlke, Christoph and Oliphant, Travis E.},
  year = 2020,
  month = sep,
  journal = {Nature},
  volume = {585},
  number = {7825},
  pages = {357--362},
  publisher = {Nature Publishing Group},
  doi = {10.1038/s41586-020-2649-2},
  copyright = {2020 The Author(s)},
  langid = {english}
}

@article{haynes_arecibo_2011,
  title = {The {{Arecibo Legacy Fast ALFA Survey}}: {{The}} {$\alpha$}.40 {{H I Source Catalog}}, {{Its Characteristics}} and {{Their Impact}} on the {{Derivation}} of the {{H I Mass Function}}},
  shorttitle = {The {{Arecibo Legacy Fast ALFA Survey}}},
  author = {Haynes, Martha P. and Giovanelli, Riccardo and Martin, Ann M. and Hess, Kelley M. and Saintonge, Am{\'e}lie and Adams, Elizabeth A. K. and Hallenbeck, Gregory and Hoffman, G. Lyle and Huang, Shan and Kent, Brian R. and Koopmann, Rebecca A. and Papastergis, Emmanouil and Stierwalt, Sabrina and Balonek, Thomas J. and Craig, David W. and Higdon, Sarah J. U. and Kornreich, David A. and Miller, Jeffrey R. and O'Donoghue, Aileen A. and Olowin, Ronald P. and Rosenberg, Jessica L. and Spekkens, Kristine and Troischt, Parker and Wilcots, Eric M.},
  year = 2011,
  month = nov,
  journal = {AJ},
  volume = {142},
  pages = {170},
  publisher = {IOP},
  doi = {10.1088/0004-6256/142/5/170}
}

@article{haynes_arecibo_2018,
  title = {The {{Arecibo Legacy Fast ALFA Survey}}: {{The ALFALFA Extragalactic H}} i {{Source Catalog}}},
  shorttitle = {The {{Arecibo Legacy Fast ALFA Survey}}},
  author = {Haynes, Martha P. and Giovanelli, Riccardo and Kent, Brian R. and Adams, Elizabeth A. K. and Balonek, Thomas J. and Craig, David W. and Fertig, Derek and Finn, Rose and Giovanardi, Carlo and Hallenbeck, Gregory and Hess, Kelley M. and Hoffman, G. Lyle and Huang, Shan and Jones, Michael G. and Koopmann, Rebecca A. and Kornreich, David A. and Leisman, Lukas and Miller, Jeffrey and Moorman, Crystal and O'Connor, Jessica and O'Donoghue, Aileen and Papastergis, Emmanouil and Troischt, Parker and Stark, David and Xiao, Li},
  year = 2018,
  month = jul,
  journal = {ApJ},
  volume = {861},
  number = {1},
  pages = {49},
  publisher = {The American Astronomical Society},
  doi = {10.3847/1538-4357/aac956},
  langid = {english}
}

@article{haynes_influence_1984,
  title = {The {{Influence}} of {{Environment}} on the {{HI Content}} of {{Galaxies}}},
  author = {Haynes, Martha P. and Giovanelli, Riccardo and Chincarini, Guido L.},
  year = 1984,
  month = sep,
  journal = {ARA\&A},
  volume = {22},
  number = {Volume 22, 1984},
  pages = {445--470},
  publisher = {Annual Reviews},
  doi = {10.1146/annurev.aa.22.090184.002305},
  langid = {english}
}

@article{hunter_identifying_2025,
  title = {Identifying {{Dwarfs}} of {{MC Analog GalaxiEs}} ({{ID-MAGE}}): {{The Search}} for {{Satellites}} around {{Low-mass Hosts}}},
  shorttitle = {Identifying {{Dwarfs}} of {{MC Analog GalaxiEs}} ({{ID-MAGE}})},
  author = {Hunter, Laura Congreve and {Mutlu-Pakd{\.I}l}, Bur{\c c}{\.I}n and Sand, David J. and Bennet, Paul and Khim, Donghyeon J. and Crnojevi{\'c}, Denija and {Doliva-Dolinsky}, Amandine and Durodola, Emmanuel and Fielder, Catherine and {Goebel-Bain}, Rowan and Jones, Michael G. and Karunakaran, Ananthan and Spekkens, Kristine and Zaritsky, Dennis},
  year = 2025,
  month = aug,
  journal = {ApJ},
  volume = {989},
  pages = {58},
  doi = {10.3847/1538-4357/ade9a4}
}

@misc{hunter_id-mage_2026,
  title = {{{ID-MAGE II}}: {{The Star Forming Satellites}} of {{Low-Mass Hosts}}},
  shorttitle = {{{ID-MAGE II}}},
  author = {Hunter, Laura Congreve and {Mutlu-Pakdil}, Bur{\c c}in and Farnell, Michael B. and Sand, David J. and Bennet, Paul and Campana, Sasha N. and Carlin, Jeffrey L. and Crnojevi{\'c}, Denija and {Doliva-Dolinsky}, Amandine and Durodola, Emmanuel and Jones, Michael G. and Khim, Donghyeon J. and Marin, Laurella and Mendez, Ricardo J. and Prabhu, Deepthi S. and Spekkens, Kristine and Zaritsky, Dennis},
  year = 2026,
  month = mar,
  publisher = {arXiv},
  doi = {10.48550/arXiv.2603.18973}
}

@article{hunter_little_2012,
  title = {Little {{Things}}},
  author = {Hunter, Deidre A. and {Ficut-Vicas}, Dana and Ashley, Trisha and Brinks, Elias and Cigan, Phil and Elmegreen, Bruce G. and Heesen, Volker and Herrmann, Kimberly A. and Johnson, Megan and Oh, Se-Heon and Rupen, Michael P. and Schruba, Andreas and Simpson, Caroline E. and Walter, Fabian and Westpfahl, David J. and Young, Lisa M. and Zhang, Hong-Xin},
  year = 2012,
  month = nov,
  journal = {AJ},
  volume = {144},
  pages = {134},
  doi = {10.1088/0004-6256/144/5/134}
}

@article{hunter_relationships_2021,
  title = {Relationships between the {{Stellar}}, {{Gaseous}}, and {{Star Formation Disks}} in {{LITTLE THINGS Dwarf Irregular Galaxies}}: {{Indirect Evidence}} for {{Substantial Fractions}} of {{Dark Molecular Gas}}},
  shorttitle = {Relationships between the {{Stellar}}, {{Gaseous}}, and {{Star Formation Disks}} in {{LITTLE THINGS Dwarf Irregular Galaxies}}},
  author = {Hunter, Deidre A. and Elmegreen, Bruce G. and Goldberger, Esther and Taylor, Hannah and Ermakov, Anton I. and Herrmann, Kimberly A. and Oh, Se-Heon and Malko, Bradley and Barandi, Brian and Jundt, Ryan},
  year = 2021,
  month = jan,
  journal = {AJ},
  volume = {161},
  number = {2},
  pages = {71},
  publisher = {The American Astronomical Society},
  doi = {10.3847/1538-3881/abd089},
  langid = {english}
}

@article{iyer_diversity_2020,
  title = {The Diversity and Variability of Star Formation Histories in Models of Galaxy Evolution},
  author = {Iyer, Kartheik G. and Tacchella, Sandro and Genel, Shy and Hayward, Christopher C. and Hernquist, Lars and Brooks, Alyson M. and Caplar, Neven and Dav{\'e}, Romeel and Diemer, Benedikt and Forbes, John C. and Gawiser, Eric and Somerville, Rachel S. and Starkenburg, Tjitske K.},
  year = 2020,
  month = oct,
  journal = {MNRAS},
  volume = {498},
  pages = {430--463},
  doi = {10.1093/mnras/staa2150}
}

@article{jahn_effects_2022,
  title = {The Effects of {{LMC-mass}} Environments on Their Dwarf Satellite Galaxies in the {{FIRE}} Simulations},
  author = {Jahn, Ethan D and Sales, Laura V and Wetzel, Andrew and Samuel, Jenna and {El-Badry}, Kareem and {Boylan-Kolchin}, Michael and Bullock, James S},
  year = 2022,
  month = jun,
  journal = {MNRAS},
  volume = {513},
  number = {2},
  pages = {2673--2688},
  doi = {10.1093/mnras/stac811}
}

@article{jones_h_2020,
  title = {The {{H I}} Mass Function of Group Galaxies in the {{ALFALFA}} Survey},
  author = {Jones, Michael G. and Hess, Kelley M. and Adams, Elizabeth A. K. and {Verdes-Montenegro}, Lourdes},
  year = 2020,
  month = may,
  journal = {MNRAS},
  volume = {494},
  pages = {2090--2108},
  publisher = {OUP},
  doi = {10.1093/mnras/staa810}
}

@article{jones_gas_2024,
  title = {Gas and {{Star Formation}} in {{Satellites}} of {{Milky Way Analogs}}},
  author = {Jones, Michael G. and Sand, David J. and Karunakaran, Ananthan and Spekkens, Kristine and Oman, Kyle A. and Bennet, Paul and Besla, Gurtina and Crnojevi{\'c}, Denija and Cuillandre, Jean-Charles and Fielder, Catherine E. and Gwyn, Stephen and {Mutlu-Pakdil}, Bur{\c c}in},
  year = 2024,
  month = may,
  journal = {ApJ},
  volume = {966},
  pages = {93},
  publisher = {IOP},
  doi = {10.3847/1538-4357/ad3076}
}

@article{jones_pavo_2025,
  title = {Pavo: {{Stellar Feedback}} in {{Action}} in a {{Low-mass Dwarf Galaxy}}},
  shorttitle = {Pavo},
  author = {Jones, Michael G. and Rey, Martin P. and Sand, David J. and Spekkens, Kristine and {Mutlu-Pakdil}, Bur{\c c}in and Adams, Elizabeth A. K. and Bennet, Paul and Crnojevi{\'c}, Denija and {Doliva-Dolinsky}, Amandine and Donnerstein, Richard and Fielder, Catherine E. and Healy, Julia and Hunter, Laura C. and Karunakaran, Ananthan and Prabhu, Deepthi S. and Zaritsky, Dennis},
  year = 2025,
  month = sep,
  journal = {ApJ},
  volume = {990},
  number = {2},
  pages = {164},
  publisher = {The American Astronomical Society},
  doi = {10.3847/1538-4357/adf6ab},
  langid = {english}
}

@article{kado-fong_tracing_2020,
  title = {Tracing the {{Intrinsic Shapes}} of {{Dwarf Galaxies Out}} to {{Four Effective Radii}}: {{Clues}} to {{Low-mass Stellar Halo Formation}}},
  shorttitle = {Tracing the {{Intrinsic Shapes}} of {{Dwarf Galaxies Out}} to {{Four Effective Radii}}},
  author = {{Kado-Fong}, Erin and Greene, Jenny E. and Huang, Song and Beaton, Rachael and Goulding, Andy D. and Komiyama, Yutaka},
  year = 2020,
  month = sep,
  journal = {ApJ},
  volume = {900},
  pages = {163},
  doi = {10.3847/1538-4357/abacc2}
}

@article{karachentsev_morphological_2018,
  title = {Morphological Properties of Galaxies in Different {{Local Volume}} Environments},
  author = {Karachentsev, I. D. and Kaisina, E. I. and Makarov, D. I.},
  year = 2018,
  month = sep,
  journal = {MNRAS},
  volume = {479},
  pages = {4136--4152},
  doi = {10.1093/mnras/sty1774}
}

@article{karachentsev_updated_2013,
  title = {{{UPDATED NEARBY GALAXY CATALOG}}},
  author = {Karachentsev, Igor D. and Makarov, Dmitry I. and Kaisina, Elena I.},
  year = 2013,
  month = mar,
  journal = {AJ},
  volume = {145},
  number = {4},
  pages = {101},
  publisher = {The American Astronomical Society},
  doi = {10.1088/0004-6256/145/4/101},
  langid = {english}
}

@article{karunakaran_h_2022,
  title = {H\,i Properties of Satellite Galaxies around Local Volume Hosts},
  author = {Karunakaran, Ananthan and Spekkens, Kristine and Carroll, Rhys and Sand, David J and Bennet, Paul and Crnojevi{\'c}, Denija and Jones, Michael G and {Mutlu-Pakd{\i}l}, Bur{\c c}{\i}n},
  year = 2022,
  month = oct,
  journal = {MNRAS},
  volume = {516},
  number = {2},
  pages = {1741--1751},
  doi = {10.1093/mnras/stac2329}
}

@article{karunakaran_quenched_2023,
  title = {The Quenched Satellite Population around {{Milky Way}} Analogues},
  author = {Karunakaran, Ananthan and Sand, David J and Jones, Michael G and Spekkens, Kristine and Bennet, Paul and Crnojevi{\'c}, Denija and {Mutlu-Pakd{\.\i}l}, Bur{\c c}{\.\i}n and Zaritsky, Dennis},
  year = 2023,
  month = oct,
  journal = {MNRAS},
  volume = {524},
  number = {4},
  pages = {5314--5326},
  doi = {10.1093/mnras/stad2208}
}

@article{kleiner_meerkat_2021,
  title = {A {{MeerKAT}} View of Pre-Processing in the {{Fornax A}} Group},
  author = {Kleiner, D. and Serra, P. and Maccagni, F. M. and Venhola, A. and {Morokuma-Matsui}, K. and Peletier, R. and Iodice, E. and Raj, M. A. and de Blok, W. J. G. and Comrie, A. and J{\'o}zsa, G. I. G. and Kamphuis, P. and Loni, A. and Loubser, S. I. and Moln{\'a}r, D. Cs and Passmoor, S. S. and Ramatsoku, M. and Sivitilli, A. and Smirnov, O. and Thorat, K. and Vitello, F.},
  year = 2021,
  month = apr,
  journal = {A\&A},
  volume = {648},
  pages = {A32},
  publisher = {EDP Sciences},
  doi = {10.1051/0004-6361/202039898},
  copyright = {\copyright{} ESO 2021},
  langid = {english}
}

@article{koribalski_neutral_2004,
  title = {Neutral Hydrogen Gas in Interacting Galaxies: The {{NGC}} 6221/6215 Galaxy Group},
  shorttitle = {Neutral Hydrogen Gas in Interacting Galaxies},
  author = {Koribalski, B{\"a}rbel and Dickey, John M.},
  year = 2004,
  month = mar,
  journal = {MNRAS},
  volume = {348},
  pages = {1255--1274},
  publisher = {OUP},
  doi = {10.1111/j.1365-2966.2004.07444.x}
}

@article{koribalski_wallaby_2020,
  title = {{{WALLABY}} -- an {{SKA Pathfinder H}}\,i Survey},
  author = {Koribalski, B{\"a}rbel S. and {Staveley-Smith}, L. and Westmeier, T. and Serra, P. and Spekkens, K. and Wong, O. I. and {Lee-Waddell}, K. and Lagos, C. D. P. and Obreschkow, D. and {Ryan-Weber}, E. V. and Zwaan, M. and Kilborn, V. and Bekiaris, G. and Bekki, K. and Bigiel, F. and Boselli, A. and Bosma, A. and Catinella, B. and Chauhan, G. and Cluver, M. E. and Colless, M. and Courtois, H. M. and Crain, R. A. and {de Blok}, W. J. G. and D{\'e}nes, H. and Duffy, A. R. and Elagali, A. and Fluke, C. J. and For, B.-Q. and Heald, G. and Henning, P. A. and Hess, K. M. and Holwerda, B. W. and Howlett, C. and Jarrett, T. and Jones, D. H. and Jones, M. G. and J{\'o}zsa, G. I. G. and Jurek, R. and J{\"u}tte, E. and Kamphuis, P. and Karachentsev, I. and Kerp, J. and Kleiner, D. and {Kraan-Korteweg}, R. C. and {L{\'o}pez-S{\'a}nchez}, {\'A}. R. and Madrid, J. and Meyer, M. and Mould, J. and Murugeshan, C. and Norris, R. P. and Oh, S.-H. and Oosterloo, T. A. and Popping, A. and Putman, M. and Reynolds, T. N. and Rhee, J. and Robotham, A. S. G. and Ryder, S. and Schr{\"o}der, A. C. and Shao, Li and Stevens, A. R. H. and Taylor, E. N. and {van~der Hulst}, J. M. and {Verdes-Montenegro}, L. and Wakker, B. P. and Wang, J. and Whiting, M. and Winkel, B. and Wolf, C.},
  year = 2020,
  month = jul,
  journal = {Astrophys Space Sci},
  volume = {365},
  number = {7},
  pages = {118},
  doi = {10.1007/s10509-020-03831-4},
  langid = {english}
}

@article{kourkchi_cosmicflows-4_2020,
  title = {Cosmicflows-4: {{The Catalog}} of {$\sim$}10,000 {{Tully}}--{{Fisher Distances}}},
  shorttitle = {Cosmicflows-4},
  author = {Kourkchi, Ehsan and Tully, R. Brent and Eftekharzadeh, Sarah and Llop, Jordan and Courtois, H{\'e}l{\`e}ne M. and Guinet, Daniel and Dupuy, Alexandra and Neill, James D. and Seibert, Mark and Andrews, Michael and Chuang, Juana and Danesh, Arash and Gonzalez, Randy and Holthaus, Alexandria and Mokelke, Amber and Schoen, Devin and Urasaki, Chase},
  year = 2020,
  month = oct,
  journal = {ApJ},
  volume = {902},
  number = {2},
  pages = {145},
  publisher = {The American Astronomical Society},
  doi = {10.3847/1538-4357/abb66b},
  langid = {english}
}

@article{lelli_gas_2022,
  title = {Gas Dynamics in Dwarf Galaxies as Testbeds for Dark Matter and Galaxy Evolution},
  author = {Lelli, Federico},
  year = 2022,
  month = jan,
  journal = {Nat Astron},
  volume = {6},
  number = {1},
  pages = {35--47},
  publisher = {Nature Publishing Group},
  doi = {10.1038/s41550-021-01562-2},
  copyright = {2022 Springer Nature Limited},
  langid = {english}
}

@article{lelli_small_2016,
  title = {{{THE SMALL SCATTER OF THE BARYONIC TULLY}}--{{FISHER RELATION}}},
  author = {Lelli, Federico and McGaugh, Stacy S. and Schombert, James M.},
  year = 2016,
  month = jan,
  journal = {ApJL},
  volume = {816},
  number = {1},
  pages = {L14},
  publisher = {The American Astronomical Society},
  doi = {10.3847/2041-8205/816/1/L14},
  langid = {english}
}

@article{lelli_sparc_2016,
  title = {{{SPARC}}: {{MASS MODELS FOR}} 175 {{DISK GALAXIES WITH SPITZER PHOTOMETRY AND ACCURATE ROTATION CURVES}}},
  shorttitle = {{{SPARC}}},
  author = {Lelli, Federico and McGaugh, Stacy S. and Schombert, James M.},
  year = 2016,
  month = nov,
  journal = {AJ},
  volume = {152},
  number = {6},
  pages = {157},
  publisher = {The American Astronomical Society},
  doi = {10.3847/0004-6256/152/6/157},
  langid = {english}
}

@article{leroy_z_2019,
  title = {A z = 0 {{Multiwavelength Galaxy Synthesis}}. {{I}}. {{A WISE}} and {{GALEX Atlas}} of {{Local Galaxies}}},
  author = {Leroy, Adam K. and Sandstrom, Karin M. and Lang, Dustin and Lewis, Alexia and Salim, Samir and Behrens, Erica A. and Chastenet, J{\'e}r{\'e}my and Chiang, I.-Da and Gallagher, Molly J. and Kessler, Sarah and Utomo, Dyas},
  year = 2019,
  month = sep,
  journal = {ApJS},
  volume = {244},
  number = {2},
  pages = {24},
  publisher = {The American Astronomical Society},
  doi = {10.3847/1538-4365/ab3925},
  langid = {english}
}

@misc{li_elves-dwarf_2025,
  title = {{{ELVES-Dwarf I}}: {{Satellites Systems}} of {{Eight Isolated Dwarf Galaxies}} in the {{Local Volume}}},
  shorttitle = {{{ELVES-Dwarf I}}},
  author = {Li, Jiaxuan and Greene, Jenny E. and Danieli, Shany and Carlsten, Scott G. and Geha, Marla and Jiang, Fangzhou and Tanaka, Masayuki},
  year = 2025,
  month = apr,
  publisher = {arXiv},
  doi = {10.48550/arXiv.2504.08030}
}

@article{maccagni_mhongoose_2024,
  title = {{{MHONGOOSE}} Discovery of a Gas-Rich Low Surface Brightness Galaxy in the {{Dorado}} Group},
  author = {Maccagni, F. M. and de Blok, W. J. G. and Pi{\~n}a, P. E. Mancera and Ragusa, R. and Iodice, E. and Spavone, M. and McGaugh, S. and Oman, K. A. and Oosterloo, T. A. and Koribalski, B. S. and Kim, M. and Adams, E. a. K. and Amram, P. and Bosma, A. and Bigiel, F. and Brinks, E. and Chemin, L. and Combes, F. and Gibson, B. and Healy, J. and Holwerda, B. W. and J{\'o}zsa, G. I. G. and Kamphuis, P. and Kleiner, D. and Kurapati, S. and Marasco, A. and Spekkens, K. and Veronese, S. and Walter, F. and Zabel, N. and Zijlstra, A.},
  year = 2024,
  month = oct,
  journal = {A\&A},
  volume = {690},
  pages = {A69},
  publisher = {EDP Sciences},
  doi = {10.1051/0004-6361/202449441},
  copyright = {\copyright{} The Authors 2024},
  langid = {english}
}

@article{maddox_variation_2015,
  title = {Variation of Galactic Cold Gas Reservoirs with Stellar Mass},
  author = {Maddox, Natasha and Hess, Kelley M. and Obreschkow, Danail and Jarvis, M. J. and Blyth, S. -L.},
  year = 2015,
  month = feb,
  journal = {MNRAS},
  volume = {447},
  pages = {1610--1617},
  doi = {10.1093/mnras/stu2532}
}

@article{mao_saga_2024,
  title = {The {{SAGA Survey}}. {{III}}. {{A Census}} of 101 {{Satellite Systems}} around {{Milky Way}}--Mass {{Galaxies}}},
  author = {Mao, Yao-Yuan and Geha, Marla and Wechsler, Risa H. and Asali, Yasmeen and Wang, Yunchong and {Kado-Fong}, Erin and Kallivayalil, Nitya and Nadler, Ethan O. and Tollerud, Erik J. and Weiner, Benjamin and {de los Reyes}, Mithi A. C. and Wu, John F.},
  year = 2024,
  month = nov,
  journal = {ApJ},
  volume = {976},
  number = {1},
  pages = {117},
  publisher = {The American Astronomical Society},
  doi = {10.3847/1538-4357/ad64c4},
  langid = {english}
}

@article{mateo_velocity_2008,
  title = {The {{Velocity Dispersion Profile}} of the {{Remote Dwarf Spheroidal Galaxy Leo I}}: {{A Tidal Hit}} and {{Run}}?},
  shorttitle = {The {{Velocity Dispersion Profile}} of the {{Remote Dwarf Spheroidal Galaxy Leo I}}},
  author = {Mateo, Mario and Olszewski, Edward W. and Walker, Matthew G.},
  year = 2008,
  month = mar,
  journal = {ApJ},
  volume = {675},
  number = {1},
  pages = {201},
  publisher = {IOP Publishing},
  doi = {10.1086/522326},
  langid = {english}
}

@article{mayer_environmental_2010,
  title = {Environmental {{Mechanisms Shaping}} the {{Nature}} of {{Dwarf Spheroidal Galaxies}}: {{The View}} of {{Computer Simulations}}},
  shorttitle = {Environmental {{Mechanisms Shaping}} the {{Nature}} of {{Dwarf Spheroidal Galaxies}}},
  author = {Mayer, Lucio},
  year = 2010,
  month = jan,
  journal = {Advances in Astronomy},
  volume = {2010},
  pages = {278434},
  doi = {10.1155/2010/278434}
}

@article{mayer_simultaneous_2006,
  title = {Simultaneous Ram Pressure and Tidal Stripping; How Dwarf Spheroidals Lost Their Gas},
  author = {Mayer, Lucio and Mastropietro, Chiara and Wadsley, James and Stadel, Joachim and Moore, Ben},
  year = 2006,
  month = jul,
  journal = {MNRAS},
  volume = {369},
  pages = {1021--1038},
  doi = {10.1111/j.1365-2966.2006.10403.x}
}

@article{mayer_tidal_2001,
  title = {Tidal {{Stirring}} and the {{Origin}} of {{Dwarf Spheroidals}} in the {{Local Group}}},
  author = {Mayer, Lucio and Governato, Fabio and Colpi, Monica and Moore, Ben and Quinn, Thomas and Wadsley, James and Stadel, Joachim and Lake, George},
  year = 2001,
  month = feb,
  journal = {ApJ},
  volume = {547},
  pages = {L123-L127},
  doi = {10.1086/318898}
}

@article{mcconnachie_observed_2012,
  title = {The {{Observed Properties}} of {{Dwarf Galaxies}} in and around the {{Local Group}}},
  author = {McConnachie, Alan W.},
  year = 2012,
  month = jul,
  journal = {AJ},
  volume = {144},
  pages = {4},
  publisher = {IOP},
  doi = {10.1088/0004-6256/144/1/4}
}

@article{mcgaugh_baryonic_2000,
  title = {The {{Baryonic Tully-Fisher Relation}}},
  author = {McGaugh, S. S. and Schombert, J. M. and Bothun, G. D. and de Blok, W. J. G.},
  year = 2000,
  month = mar,
  journal = {ApJ},
  volume = {533},
  number = {2},
  pages = {L99},
  publisher = {IOP Publishing},
  doi = {10.1086/312628},
  langid = {english}
}

@article{mcgaugh_dark_2021,
  title = {Dark {{Matter Halo Masses}} from {{Abundance Matching}} and {{Kinematics}}: {{Tensions}} for the {{Milky Way}} and {{M31}}},
  shorttitle = {Dark {{Matter Halo Masses}} from {{Abundance Matching}} and {{Kinematics}}},
  author = {McGaugh, Stacy S. and {van Dokkum}, Pieter},
  year = 2021,
  month = feb,
  journal = {Research Notes of the American Astronomical Society},
  volume = {5},
  pages = {23},
  doi = {10.3847/2515-5172/abe1ba}
}

@article{mcgaugh_star-forming_2017,
  title = {The {{Star-forming Main Sequence}} of {{Dwarf Low Surface Brightness Galaxies}}},
  author = {McGaugh, Stacy S. and Schombert, James M. and Lelli, Federico},
  year = 2017,
  month = dec,
  journal = {ApJ},
  volume = {851},
  number = {1},
  pages = {22},
  publisher = {The American Astronomical Society},
  doi = {10.3847/1538-4357/aa9790},
  langid = {english}
}

@article{mcquinn_accurate_2017,
  title = {Accurate {{Distances}} to {{Important Spiral Galaxies}}: {{M63}}, {{M74}}, {{NGC}} 1291, {{NGC}} 4559, {{NGC}} 4625, and {{NGC}} 5398},
  shorttitle = {Accurate {{Distances}} to {{Important Spiral Galaxies}}},
  author = {McQuinn, {\relax Kristen}. B. W. and Skillman, Evan D. and Dolphin, Andrew E. and Berg, Danielle and Kennicutt, Robert},
  year = 2017,
  month = aug,
  journal = {AJ},
  volume = {154},
  pages = {51},
  publisher = {IOP},
  doi = {10.3847/1538-3881/aa7aad}
}

@article{mcquinn_nature_2010,
  title = {{{THE NATURE OF STARBURSTS}}. {{II}}. {{THE DURATION OF STARBURSTS IN DWARF GALAXIES}}*},
  author = {McQuinn, Kristen B. W. and Skillman, Evan D. and Cannon, John M. and Dalcanton, Julianne and Dolphin, Andrew and {Hidalgo-Rodr{\'i}guez}, Sebastian and Holtzman, Jon and Stark, David and Weisz, Daniel and Williams, Benjamin},
  year = 2010,
  month = oct,
  journal = {ApJ},
  volume = {724},
  number = {1},
  pages = {49},
  publisher = {The American Astronomical Society},
  doi = {10.1088/0004-637X/724/1/49},
  langid = {english}
}

@article{mcquinn_nature_2010-1,
  title = {{{THE NATURE OF STARBURSTS}}. {{I}}. {{THE STAR FORMATION HISTORIES OF EIGHTEEN NEARBY STARBURST DWARF GALAXIES}}*},
  author = {McQuinn, Kristen B. W. and Skillman, Evan D. and Cannon, John M. and Dalcanton, Julianne and Dolphin, Andrew and {Hidalgo-Rodr{\'i}guez}, Sebastian and Holtzman, Jon and Stark, David and Weisz, Daniel and Williams, Benjamin},
  year = 2010,
  month = aug,
  journal = {ApJ},
  volume = {721},
  number = {1},
  pages = {297},
  publisher = {The American Astronomical Society},
  doi = {10.1088/0004-637X/721/1/297},
  langid = {english}
}

@article{meurer_survey_2006,
  title = {The {{Survey}} for {{Ionization}} in {{Neutral Gas Galaxies}}. {{I}}. {{Description}} and {{Initial Results}}},
  author = {Meurer, Gerhardt R. and Hanish, D. J. and Ferguson, H. C. and Knezek, P. M. and Kilborn, V. A. and Putman, M. E. and Smith, R. C. and Koribalski, B. and Meyer, M. and Oey, M. S. and {Ryan-Weber}, E. V. and Zwaan, M. A. and Heckman, T. M. and R. C. Kennicutt, Jr and Lee, J. C. and Webster, R. L. and {Bland-Hawthorn}, J. and Dopita, M. A. and Freeman, K. C. and Doyle, M. T. and Drinkwater, M. J. and {Staveley-Smith}, L. and Werk, J.},
  year = 2006,
  month = jul,
  journal = {ApJS},
  volume = {165},
  number = {1},
  pages = {307},
  publisher = {IOP Publishing},
  doi = {10.1086/504685},
  langid = {english}
}

@article{meyer_hipass_2004,
  title = {The {{HIPASS}} Catalogue - {{I}}. {{Data}} Presentation},
  author = {Meyer, M. J. and Zwaan, M. A. and Webster, R. L. and {Staveley-Smith}, L. and {Ryan-Weber}, E. and Drinkwater, M. J. and Barnes, D. G. and Howlett, M. and Kilborn, V. A. and Stevens, J. and Waugh, M. and Pierce, M. J. and Bhathal, R. and {de Blok}, W. J. G. and Disney, M. J. and Ekers, R. D. and Freeman, K. C. and Garcia, D. A. and Gibson, B. K. and Harnett, J. and Henning, P. A. and Jerjen, H. and Kesteven, M. J. and Knezek, P. M. and Koribalski, B. S. and Mader, S. and Marquarding, M. and Minchin, R. F. and O'Brien, J. and Oosterloo, T. and Price, R. M. and Putman, M. E. and Ryder, S. D. and Sadler, E. M. and Stewart, I. M. and Stootman, F. and Wright, A. E.},
  year = 2004,
  month = jun,
  journal = {MNRAS},
  volume = {350},
  pages = {1195--1209},
  publisher = {OUP},
  doi = {10.1111/j.1365-2966.2004.07710.x}
}

@article{meyer_tracing_2017,
  title = {Tracing {{HI Beyond}} the {{Local Universe}}},
  author = {Meyer, Martin and Robotham, Aaron and Obreschkow, Danail and Westmeier, Tobias and Duffy, Alan and {Staveley-Smith}, Lister},
  year = 2017,
  journal = {Publ. Astron. Soc. Aust.},
  volume = {34},
  eprint = {1705.04210},
  primaryclass = {astro-ph},
  pages = {e052},
  doi = {10.1017/pasa.2017.31},
  archiveprefix = {arXiv}
}

@misc{mintz_taking_2025,
  title = {Taking a {{Break}} at {{Cosmic Noon}}: {{Continuum-selected Low-mass Galaxies Require Long Burst Cycles}}},
  shorttitle = {Taking a {{Break}} at {{Cosmic Noon}}},
  author = {Mintz, Abby and Setton, David J. and Greene, Jenny E. and Leja, Joel and Wang, Bingjie and Burnham, Emilie and Suess, Katherine A. and Atek, Hakim and Bezanson, Rachel and Brammer, Gabriel and Cutler, Sam E. and Dayal, Pratika and Feldmann, Robert and Furtak, Lukas J. and Glazebrook, Karl and Khullar, Gourav and Kokorev, Vasily and Labb{\'e}, Ivo and Maseda, Michael V. and Miller, Tim B. and Mitsuhashi, Ikki and Nanayakkara, Themiya and Pan, Richard and Price, Sedona H. and Weaver, John R. and Whitaker, Katherine E.},
  year = 2025,
  month = jun,
  number = {arXiv:2506.16510},
  eprint = {2506.16510},
  primaryclass = {astro-ph},
  publisher = {arXiv},
  doi = {10.48550/arXiv.2506.16510},
  archiveprefix = {arXiv}
}

@article{muller_dwarf_2019,
  title = {The Dwarf Galaxy Satellite System of {{Centaurus A}}},
  author = {M{\"u}ller, Oliver and Rejkuba, Marina and Pawlowski, Marcel S. and Ibata, Rodrigo and Lelli, Federico and Hilker, Michael and Jerjen, Helmut},
  year = 2019,
  month = sep,
  journal = {A\&A},
  volume = {629},
  pages = {A18},
  publisher = {EDP Sciences},
  doi = {10.1051/0004-6361/201935807},
  copyright = {\copyright{} O. M\"uller et al. 2019},
  langid = {english}
}

@article{muller_too-many-dwarf-galaxy-satellites_2024,
  title = {A Too-Many-Dwarf-Galaxy-Satellites Problem in the {{M}} 83 Group},
  author = {M{\"u}ller, Oliver and Pawlowski, Marcel S. and Revaz, Yves and Venhola, Aku and Rejkuba, Marina and Hilker, Michael and Lutz, Katharina},
  year = 2024,
  month = apr,
  journal = {A\&A},
  volume = {684},
  pages = {L6},
  publisher = {EDP Sciences},
  doi = {10.1051/0004-6361/202348969},
  copyright = {\copyright{} The Authors 2024},
  langid = {english}
}

@article{munshi_quantifying_2021,
  title = {Quantifying {{Scatter}} in {{Galaxy Formation}} at the {{Lowest Masses}}},
  author = {Munshi, Ferah and Brooks, Alyson M. and Applebaum, Elaad and Christensen, Charlotte R. and Quinn, T. and Sligh, Serena},
  year = 2021,
  month = dec,
  journal = {ApJ},
  volume = {923},
  pages = {35},
  publisher = {IOP},
  doi = {10.3847/1538-4357/ac0db6}
}

@article{murugeshan_wallaby_2024,
  title = {{{WALLABY Pilot Survey}}: {{Public}} Data Release of {$\sim$} 1800 {{H I}} Sources and High-Resolution Cut-Outs from {{Pilot Survey Phase}} 2},
  shorttitle = {{{WALLABY Pilot Survey}}},
  author = {Murugeshan, C. and Deg, N. and Westmeier, T. and Shen, A. X. and For, B. -Q. and Spekkens, K. and Wong, O. I. and {Staveley-Smith}, L. and Catinella, B. and {Lee-Waddell}, K. and D{\'e}nes, H. and Rhee, J. and Cortese, L. and Goliath, S. and Halloran, R. and {van der Hulst}, J. M. and Kamphuis, P. and Koribalski, B. S. and {Kraan-Korteweg}, R. C. and Lelli, F. and Venkataraman, P. and {Verdes-Montenegro}, L. and Yu, N.},
  year = 2024,
  month = nov,
  journal = {PASA},
  volume = {41},
  pages = {e088},
  doi = {10.1017/pasa.2024.91}
}

@article{mutlu-pakdil_faint_2024,
  title = {The {{Faint Satellite System}} of {{NGC}} 253: {{Insights}} into {{Low-density Environments}} and {{No Satellite Plane}}*},
  shorttitle = {The {{Faint Satellite System}} of {{NGC}} 253},
  author = {{Mutlu-Pakdil}, Bur{\c c}in and Sand, David J. and Crnojevi{\'c}, Denija and Bennet, Paul and Jones, Michael G. and Spekkens, Kristine and Karunakaran, Ananthan and Zaritsky, Dennis and Caldwell, Nelson and Fielder, Catherine E. and Guhathakurta, Puragra and Seth, Anil C. and Simon, Joshua D. and Strader, Jay and Toloba, Elisa},
  year = 2024,
  month = may,
  journal = {ApJ},
  volume = {966},
  number = {2},
  pages = {188},
  publisher = {The American Astronomical Society},
  doi = {10.3847/1538-4357/ad36c4},
  langid = {english}
}

@article{mutlu-pakdil_hubble_2022,
  title = {Hubble {{Space Telescope Observations}} of {{NGC}} 253 {{Dwarf Satellites}}: {{Three Ultra-faint Dwarf Galaxies}}*},
  shorttitle = {Hubble {{Space Telescope Observations}} of {{NGC}} 253 {{Dwarf Satellites}}},
  author = {{Mutlu-Pakdil}, Bur{\c c}in and Sand, David J. and Crnojevi{\'c}, Denija and Jones, Michael G. and Caldwell, Nelson and Guhathakurta, Puragra and Seth, Anil C. and Simon, Joshua D. and Spekkens, Kristine and Strader, Jay and Toloba, Elisa},
  year = 2022,
  month = feb,
  journal = {ApJ},
  volume = {926},
  number = {1},
  pages = {77},
  publisher = {The American Astronomical Society},
  doi = {10.3847/1538-4357/ac4418},
  langid = {english}
}

@article{nadler_milky_2020,
  title = {Milky {{Way Satellite Census}}. {{II}}. {{Galaxy}}--{{Halo Connection Constraints Including}} the {{Impact}} of the {{Large Magellanic Cloud}}},
  author = {Nadler, E. O. and Wechsler, R. H. and Bechtol, K. and Mao, Y.-Y. and Green, G. and {Drlica-Wagner}, A. and McNanna, M. and Mau, S. and Pace, A. B. and Simon, J. D. and Kravtsov, A. and Dodelson, S. and Li, T. S. and Riley, A. H. and Wang, M. Y. and Abbott, T. M. C. and Aguena, M. and Allam, S. and Annis, J. and Avila, S. and Bernstein, G. M. and Bertin, E. and Brooks, D. and Burke, D. L. and Rosell, A. Carnero and Kind, M. Carrasco and Carretero, J. and Costanzi, M. and da Costa, L. N. and Vicente, J. De and Desai, S. and Evrard, A. E. and Flaugher, B. and Fosalba, P. and Frieman, J. and {Garc{\'i}a-Bellido}, J. and Gaztanaga, E. and Gerdes, D. W. and Gruen, D. and Gschwend, J. and Gutierrez, G. and Hartley, W. G. and Hinton, S. R. and Honscheid, K. and Krause, E. and Kuehn, K. and Kuropatkin, N. and Lahav, O. and Maia, M. A. G. and Marshall, J. L. and Menanteau, F. and Miquel, R. and Palmese, A. and {Paz-Chinch{\'o}n}, F. and Plazas, A. A. and Romer, A. K. and Sanchez, E. and Santiago, B. and Scarpine, V. and Serrano, S. and Smith, M. and {Soares-Santos}, M. and Suchyta, E. and Tarle, G. and Thomas, D. and Varga, T. N. and Walker, A. R. and Collaboration), (DES},
  year = 2020,
  month = apr,
  journal = {ApJ},
  volume = {893},
  number = {1},
  pages = {48},
  publisher = {The American Astronomical Society},
  doi = {10.3847/1538-4357/ab846a},
  langid = {english}
}

@article{nadler_symphony_2023,
  title = {Symphony: {{Cosmological Zoom-in Simulation Suites}} over {{Four Decades}} of {{Host Halo Mass}}},
  shorttitle = {Symphony},
  author = {Nadler, Ethan O. and Mansfield, Philip and Wang, Yunchong and Du, Xiaolong and Adhikari, Susmita and Banerjee, Arka and Benson, Andrew and {Darragh-Ford}, Elise and Mao, Yao-Yuan and {Wagner-Carena}, Sebastian and Wechsler, Risa H. and Wu, Hao-Yi},
  year = 2023,
  month = mar,
  journal = {ApJ},
  volume = {945},
  pages = {159},
  publisher = {IOP},
  doi = {10.3847/1538-4357/acb68c}
}

@article{navarro_assembly_1995,
  title = {The Assembly of Galaxies in a Hierarchically Clustering Universe},
  author = {Navarro, Julio F. and Frenk, Carlos S. and White, Simon D. M.},
  year = 1995,
  month = jul,
  journal = {MNRAS},
  volume = {275},
  pages = {56--66},
  doi = {10.1093/mnras/275.1.56}
}

@article{oh_high-resolution_2015,
  title = {High-Resolution {{Mass Models}} of {{Dwarf Galaxies}} from {{LITTLE THINGS}}},
  author = {Oh, Se-Heon and Hunter, Deidre A. and Brinks, Elias and Elmegreen, Bruce G. and Schruba, Andreas and Walter, Fabian and Rupen, Michael P. and Young, Lisa M. and Simpson, Caroline E. and Johnson, Megan C. and Herrmann, Kimberly A. and {Ficut-Vicas}, Dana and Cigan, Phil and Heesen, Volker and Ashley, Trisha and Zhang, Hong-Xin},
  year = 2015,
  month = jun,
  journal = {AJ},
  volume = {149},
  pages = {180},
  doi = {10.1088/0004-6256/149/6/180}
}

@article{oman_satellite_2016,
  title = {Satellite Quenching Time-Scales in Clusters from Projected Phase Space Measurements Matched to Simulated Orbits},
  author = {Oman, Kyle A. and Hudson, Michael J.},
  year = 2016,
  month = dec,
  journal = {MNRAS},
  volume = {463},
  number = {3},
  pages = {3083--3095},
  doi = {10.1093/mnras/stw2195}
}

@article{pace_local_2025,
  title = {The {{Local Volume Database}}: A Library of the Observed Properties of Nearby Dwarf Galaxies and Star Clusters},
  shorttitle = {The {{Local Volume Database}}},
  author = {Pace, Andrew B},
  year = 2025,
  month = sep,
  journal = {The Open Journal of Astrophysics},
  volume = {8},
  pages = {142},
  doi = {10.33232/001c.144859}
}

@article{pace_proper_2022,
  title = {Proper {{Motions}}, {{Orbits}}, and {{Tidal Influences}} of {{Milky Way Dwarf Spheroidal Galaxies}}},
  author = {Pace, Andrew B. and Erkal, Denis and Li, Ting S.},
  year = 2022,
  month = nov,
  journal = {ApJ},
  volume = {940},
  number = {2},
  pages = {136},
  publisher = {The American Astronomical Society},
  doi = {10.3847/1538-4357/ac997b},
  langid = {english}
}

@article{paturel_hyperleda_2003,
  title = {{{HYPERLEDA}}.  {{I}}. {{Identification}} and Designation of Galaxies},
  author = {Paturel, G. and Petit, C. and Prugniel, {\relax Ph}. and Theureau, G. and Rousseau, J. and Brouty, M. and Dubois, P. and Cambr{\'e}sy, L.},
  year = 2003,
  month = dec,
  journal = {A\&A},
  volume = {412},
  pages = {45--55},
  publisher = {EDP},
  doi = {10.1051/0004-6361:20031411}
}

@article{penarrubia_tidal_2008,
  title = {The {{Tidal Evolution}} of {{Local Group Dwarf Spheroidals}}},
  author = {Pe{\~n}arrubia, Jorge and Navarro, Julio F. and McConnachie, Alan W.},
  year = 2008,
  month = jan,
  journal = {ApJ},
  volume = {673},
  pages = {226--240},
  publisher = {IOP},
  doi = {10.1086/523686}
}

@misc{piacitelli_marvelous_2025,
  title = {Marvelous {{Metals}}: {{Surveying}} the {{Circumgalactic Medium}} of {{Simulated Dwarf Galaxies}}},
  shorttitle = {Marvelous {{Metals}}},
  author = {Piacitelli, Daniel R. and Brooks, Alyson M. and Christensen, Charlotte and Sanchez, N. Nicole and Faerman, Yakov and Shen, Sijing and Cruz, Akaxia and Keller, Ben and Quinn, Thomas R. and Wadsley, James},
  year = 2025,
  month = may,
  publisher = {arXiv},
  doi = {10.48550/arXiv.2505.08861}
}

@article{pisano_h_2011,
  title = {{{AN H}} i {{SURVEY OF SIX LOCAL GROUP ANALOGS}}. {{II}}. {{H}} i {{PROPERTIES OF GROUP GALAXIES}}},
  author = {Pisano, D. J. and Barnes, David G. and {Staveley-Smith}, Lister and Gibson, Brad K. and Kilborn, Virginia A. and Freeman, Ken C.},
  year = 2011,
  month = nov,
  journal = {ApJS},
  volume = {197},
  number = {2},
  pages = {28},
  publisher = {The American Astronomical Society},
  doi = {10.1088/0067-0049/197/2/28},
  langid = {english}
}

@article{putman_gas_2021,
  title = {The {{Gas Content}} and {{Stripping}} of {{Local Group Dwarf Galaxies}}},
  author = {Putman, Mary E. and Zheng, Yong and {Price-Whelan}, Adrian M. and Grcevich, Jana and Johnson, Amalya C. and Tollerud, Erik and Peek, Joshua E. G.},
  year = 2021,
  month = may,
  journal = {ApJ},
  volume = {913},
  pages = {53},
  publisher = {IOP},
  doi = {10.3847/1538-4357/abe391}
}

@article{read_stellar_2017,
  title = {The Stellar Mass--Halo Mass Relation of Isolated Field Dwarfs: A Critical Test of {{$\Lambda$CDM}} at the Edge of Galaxy Formation},
  shorttitle = {The Stellar Mass--Halo Mass Relation of Isolated Field Dwarfs},
  author = {Read, J. I. and Iorio, G. and Agertz, O. and Fraternali, F.},
  year = 2017,
  month = may,
  journal = {MNRAS},
  volume = {467},
  number = {2},
  pages = {2019--2038},
  doi = {10.1093/mnras/stx147}
}

@article{rhee_phase-space_2017,
  title = {Phase-Space {{Analysis}} in the {{Group}} and {{Cluster Environment}}: {{Time Since Infall}} and {{Tidal Mass Loss}}},
  shorttitle = {Phase-Space {{Analysis}} in the {{Group}} and {{Cluster Environment}}},
  author = {Rhee, Jinsu and Smith, Rory and Choi, Hoseung and Yi, Sukyoung K. and Jaff{\'e}, Yara and Candlish, Graeme and {S{\'a}nchez-J{\'a}nssen}, Ruben},
  year = 2017,
  month = jul,
  journal = {ApJ},
  volume = {843},
  number = {2},
  pages = {128},
  publisher = {The American Astronomical Society},
  doi = {10.3847/1538-4357/aa6d6c},
  langid = {english}
}

@article{riley_auriga_2025,
  title = {Auriga {{Streams}} -- {{I}}: Disrupting Satellites Surrounding {{Milky Way-mass}} Haloes at Multiple Resolutions},
  shorttitle = {Auriga {{Streams}} -- {{I}}},
  author = {Riley, Alexander H. and Shipp, Nora and Simpson, Christine M. and Bieri, Rebekka and Fattahi, Azadeh and Brown, Shaun T. and Oman, Kyle A. and Fragkoudi, Francesca and G{\'o}mez, Facundo A. and Grand, Robert J. J. and Marinacci, Federico},
  year = 2025,
  month = sep,
  journal = {MNRAS},
  volume = {542},
  pages = {2443--2463},
  publisher = {OUP},
  doi = {10.1093/mnras/staf1350}
}

@article{robotham_galaxy_2012,
  title = {Galaxy {{And Mass Assembly}} ({{GAMA}}): In Search of {{Milky Way Magellanic Cloud}} Analogues},
  shorttitle = {Galaxy {{And Mass Assembly}} ({{GAMA}})},
  author = {Robotham, A. S. G. and Baldry, I. K. and {Bland-Hawthorn}, J. and Driver, S. P. and Loveday, J. and Norberg, P. and Bauer, A. E. and Bekki, K. and Brough, S. and Brown, M. and Graham, A. and Hopkins, A. M. and Phillipps, S. and Power, C. and Sansom, A. and {Staveley-Smith}, L.},
  year = 2012,
  month = aug,
  journal = {MNRAS},
  volume = {424},
  pages = {1448--1453},
  doi = {10.1111/j.1365-2966.2012.21332.x}
}

@article{rocha_infall_2012,
  title = {Infall Times for {{Milky Way}} Satellites from Their Present-Day Kinematics},
  author = {Rocha, Miguel and Peter, Annika H. G. and Bullock, James},
  year = 2012,
  month = sep,
  journal = {MNRAS},
  volume = {425},
  pages = {231--244},
  doi = {10.1111/j.1365-2966.2012.21432.x}
}

@article{rodriguez-cardoso_agora_2025,
  title = {The {{AGORA High-Resolution Galaxy Simulations Comparison Project}}: {{VII}}. {{Satellite}} Quenching in Zoom-in Simulation of a {{Milky Way-mass}} Halo},
  shorttitle = {The {{AGORA High-Resolution Galaxy Simulations Comparison Project}}},
  author = {{Rodr{\'i}guez-Cardoso}, Ram{\'o}n and {Roca-F{\`a}brega}, Santi and Jung, Minyong and Nguy{\~\^e}n, Th{\d i}nh H. and Kim, Ji-Hoon and Primack, Joel and Agertz, Oscar and Barrow, Kirk S. S. and Gallego, Jesus and Nagamine, Kentaro and Powell, Johnny W. and Revaz, Yves and Vel{\'a}zquez, Hector and Genina, Anna and Kim, Hyeonyong and Lupi, Alessandro and Abel, Tom and Cen, Renyue and Ceverino, Daniel and Dekel, Avishai and Oh, Boon Kiat and Quinn, Thomas R. and {The Agora Collaboration}},
  year = 2025,
  month = jun,
  journal = {A\&A},
  volume = {698},
  pages = {A303},
  publisher = {EDP},
  doi = {10.1051/0004-6361/202453639}
}

@article{salem_ram_2015,
  title = {Ram {{Pressure Stripping}} of the {{Large Magellanic Cloud}}'s {{Disk}} as a {{Probe}} of the {{Milky Way}}'s {{Circumgalactic Medium}}},
  author = {Salem, Munier and Besla, Gurtina and Bryan, Greg and Putman, Mary and {van der Marel}, Roeland P. and Tonnesen, Stephanie},
  year = 2015,
  month = dec,
  journal = {ApJ},
  volume = {815},
  pages = {77},
  doi = {10.1088/0004-637X/815/1/77}
}

@article{sales_baryonic_2022,
  title = {Baryonic Solutions and Challenges for Cosmological Models of Dwarf Galaxies},
  author = {Sales, Laura V. and Wetzel, Andrew and Fattahi, Azadeh},
  year = 2022,
  month = jun,
  journal = {Nature Astronomy},
  volume = {6},
  pages = {897--910},
  doi = {10.1038/s41550-022-01689-w}
}

@article{samuel_extinguishing_2022,
  title = {Extinguishing the {{FIRE}}: Environmental Quenching of Satellite Galaxies around {{Milky Way-mass}} Hosts in Simulations},
  shorttitle = {Extinguishing the {{FIRE}}},
  author = {Samuel, Jenna and Wetzel, Andrew and Santistevan, Isaiah and Tollerud, Erik and Moreno, Jorge and {Boylan-Kolchin}, Michael and Bailin, Jeremy and Pardasani, Bhavya},
  year = 2022,
  month = aug,
  journal = {MNRAS},
  volume = {514},
  pages = {5276--5295},
  doi = {10.1093/mnras/stac1706}
}

@misc{santos-santos_unabridged_2025,
  title = {The Unabridged Satellite Luminosity Function of {{Milky Way-like}} Galaxies in \${{$\Lambda$}}\${{CDM}}: The Contribution of "Orphan" Satellites},
  shorttitle = {The Unabridged Satellite Luminosity Function of {{Milky Way-like}} Galaxies in \${{$\Lambda$}}\${{CDM}}},
  author = {{Santos-Santos}, Isabel and Frenk, Carlos and Navarro, Julio and Cole, Shaun and Helly, John},
  year = 2025,
  month = may,
  number = {arXiv:2410.19475},
  eprint = {2410.19475},
  primaryclass = {astro-ph},
  publisher = {arXiv},
  doi = {10.48550/arXiv.2410.19475},
  archiveprefix = {arXiv}
}

@article{scholte_atomic_2024,
  title = {The Atomic Gas Sequence and Mass-Metallicity Relation from Dwarfs to Massive Galaxies},
  author = {Scholte, Dirk and Saintonge, Am{\'e}lie and Moustakas, John and Catinella, Barbara and Zou, Hu and Dey, Biprateep and Aguilar, J. and Ahlen, S. and Anand, A. and Blum, R. and Brooks, D. and Circosta, C. and Claybaugh, T. and {de la Macorra}, A. and Doel, P. and {Font-Ribera}, A. and F{\"o}rster, P. U. and {Forero-Romero}, J. E. and Gazta{\~n}aga, E. and Gontcho A Gontcho, S. and Juneau, S. and Kehoe, R. and Kisner, T. and Koposov, S. E. and Kremin, A. and Lambert, A. and Landriau, M. and Maraston, C. and Martini, P. and Meisner, A. and Mighty, A. S. and Miquel, R. and Myers, A. D. and Nie, J. and Poppett, C. and Prada, F. and Rezaie, M. and Rossi, G. and Sanchez, E. and Schubnell, M. and Silber, J. and Sprayberry, D. and Siudek, M. and Speranza, F. and Tarl{\'e}, G. and Tojeiro, R. and Weaver, B. A.},
  year = 2024,
  month = dec,
  journal = {MNRAS},
  volume = {535},
  pages = {2341--2356},
  publisher = {OUP},
  doi = {10.1093/mnras/stae2477}
}

@article{schombert_gas_2001,
  title = {Gas {{Mass Fractions}} and the {{Evolution}} of {{Low Surface Brightness Dwarf Galaxies}}},
  author = {Schombert, James M. and McGaugh, Stacy S. and Eder, Jo Ann},
  year = 2001,
  month = may,
  journal = {AJ},
  volume = {121},
  pages = {2420--2430},
  doi = {10.1086/320398}
}

@article{serra_meerkat_2023,
  title = {The {{MeerKAT Fornax Survey}}. {{I}}. {{Survey}} Description and First Evidence of Ram Pressure in the {{Fornax}} Galaxy Cluster},
  author = {Serra, P. and Maccagni, F. M. and Kleiner, D. and Moln{\'a}r, D. and Ramatsoku, M. and Loni, A. and Loi, F. and {de Blok}, W. J. G. and Bryan, G. L. and Dettmar, R. J. and Frank, B. S. and {van Gorkom}, J. H. and Govoni, F. and Iodice, E. and J{\'o}zsa, G. I. G. and Kamphuis, P. and {Kraan-Korteweg}, R. and Loubser, S. I. and Murgia, M. and Oosterloo, T. A. and Peletier, R. and Pisano, D. J. and Smith, M. W. L. and Trager, S. C. and Verheijen, M. A. W.},
  year = 2023,
  month = may,
  journal = {A\&A},
  volume = {673},
  pages = {A146},
  doi = {10.1051/0004-6361/202346071}
}

@article{serra_sofia_2015,
  title = {{{SoFiA}}: A Flexible Source Finder for {{3D}} Spectral Line Data},
  shorttitle = {{{SoFiA}}},
  author = {Serra, Paolo and Westmeier, Tobias and Giese, Nadine and Jurek, Russell and Fl{\"o}er, Lars and Popping, Attila and Winkel, Benjamin and {van der Hulst}, Thijs and Meyer, Martin and Koribalski, B{\"a}rbel S. and {Staveley-Smith}, Lister and Courtois, H{\'e}l{\`e}ne},
  year = 2015,
  month = apr,
  journal = {MNRAS},
  volume = {448},
  number = {2},
  pages = {1922--1929},
  doi = {10.1093/mnras/stv079}
}

@article{serra_using_2012,
  title = {Using {{Negative Detections}} to {{Estimate Source-Finder Reliability}}},
  author = {Serra, P. and Jurek, R. and Fl{\"o}er, L.},
  year = 2012,
  month = feb,
  journal = {PASA},
  volume = {29},
  pages = {296--300},
  doi = {10.1071/AS11065}
}

@article{shappee_young_2016,
  title = {The {{Young}} and {{Bright Type Ia Supernova ASASSN-14lp}}: {{Discovery}}, {{Early-time Observations}}, {{First-light Time}}, {{Distance}} to {{NGC}} 4666, and {{Progenitor Constraints}}},
  shorttitle = {The {{Young}} and {{Bright Type Ia Supernova ASASSN-14lp}}},
  author = {Shappee, B. J. and Piro, A. L. and Holoien, T. W. -S. and Prieto, J. L. and Contreras, C. and Itagaki, K. and Burns, C. R. and Kochanek, C. S. and Stanek, K. Z. and Alper, E. and Basu, U. and Beacom, J. F. and Bersier, D. and Brimacombe, J. and Conseil, E. and Danilet, A. B. and Dong, Subo and Falco, E. and Grupe, D. and Hsiao, E. Y. and Kiyota, S. and Morrell, N. and Nicolas, J. and Phillips, M. M. and Pojmanski, G. and Simonian, G. and Stritzinger, M. and Szczygie{\l}, D. M. and Taddia, F. and Thompson, T. A. and Thorstensen, J. and Wagner, M. R. and Wo{\'z}niak, P. R.},
  year = 2016,
  month = aug,
  journal = {ApJ},
  volume = {826},
  pages = {144},
  publisher = {IOP},
  doi = {10.3847/0004-637X/826/2/144}
}

@misc{siljeg_photometry_2024,
  title = {Photometry and Kinematics of Dwarf Galaxies from the {{Apertif HI}} Survey},
  author = {{\v S}iljeg, Barbara and Adams, Elizabeth A. K. and Fraternali, Filippo and Hess, Kelley M. and Oosterloo, Tom A. and Marasco, Antonino and Adebahr, Bj{\"o}rn and D{\'e}nes, Helga and Lucero, Danielle M. and Pi{\~n}a, Pavel E. Mancera and Moss, Vanessa A. and Ponomareva, Anastasia A. and {van der Hulst}, J. M.},
  year = 2024,
  month = sep,
  number = {arXiv:2409.18825},
  eprint = {2409.18825},
  primaryclass = {astro-ph},
  publisher = {arXiv},
  doi = {10.48550/arXiv.2409.18825},
  archiveprefix = {arXiv}
}

@article{simons_figuring_2020,
  title = {Figuring {{Out Gas}} \& {{Galaxies}} in {{Enzo}} ({{FOGGIE}}). {{IV}}. {{The Stochasticity}} of {{Ram Pressure Stripping}} in {{Galactic Halos}}},
  author = {Simons, Raymond C. and Peeples, Molly S. and Tumlinson, Jason and O'Shea, Brian W. and Smith, Britton D. and Corlies, Lauren and Lochhaas, Cassandra and Zheng, Yong and Augustin, Ramona and Prasad, Deovrat and Snyder, Gregory F. and Tollerud, Erik},
  year = 2020,
  month = dec,
  journal = {ApJ},
  volume = {905},
  pages = {167},
  doi = {10.3847/1538-4357/abc5b8}
}

@article{simpson_quenching_2018,
  title = {Quenching and Ram Pressure Stripping of Simulated {{Milky Way}} Satellite Galaxies},
  author = {Simpson, Christine M. and Grand, Robert J. J. and G{\'o}mez, Facundo A. and Marinacci, Federico and Pakmor, R{\"u}diger and Springel, Volker and Campbell, David J. R. and Frenk, Carlos S.},
  year = 2018,
  month = jul,
  journal = {MNRAS},
  volume = {478},
  pages = {548--567},
  publisher = {OUP},
  doi = {10.1093/mnras/sty774}
}

@article{smercina_lonely_2018,
  title = {A {{Lonely Giant}}: {{The Sparse Satellite Population}} of {{M94 Challenges Galaxy Formation}}},
  shorttitle = {A {{Lonely Giant}}},
  author = {Smercina, Adam and Bell, Eric F. and Price, Paul A. and D'Souza, Richard and Slater, Colin T. and Bailin, Jeremy and Monachesi, Antonela and Nidever, David},
  year = 2018,
  month = aug,
  journal = {ApJ},
  volume = {863},
  pages = {152},
  publisher = {IOP},
  doi = {10.3847/1538-4357/aad2d6}
}

@article{smercina_relating_2022,
  title = {Relating the {{Diverse Merger Histories}} and {{Satellite Populations}} of {{Nearby Galaxies}}},
  author = {Smercina, Adam and Bell, Eric F. and Samuel, Jenna and D'Souza, Richard},
  year = 2022,
  month = may,
  journal = {ApJ},
  volume = {930},
  pages = {69},
  doi = {10.3847/1538-4357/ac5d56}
}

@article{smercina_saga_2020,
  title = {The {{Saga}} of {{M81}}: {{Global View}} of a {{Massive Stellar Halo}} in {{Formation}}},
  shorttitle = {The {{Saga}} of {{M81}}},
  author = {Smercina, Adam and Bell, Eric F. and Price, Paul A. and Slater, Colin T. and D'Souza, Richard and Bailin, Jeremy and {de Jong}, Roelof S. and Jang, In Sung and Monachesi, Antonela and Nidever, David},
  year = 2020,
  month = dec,
  journal = {ApJ},
  volume = {905},
  pages = {60},
  doi = {10.3847/1538-4357/abc485}
}

@article{somerville_can_2002,
  title = {Can {{Photoionization Squelching Resolve}} the {{Substructure Crisis}}?},
  author = {Somerville, Rachel S.},
  year = 2002,
  month = jun,
  journal = {ApJ},
  volume = {572},
  pages = {L23-L26},
  doi = {10.1086/341444}
}

@article{somerville_physical_2015,
  title = {Physical {{Models}} of {{Galaxy Formation}} in a {{Cosmological Framework}}},
  author = {Somerville, Rachel S. and Dav{\'e}, Romeel},
  year = 2015,
  month = aug,
  journal = {ARA\&A},
  volume = {53},
  pages = {51--113},
  doi = {10.1146/annurev-astro-082812-140951}
}

@article{spekkens_dearth_2014,
  title = {{{THE DEARTH OF NEUTRAL HYDROGEN IN GALACTIC DWARF SPHEROIDAL GALAXIES}}},
  author = {Spekkens, Kristine and Urbancic, Natasha and Mason, Brian S. and Willman, Beth and Aguirre, James E.},
  year = 2014,
  month = oct,
  journal = {ApJL},
  volume = {795},
  number = {1},
  pages = {L5},
  publisher = {The American Astronomical Society},
  doi = {10.1088/2041-8205/795/1/L5},
  langid = {english}
}

@article{spencer_survey_2014,
  title = {A {{Survey}} of {{Satellite Galaxies}} around {{NGC}} 4258},
  author = {Spencer, Meghin and Loebman, Sarah and Yoachim, Peter},
  year = 2014,
  month = jun,
  journal = {ApJ},
  volume = {788},
  pages = {146},
  doi = {10.1088/0004-637X/788/2/146}
}

@article{tamm_stellar_2012,
  title = {Stellar Mass Map and Dark Matter Distribution in {{M}} 31},
  author = {Tamm, A. and Tempel, E. and Tenjes, P. and Tihhonova, O. and Tuvikene, T.},
  year = 2012,
  month = oct,
  journal = {A\&A},
  volume = {546},
  pages = {A4},
  publisher = {EDP Sciences},
  doi = {10.1051/0004-6361/201220065},
  copyright = {\copyright{} ESO, 2012},
  langid = {english}
}

@article{teich_shield_2016,
  title = {{{SHIELD}}: {{COMPARING GAS AND STAR FORMATION IN LOW-MASS GALAXIES}}},
  shorttitle = {{{SHIELD}}},
  author = {Teich, Yaron G. and McNichols, Andrew T. and Nims, Elise and Cannon, John M. and Adams, Elizabeth A. K. and Giovanelli, Riccardo and Haynes, Martha P. and McQuinn, Kristen B. W. and Salzer, John J. and Skillman, Evan D. and {Bernstein-Cooper}, Elijah Z. and Dolphin, Andrew and Elson, E. C. and Haurberg, Nathalie and J{\'o}zsa, Gyula I. G. and Ott, J{\"u}rgen and Saintonge, Amelie and Warren, Steven R. and Cave, Ian and Hagen, Cedric and Huang, Shan and Janowiecki, Steven and Marshall, Melissa V. and Thomann, Clara M. and Sistine, Angela Van},
  year = 2016,
  month = nov,
  journal = {ApJ},
  volume = {832},
  number = {1},
  pages = {85},
  publisher = {The American Astronomical Society},
  doi = {10.3847/0004-637X/832/1/85},
  langid = {english}
}

@article{tollerud_small-scale_2011,
  title = {{{SMALL-SCALE STRUCTURE IN THE SLOAN DIGITAL SKY SURVEY AND $\Lambda$CDM}}: {{ISOLATED}} {$\sim$}{{L}}* {{GALAXIES WITH BRIGHT SA}}TEL{{LITES}}},
  shorttitle = {{{SMALL-SCALE STRUCTURE IN THE SLOAN DIGITAL SKY SURVEY AND $\Lambda$CDM}}},
  author = {Tollerud, Erik J. and {Boylan-Kolchin}, Michael and Barton, Elizabeth J. and Bullock, James S. and Trinh, Christopher Q.},
  year = 2011,
  month = aug,
  journal = {ApJ},
  volume = {738},
  number = {1},
  pages = {102},
  publisher = {The American Astronomical Society},
  doi = {10.1088/0004-637X/738/1/102},
  langid = {english}
}

@article{trentham_dwarf_2009,
  title = {Dwarf Galaxies in the {{NGC}} 1023 {{Group}}},
  author = {Trentham, Neil and Tully, R. Brent},
  year = 2009,
  month = sep,
  journal = {MNRAS},
  volume = {398},
  number = {2},
  pages = {722--734},
  doi = {10.1111/j.1365-2966.2009.15189.x}
}

@article{tully_cosmicflows-2_2013,
  title = {{{COSMICFLOWS-2}}: {{THE DATA}}},
  shorttitle = {{{COSMICFLOWS-2}}},
  author = {Tully, R. Brent and Courtois, H{\'e}l{\`e}ne M. and Dolphin, Andrew E. and Fisher, J. Richard and H{\'e}raudeau, Philippe and Jacobs, Bradley A. and Karachentsev, Igor D. and Makarov, Dmitry and Makarova, Lidia and Mitronova, Sofia and Rizzi, Luca and Shaya, Edward J. and Sorce, Jenny G. and Wu, Po-Feng},
  year = 2013,
  month = sep,
  journal = {AJ},
  volume = {146},
  number = {4},
  pages = {86},
  publisher = {The American Astronomical Society},
  doi = {10.1088/0004-6256/146/4/86},
  langid = {english}
}

@article{tully_cosmicflows-3_2016,
  title = {Cosmicflows-3},
  author = {Tully, R. Brent and Courtois, H{\'e}l{\`e}ne M. and Sorce, Jenny G.},
  year = 2016,
  month = aug,
  journal = {AJ},
  volume = {152},
  pages = {50},
  doi = {10.3847/0004-6256/152/2/50}
}

@article{tully_extragalactic_2009,
  title = {The {{Extragalactic Distance Database}}},
  author = {Tully, R. Brent and Rizzi, Luca and Shaya, Edward J. and Courtois, H{\'e}l{\`e}ne M. and Makarov, Dmitry I. and Jacobs, Bradley A.},
  year = 2009,
  month = aug,
  journal = {AJ},
  volume = {138},
  pages = {323--331},
  publisher = {IOP},
  doi = {10.1088/0004-6256/138/2/323}
}

@article{tully_new_1977,
  title = {A New Method of Determining Distances to Galaxies.},
  author = {Tully, R. B. and Fisher, J. R.},
  year = 1977,
  month = feb,
  journal = {A\&A},
  volume = {54},
  pages = {661--673},
  publisher = {EDP}
}

@article{tumlinson_circumgalactic_2017,
  title = {The {{Circumgalactic Medium}}},
  author = {Tumlinson, Jason and Peeples, Molly S. and Werk, Jessica K.},
  year = 2017,
  month = aug,
  journal = {ARA\&A},
  volume = {55},
  pages = {389--432},
  doi = {10.1146/annurev-astro-091916-055240}
}

@article{van_sistine_alfalfa_2016,
  title = {{{THE ALFALFA H$\alpha$ SURVEY}}. {{I}}. {{PROJECT DESCRIPTION AND THE LOCAL STAR FORMATION RATE DENSITY FROM THE FALL SAMPLE}}},
  author = {Van Sistine, Angela and Salzer, John J. and Sugden, Arthur and Giovanelli, Riccardo and Haynes, Martha P. and Janowiecki, Steven and Jaskot, Anne E. and Wilcots, Eric M.},
  year = 2016,
  month = jun,
  journal = {ApJ},
  volume = {824},
  number = {1},
  pages = {25},
  publisher = {The American Astronomical Society},
  doi = {10.3847/0004-637X/824/1/25},
  langid = {english}
}

@article{walter_superwind_2004,
  title = {The {{Superwind Galaxy NGC}} 4666: {{Gravitational Interactions}} and the {{Influence}} of the {{Resulting Starburst}} on the {{Interstellar Medium}}},
  shorttitle = {The {{Superwind Galaxy NGC}} 4666},
  author = {Walter, Fabian and Dahlem, Michael and Lisenfeld, Ute},
  year = 2004,
  month = may,
  journal = {ApJ},
  volume = {606},
  pages = {258--270},
  publisher = {IOP},
  doi = {10.1086/382774}
}

@article{wang_saga_2024,
  title = {The {{SAGA Survey}}. {{V}}. {{Modeling Satellite Systems}} around {{Milky Way}}--{{Mass Galaxies}} with {{Updated UNIVERSEMACHINE}}},
  author = {Wang, Yunchong and Nadler, Ethan O. and Mao, Yao-Yuan and Wechsler, Risa H. and Abel, Tom and Behroozi, Peter and Geha, Marla and Asali, Yasmeen and {de los Reyes}, Mithi A. C. and {Kado-Fong}, Erin and Kallivayalil, Nitya and Tollerud, Erik J. and Weiner, Benjamin and Wu, John F.},
  year = 2024,
  month = nov,
  journal = {ApJ},
  volume = {976},
  pages = {119},
  publisher = {IOP},
  doi = {10.3847/1538-4357/ad7f4c}
}

@article{weisz_acs_2011,
  title = {{{THE ACS NEARBY GALAXY SURVEY TREASURY}}. {{VIII}}. {{THE GLOBAL STAR FORMATION HISTORIES OF}} 60 {{DWARF GALAXIES IN THE LOCAL VOLUME}}*},
  author = {Weisz, Daniel R. and Dalcanton, Julianne J. and Williams, Benjamin F. and Gilbert, Karoline M. and Skillman, Evan D. and Seth, Anil C. and Dolphin, Andrew E. and McQuinn, Kristen B. W. and Gogarten, Stephanie M. and Holtzman, Jon and Rosema, Keith and Cole, Andrew and Karachentsev, Igor D. and Zaritsky, Dennis},
  year = 2011,
  month = aug,
  journal = {ApJ},
  volume = {739},
  number = {1},
  pages = {5},
  publisher = {The American Astronomical Society},
  doi = {10.1088/0004-637X/739/1/5},
  langid = {english}
}

@article{weisz_modeling_2011,
  title = {{{MODELING THE EFFECTS OF STAR FORMATION HISTORIES ON H$\alpha$ AND ULTRAVIOLET FLUXES IN NEARBY DWARF GALAXIES}}*},
  author = {Weisz, Daniel R. and Johnson, Benjamin D. and Johnson, L. Clifton and Skillman, Evan D. and Lee, Janice C. and Kennicutt, Robert C. and Calzetti, Daniela and {van Zee}, Liese and Bothwell, Matthew S. and Dalcanton, Julianne J. and Dale, Daniel A. and Williams, Benjamin F.},
  year = 2011,
  month = dec,
  journal = {ApJ},
  volume = {744},
  number = {1},
  pages = {44},
  publisher = {The American Astronomical Society},
  doi = {10.1088/0004-637X/744/1/44},
  langid = {english}
}

@article{weisz_star_2014-1,
  title = {The {{Star Formation Histories}} of {{Local Group Dwarf Galaxies}}. {{I}}. {{Hubble Space Telescope}}/{{Wide Field Planetary Camera}} 2 {{Observations}}},
  author = {Weisz, Daniel R. and Dolphin, Andrew E. and Skillman, Evan D. and Holtzman, Jon and Gilbert, Karoline M. and Dalcanton, Julianne J. and Williams, Benjamin F.},
  year = 2014,
  month = jul,
  journal = {ApJ},
  volume = {789},
  pages = {147},
  doi = {10.1088/0004-637X/789/2/147}
}

@article{westmeier_sofia_2021,
  title = {{{SOFIA}} 2 - an Automated, Parallel {{H I}} Source Finding Pipeline for the {{WALLABY}} Survey},
  author = {Westmeier, T. and Kitaeff, S. and Pallot, D. and Serra, P. and {van der Hulst}, J. M. and Jurek, R. J. and Elagali, A. and For, B. -Q. and Kleiner, D. and Koribalski, B. S. and {Lee-Waddell}, K. and Mould, J. R. and Reynolds, T. N. and Rhee, J. and {Staveley-Smith}, L.},
  year = 2021,
  month = sep,
  journal = {MNRAS},
  volume = {506},
  pages = {3962--3976},
  doi = {10.1093/mnras/stab1881}
}

@article{westmeier_wallaby_2022,
  title = {{{WALLABY}} Pilot Survey: {{Public}} Release of {{H I}} Data for Almost 600 Galaxies from Phase 1 of {{ASKAP}} Pilot Observations},
  shorttitle = {{{WALLABY}} Pilot Survey},
  author = {Westmeier, T. and Deg, N. and Spekkens, K. and Reynolds, T. N. and Shen, A. X. and Gaudet, S. and Goliath, S. and Huynh, M. T. and Venkataraman, P. and Lin, X. and O'Beirne, T. and Catinella, B. and Cortese, L. and D{\'e}nes, H. and Elagali, A. and For, B. -Q. and J{\'o}zsa, G. I. G. and Howlett, C. and {van der Hulst}, J. M. and Jurek, R. J. and Kamphuis, P. and Kilborn, V. A. and Kleiner, D. and Koribalski, B. S. and {Lee-Waddell}, K. and Murugeshan, C. and Rhee, J. and Serra, P. and Shao, L. and {Staveley-Smith}, L. and Wang, J. and Wong, O. I. and Zwaan, M. A. and Allison, J. R. and Anderson, C. S. and Ball, Lewis and Bock, D. C. -J. and Brodrick, D. and Bunton, J. D. and Cooray, F. R. and Gupta, N. and Hayman, D. B. and Mahony, E. K. and Moss, V. A. and Ng, A. and Pearce, S. E. and Raja, W. and Roxby, D. N. and Voronkov, M. A. and Warhurst, K. A. and Courtois, H. M. and Said, K.},
  year = 2022,
  month = nov,
  journal = {PASA},
  volume = {39},
  pages = {e058},
  doi = {10.1017/pasa.2022.50}
}

@article{wetzel_orbits_2011,
  title = {On the Orbits of Infalling Satellite Haloes},
  author = {Wetzel, Andrew R.},
  year = 2011,
  month = mar,
  journal = {MNRAS},
  volume = {412},
  pages = {49--58},
  doi = {10.1111/j.1365-2966.2010.17877.x}
}

@article{wetzel_rapid_2015,
  title = {{{RAPID ENVIRONMENTAL QUENCHING OF SA}}TEL{{LITE DWARF GALAXIES IN THE LOCAL GROUP}}},
  author = {Wetzel, Andrew R. and Tollerud, Erik J. and Weisz, Daniel R.},
  year = 2015,
  month = jul,
  journal = {ApJL},
  volume = {808},
  number = {1},
  pages = {L27},
  publisher = {The American Astronomical Society},
  doi = {10.1088/2041-8205/808/1/L27},
  langid = {english}
}

@article{white_galaxy_1991,
  title = {Galaxy {{Formation}} through {{Hierarchical Clustering}}},
  author = {White, Simon D. M. and Frenk, Carlos S.},
  year = 1991,
  month = sep,
  journal = {ApJ},
  volume = {379},
  pages = {52},
  publisher = {IOP},
  doi = {10.1086/170483}
}

@article{white_mass_2001,
  title = {The Mass of a Halo},
  author = {White, M.},
  year = 2001,
  month = feb,
  journal = {A\&A},
  volume = {367},
  number = {1},
  pages = {27--32},
  publisher = {EDP Sciences},
  doi = {10.1051/0004-6361:20000357},
  copyright = {\copyright{} ESO, 2001},
  langid = {english}
}

@article{wright_baryon_2024,
  title = {The Baryon Cycle in Modern Cosmological Hydrodynamical Simulations},
  author = {Wright, Ruby J and Somerville, Rachel S and Lagos, Claudia del P and Schaller, Matthieu and Dav{\'e}, Romeel and {Angl{\'e}s-Alc{\'a}zar}, Daniel and Genel, Shy},
  year = 2024,
  month = aug,
  journal = {MNRAS},
  volume = {532},
  number = {3},
  pages = {3417--3440},
  doi = {10.1093/mnras/stae1688}
}

@article{young_neutral_1997,
  title = {The {{Neutral Interstellar Medium}} in {{Nearby Dwarf Galaxies}}. {{II}}. {{NGC}} 185, {{NGC}} 205, and {{NGC}} 147},
  author = {Young, L. M. and Lo, K. Y.},
  year = 1997,
  month = feb,
  journal = {ApJ},
  volume = {476},
  number = {1},
  pages = {127},
  publisher = {IOP Publishing},
  doi = {10.1086/303618},
  langid = {english}
}

@article{zheng_comprehensive_2024,
  title = {A {{Comprehensive Investigation}} of {{Metals}} in the {{Circumgalactic Medium}} of {{Nearby Dwarf Galaxies}}},
  author = {Zheng, Yong and Faerman, Yakov and Oppenheimer, Benjamin D. and Putman, Mary E. and McQuinn, Kristen B. W. and Kirby, Evan N. and Burchett, Joseph N. and Telford, O. Grace and Werk, Jessica K. and Kim, Doyeon A.},
  year = 2024,
  month = jan,
  journal = {ApJ},
  volume = {960},
  pages = {55},
  doi = {10.3847/1538-4357/acfe6b}
}

@article{zhu_census_2023,
  title = {Census of Gaseous Satellites around Local Spiral Galaxies},
  author = {Zhu, Jingyao and Putman, Mary E.},
  year = 2023,
  month = may,
  journal = {MNRAS},
  volume = {521},
  pages = {3765--3783},
  doi = {10.1093/mnras/stad695}
}

@article{zhu_its_2024,
  title = {It's a {{Breeze}}: {{The Circumgalactic Medium}} of a {{Dwarf Galaxy Is Easy}} to {{Strip}}},
  shorttitle = {It's a {{Breeze}}},
  author = {Zhu, Jingyao and Tonnesen, Stephanie and Bryan, Greg L. and Putman, Mary E.},
  year = 2024,
  month = oct,
  journal = {ApJ},
  volume = {974},
  number = {1},
  pages = {142},
  publisher = {The American Astronomical Society},
  doi = {10.3847/1538-4357/ad6c3f},
  langid = {english}
}

@article{zhu_too_2026,
  title = {Too {{Big}} to {{Quench}}? {{I}}. {{Constraining Interstellar Medium Stripping}} of {{Dwarf Satellites}} in {{Milky Way}}--like {{Halos}}},
  shorttitle = {Too {{Big}} to {{Quench}}?},
  author = {Zhu, Jingyao and Tonnesen, Stephanie and Bryan, Greg L. and Putman, Mary E.},
  year = 2026,
  month = aug,
  journal = {\apj},
  volume = {1007},
  number = {2},
  pages = {158},
  publisher = {The American Astronomical Society},
  doi = {10.3847/1538-4357/ae8777},
  langid = {english}
}

@article{zhu_when_2024,
  title = {When and {{How Ram Pressure Stripping}} in {{Low-mass Satellite Galaxies Enhances Star Formation}}},
  author = {Zhu, Jingyao and Tonnesen, Stephanie and Bryan, Greg L.},
  year = 2024,
  month = jan,
  journal = {ApJ},
  volume = {960},
  number = {1},
  pages = {54},
  doi = {10.3847/1538-4357/acfe6f},
  langid = {english}
}

@ARTICLE{dey_2019AJ....157..168D,
       author = {{Dey}, Arjun and {Schlegel}, David J. and {Lang}, Dustin and {Blum}, Robert and {Burleigh}, Kaylan and {Fan}, Xiaohui and {Findlay}, Joseph R. and {Finkbeiner}, Doug and {Herrera}, David and {Juneau}, St{\'e}phanie and {Landriau}, Martin and {Levi}, Michael and {McGreer}, Ian and {Meisner}, Aaron and {Myers}, Adam D. and {Moustakas}, John and {Nugent}, Peter and {Patej}, Anna and {Schlafly}, Edward F. and {Walker}, Alistair R. and {Valdes}, Francisco and {Weaver}, Benjamin A. and {Y{\`e}che}, Christophe and {Zou}, Hu and {Zhou}, Xu and {Abareshi}, Behzad and {Abbott}, T.~M.~C. and {Abolfathi}, Bela and {Aguilera}, C. and {Alam}, Shadab and {Allen}, Lori and {Alvarez}, A. and {Annis}, James and {Ansarinejad}, Behzad and {Aubert}, Marie and {Beechert}, Jacqueline and {Bell}, Eric F. and {BenZvi}, Segev Y. and {Beutler}, Florian and {Bielby}, Richard M. and {Bolton}, Adam S. and {Brice{\~n}o}, C{\'e}sar and {Buckley-Geer}, Elizabeth J. and {Butler}, Karen and {Calamida}, Annalisa and {Carlberg}, Raymond G. and {Carter}, Paul and {Casas}, Ricard and {Castander}, Francisco J. and {Choi}, Yumi and {Comparat}, Johan and {Cukanovaite}, Elena and {Delubac}, Timoth{\'e}e and {DeVries}, Kaitlin and {Dey}, Sharmila and {Dhungana}, Govinda and {Dickinson}, Mark and {Ding}, Zhejie and {Donaldson}, John B. and {Duan}, Yutong and {Duckworth}, Christopher J. and {Eftekharzadeh}, Sarah and {Eisenstein}, Daniel J. and {Etourneau}, Thomas and {Fagrelius}, Parker A. and {Farihi}, Jay and {Fitzpatrick}, Mike and {Font-Ribera}, Andreu and {Fulmer}, Leah and {G{\"a}nsicke}, Boris T. and {Gaztanaga}, Enrique and {George}, Koshy and {Gerdes}, David W. and {Gontcho}, Satya Gontcho A. and {Gorgoni}, Claudio and {Green}, Gregory and {Guy}, Julien and {Harmer}, Diane and {Hernandez}, M. and {Honscheid}, Klaus and {Huang}, Lijuan Wendy and {James}, David J. and {Jannuzi}, Buell T. and {Jiang}, Linhua and {Joyce}, Richard and {Karcher}, Armin and {Karkar}, Sonia and {Kehoe}, Robert and {Kneib}, Jean-Paul and {Kueter-Young}, Andrea and {Lan}, Ting-Wen and {Lauer}, Tod R. and {Le Guillou}, Laurent and {Le Van Suu}, Auguste and {Lee}, Jae Hyeon and {Lesser}, Michael and {Perreault Levasseur}, Laurence and {Li}, Ting S. and {Mann}, Justin L. and {Marshall}, Robert and {Mart{\'\i}nez-V{\'a}zquez}, C.~E. and {Martini}, Paul and {du Mas des Bourboux}, H{\'e}lion and {McManus}, Sean and {Meier}, Tobias Gabriel and {M{\'e}nard}, Brice and {Metcalfe}, Nigel and {Mu{\~n}oz-Guti{\'e}rrez}, Andrea and {Najita}, Joan and {Napier}, Kevin and {Narayan}, Gautham and {Newman}, Jeffrey A. and {Nie}, Jundan and {Nord}, Brian and {Norman}, Dara J. and {Olsen}, Knut A.~G. and {Paat}, Anthony and {Palanque-Delabrouille}, Nathalie and {Peng}, Xiyan and {Poppett}, Claire L. and {Poremba}, Megan R. and {Prakash}, Abhishek and {Rabinowitz}, David and {Raichoor}, Anand and {Rezaie}, Mehdi and {Robertson}, A.~N. and {Roe}, Natalie A. and {Ross}, Ashley J. and {Ross}, Nicholas P. and {Rudnick}, Gregory and {Safonova}, Sasha and {Saha}, Abhijit and {S{\'a}nchez}, F. Javier and {Savary}, Elodie and {Schweiker}, Heidi and {Scott}, Adam and {Seo}, Hee-Jong and {Shan}, Huanyuan and {Silva}, David R. and {Slepian}, Zachary and {Soto}, Christian and {Sprayberry}, David and {Staten}, Ryan and {Stillman}, Coley M. and {Stupak}, Robert J. and {Summers}, David L. and {Sien Tie}, Suk and {Tirado}, H. and {Vargas-Maga{\~n}a}, Mariana and {Vivas}, A. Katherina and {Wechsler}, Risa H. and {Williams}, Doug and {Yang}, Jinyi and {Yang}, Qian and {Yapici}, Tolga and {Zaritsky}, Dennis and {Zenteno}, A. and {Zhang}, Kai and {Zhang}, Tianmeng and {Zhou}, Rongpu and {Zhou}, Zhimin},
        title = "{Overview of the DESI Legacy Imaging Surveys}",
      journal = {\aj},
         year = 2019,
        month = may,
       volume = {157},
       number = {5},
          eid = {168},
        pages = {168},
          doi = {10.3847/1538-3881/ab089d},
archivePrefix = {arXiv},
       eprint = {1804.08657},
 primaryClass = {astro-ph.IM},
       adsurl = {https://ui.adsabs.harvard.edu/abs/2019AJ....157..168D}
}

@misc{lang_2016ascl.soft04008L,
       author = {{Lang}, Dustin and {Hogg}, David W. and {Mykytyn}, David},
        title = "{The Tractor: Probabilistic astronomical source detection and measurement}",
 howpublished = {Astrophysics Source Code Library, record ascl:1604.008},
         year = 2016,
        month = apr,
          eid = {ascl:1604.008},
       adsurl = {https://ui.adsabs.harvard.edu/abs/2016ascl.soft04008L}
}

@article{schlegel98,
 adsurl = {https://ui.adsabs.harvard.edu/abs/1998ApJ...500..525S},
 archiveprefix = {arXiv},
 author = {{Schlegel}, David J. and {Finkbeiner}, Douglas P. and {Davis}, Marc},
 doi = {10.1086/305772},
 eprint = {astro-ph/9710327},
 journal = {\apj},
 month = {June},
 number = {2},
 pages = {525-553},
 primaryclass = {astro-ph},
 title = {{Maps of Dust Infrared Emission for Use in Estimation of Reddening and Cosmic Microwave Background Radiation Foregrounds}},
 volume = {500},
 year = {1998}
}

@article{Schlafly2011,
 adsurl = {https://ui.adsabs.harvard.edu/abs/2011ApJ...737..103S},
 archiveprefix = {arXiv},
 author = {{Schlafly}, Edward F. and {Finkbeiner}, Douglas P.},
 doi = {10.1088/0004-637X/737/2/103},
 eid = {103},
 eprint = {1012.4804},
 journal = {\apj},
 month = {August},
 number = {2},
 pages = {103},
 primaryclass = {astro-ph.GA},
 title = {{Measuring Reddening with Sloan Digital Sky Survey Stellar Spectra and Recalibrating SFD}},
 volume = {737},
 year = {2011}
}

@ARTICLE{pasha_2023JOSS....8.5703P,
       author = {{Pasha}, Imad and {Miller}, Tim B.},
        title = "{pysersic: A Python package for determining galaxy structural properties via Bayesian inference, accelerated with jax}",
      journal = {The Journal of Open Source Software},
         year = 2023,
        month = sep,
       volume = {8},
       number = {89},
          eid = {5703},
        pages = {5703},
          doi = {10.21105/joss.05703},
archivePrefix = {arXiv},
       eprint = {2306.05454},
 primaryClass = {astro-ph.GA},
       adsurl = {https://ui.adsabs.harvard.edu/abs/2023JOSS....8.5703P}
}

@ARTICLE{bell_2001ApJ...550..212B,
       author = {{Bell}, Eric F. and {de Jong}, Roelof S.},
        title = "{Stellar Mass-to-Light Ratios and the Tully-Fisher Relation}",
      journal = {\apj},
         year = 2001,
        month = mar,
       volume = {550},
       number = {1},
        pages = {212-229},
          doi = {10.1086/319728},
archivePrefix = {arXiv},
       eprint = {astro-ph/0011493},
 primaryClass = {astro-ph},
       adsurl = {https://ui.adsabs.harvard.edu/abs/2001ApJ...550..212B}
}

@ARTICLE{bell_2003ApJS..149..289B,
       author = {{Bell}, Eric F. and {McIntosh}, Daniel H. and {Katz}, Neal and {Weinberg}, Martin D.},
        title = "{The Optical and Near-Infrared Properties of Galaxies. I. Luminosity and Stellar Mass Functions}",
      journal = {\apjs},
         year = 2003,
        month = dec,
       volume = {149},
       number = {2},
        pages = {289-312},
          doi = {10.1086/378847},
archivePrefix = {arXiv},
       eprint = {astro-ph/0302543},
 primaryClass = {astro-ph},
       adsurl = {https://ui.adsabs.harvard.edu/abs/2003ApJS..149..289B}
}

@ARTICLE{asali_2025arXiv250925335A,
       author = {{Asali}, Yasmeen and {Geha}, Marla and {Kado-Fong}, Erin and {Mao}, Yao-Yuan and {Wechsler}, Risa H. and {de los Reyes}, Mithi A.~C. and {Pasha}, Imad and {Kallivayalil}, Nitya and {Nadler}, Ethan O. and {Tollerud}, Erik J. and {Wang}, Yunchong and {Weiner}, Benjamin and {Wu}, John F.},
        title = "{The SAGA Survey. VI. The Size-Mass Relation for Low-Mass Galaxies Across Environments}",
      journal = {arXiv e-prints},
         year = 2025,
        month = sep,
          eid = {arXiv:2509.25335},
        pages = {arXiv:2509.25335},
          doi = {10.48550/arXiv.2509.25335},
archivePrefix = {arXiv},
       eprint = {2509.25335},
 primaryClass = {astro-ph.GA},
       adsurl = {https://ui.adsabs.harvard.edu/abs/2025arXiv250925335A}
}

@ARTICLE{pacifici_2023ApJ...944..141P,
       author = {{Pacifici}, Camilla and {Iyer}, Kartheik G. and {Mobasher}, Bahram and {da Cunha}, Elisabete and {Acquaviva}, Viviana and {Burgarella}, Denis and {Calistro Rivera}, Gabriela and {Carnall}, Adam C. and {Chang}, Yu-Yen and {Chartab}, Nima and {Cooke}, Kevin C. and {Fairhurst}, Ciaran and {Kartaltepe}, Jeyhan and {Leja}, Joel and {Ma{\l}ek}, Katarzyna and {Salmon}, Brett and {Torelli}, Marianna and {Vidal-Garc{\'\i}a}, Alba and {Boquien}, M{\'e}d{\'e}ric and {Brammer}, Gabriel G. and {Brown}, Michael J.~I. and {Capak}, Peter L. and {Chevallard}, Jacopo and {Circosta}, Chiara and {Croton}, Darren and {Davidzon}, Iary and {Dickinson}, Mark and {Duncan}, Kenneth J. and {Faber}, Sandra M. and {Ferguson}, Harry C. and {Fontana}, Adriano and {Guo}, Yicheng and {Haeussler}, Boris and {Hemmati}, Shoubaneh and {Jafariyazani}, Marziye and {Kassin}, Susan A. and {Larson}, Rebecca L. and {Lee}, Bomee and {Mantha}, Kameswara Bharadwaj and {Marchi}, Francesca and {Nayyeri}, Hooshang and {Newman}, Jeffrey A. and {Pandya}, Viraj and {Pforr}, Janine and {Reddy}, Naveen and {Sanders}, Ryan and {Shah}, Ekta and {Shahidi}, Abtin and {Stevans}, Matthew L. and {Triani}, Dian Puspita and {Tyler}, Krystal D. and {Vanderhoof}, Brittany N. and {de la Vega}, Alexander and {Wang}, Weichen and {Weston}, Madalyn E.},
        title = "{The Art of Measuring Physical Parameters in Galaxies: A Critical Assessment of Spectral Energy Distribution Fitting Techniques}",
      journal = {\apj},
         year = 2023,
        month = feb,
       volume = {944},
       number = {2},
          eid = {141},
        pages = {141},
          doi = {10.3847/1538-4357/acacff},
archivePrefix = {arXiv},
       eprint = {2212.01915},
 primaryClass = {astro-ph.GA},
       adsurl = {https://ui.adsabs.harvard.edu/abs/2023ApJ...944..141P}
}
\bibliographystyle{aasjournal}



\end{document}